\documentclass[12pt,notitlepage,a4paper]{article}

\pdfoutput=1

\usepackage{a4wide}
\usepackage{epsfig}
\usepackage{color,graphicx}
\usepackage{cite}
\usepackage{amssymb,amsmath}
\usepackage{changepage}
\usepackage{multirow}
\usepackage{braket}
\usepackage{youngtab}

\newcommand{\be}{\begin{equation}}
\newcommand{\ee}{\end{equation}}
\newcommand{\bea}{\begin{eqnarray}}
\newcommand{\eea}{\end{eqnarray}}

\begin{document}

\begin{center}  

\vskip 2cm 

\centerline{\Large {\bf Brane webs, $5d$ gauge theories and $6d$ $\mathcal{N}$$=(1,0)$ SCFT's}}
\vskip 0.5cm

\renewcommand{\thefootnote}{\fnsymbol{footnote}}

   \centerline{
    {\large \bf Gabi Zafrir${}^{a}$} \footnote{gabizaf@techunix.technion.ac.il}}

\vspace{1cm}
\centerline{{\it ${}^a$ Department of Physics, Technion, Israel Institute of Technology}} \centerline{{\it Haifa, 32000, Israel}}
\vspace{1cm}

\end{center}

\vskip 0.3 cm

\setcounter{footnote}{0}
\renewcommand{\thefootnote}{\arabic{footnote}}   
   
      \begin{abstract}
     
		We study $5d$ gauge theories that go in the UV to $6d$ $\mathcal{N}$$=(1,0)$ SCFT. We focus on these theories that can be engineered in string theory by brane webs. Given a theory in this class, we propose a method to determine the $6d$ SCFT it goes to. We also discuss the implication of this to the compactification of the resulting $6d$ SCFT on a torus to $4d$. We test and demonstrate this method with a variety of examples.  
		
      \end{abstract}

\tableofcontents

\section{Introduction}

One interesting implication of string theory is the existence of interacting superconformal field theories in $5d$ and $6d$. This is quite surprising as most interaction terms in these dimensions are non-renormalizable. Even more surprising, string theory suggests that these SCFT's are connected to gauge theories in these dimensions providing, in some sense, a UV completion to $5d$ and $6d$ gauge theories that are non-renormalizable. In this article we concentrate on theories with minimal supersymmetry, meaning $8$ supercharges, which is usually denoted as $\mathcal{N}$$=1$ in $5d$ and $\mathcal{N}$$=(1,0)$ in $6d$.

The study of $\mathcal{N}$$=1$ supersymmetric $5d$ gauge theories originated in \cite{SEI,SM,SMI}. These theories can also be realized in string theory using brane webs and geometric engineering\cite{DKV,HA,HAK,DHIK}. The picture emerging from these methods is that $5d$ SCFT's exist and that they sometimes posses mass deformations leading to $5d$ gauge theories, with the mass identified as the inverse gauge coupling squared, $g^{-2}$. These theories also posses some quite interesting non-perturbative behavior. One such phenomenon is the occurrence of enhancement of symmetry, in which the fixed point has a larger global symmetry than that perturbatively exhibited in the gauge theory. An important ingredient in this is the existence of a topological $U(1)$ conserved current, $j_T = *\mbox{Tr}(F\wedge F)$, associated with every non-abelian gauge group. The particles charged under this current are instantons. 

 These instantonic particles sometimes provide additional conserved currents leading to an enhancement of the perturbative global symmetry. A simple example is $SU(2)$ gauge theory with $N_f$ hypermultiplets in the doublet of $SU(2)$. For $N_f\leq 7$, this theory is known to flow to a $5d$ fixed point, where the global symmetry is enhanced from $U(1)\times SO(2N_f)$ to $E_{N_f+1}$ by instantonic particles\cite{SEI}. This can be argued from string theory constructions, and is further supported by the superconformal index\cite{KKL,HKKP}.

In many cases, a single $5d$ SCFT may have many different gauge theory deformations. This is a type of duality in which different IR gauge theories go to the same underlying $5d$ SCFT. An example of this is $SU_0(3)+2F$ gauge theory and $SU_{\pi}(2)\times SU_{\pi}(2)$ quiver theory \cite{HA,BGZ}\footnote{In $5d$ one can add a Chern-Simons (CS) term to any $SU(N)$ gauge theory, for $N>2$, and we use a subscript under the gauge group to denote the CS level. For $USp(2N)$ groups, a CS term is not possible, but there is a discrete $Z_2$ parameter, called the $\theta$ angle, which can be either $0$ or $\pi$\cite{SM}. We again use a subscript under the gauge group to denote it. Also, when denoting gauge theories we use $F$ for matter in the fundamental representation and $AS$ for matter in the antisymmetric representation. When writing quiver theories, we use the notation $G_1\times G_2\times...$ where it is understood that there is a single bifundamental hyper associated with every $\times$.}. By now a great many examples of this are known, see \cite{BPTY,BGZ,Zaf,BZ,BZ1,GC}.

String theory methods, such as brane constructions, also suggest the existence of interacting $6d$ $\mathcal{N}$$=(1,0)$ SCFT's\cite{KB,KB1,HaZ}. These theories include massless tensor multiplets, in addition to hyper and vector multiplets. The tensor multiplets contain a scalar leading to a moduli space of vacua. In some cases, the low energy theory around a generic point in this space is a $6d$ gauge theory, where $g^{-2}$ is identified with the scalar vev\cite{SEI1}. By now, a large number of such SCFT's are known. In fact, there exists a classification of $\mathcal{N}$$=(1,0)$ SCFT's using F-theory\cite{HMV,HMRV}. See also \cite{Bh} for a classification of $\mathcal{N}$$=(1,0)$ gauge theories. 

There is an interesting relationship between $5d$ gauge theories and $6d$ $\mathcal{N}$$=(1,0)$ theories, where, in some cases, a $5d$ gauge theory has a $6d$ $\mathcal{N}$$=(1,0)$ UV completion. The best known example is $5d$ maximally supersymmetric Yang-Mills theory, which is believed to lift to the $6d$ $(2,0)$ theory \cite{Dou,LPS}. Yet another notable example is the $5d$ gauge theory with a $USp(2N)$ gauge group, a hypermultiplet in the antisymmetric representation, and $8$ hypermultiplets in the fundamental representation, which is believed to lift to the $6d$ rank $N$ E-string theory\cite{GMS}. Recently, another example was given in \cite{HKLTY}. There the $6d$ theory in question is known as the $(D_{N+4},D_{N+4})$ conformal matter\cite{ZHTV}, which has a $6d$ gauge theory description as $USp(2N)+(2N+8)F$. This theory is suspected to be the UV completion of the $5d$ gauge theory $SU_0(N+2)+(2N+8)F$. 

The purpose of this paper is to extend these results to a large class of $5d$ gauge theories with an expected $6d$ $\mathcal{N}$$=(1,0)$ SCFT UV completion. We consider theories which can be represented as ordinary $5$-brane webs. The starting point is to generalize the discussion of \cite{HKLTY} to the class of $5d$ gauge theories of the form $(N+2)F+SU_0(N)^k+(N+2)F$. These were recently conjectured to lift to $6d$ SCFT\cite{KTY}. Furthermore, in \cite{Yon} a conjecture for this $6d$ SCFT appeared. We start by generalizing the method of \cite{HKLTY} to give evidence for this conjecture. 

Using this result we then go on to propose a technique to determine the answer for other $5d$ gauge theories, by thinking of them as a limit on the Higgs branch of a $5d$ gauge theory $(N+2)F+SU_0(N)^k+(N+2)F$ for some $N$ and $k$. Then we can determine the $6d$ SCFT by mapping the appropriate limit of the $5d$ Higgs branch to the corresponding one of the $6d$ theory. We consider a variety of examples, exhibiting both the advantages and limitations of this technique.  




As an application of these results, we also consider the compactification on a torus of the $6d$ SCFT's appearing as the lift of $5d$ gauge theories. For example, consider the compactification of the rank $1$ $E$ string theory on a torus, where we take the limit of zero area, keeping the $6d$ global symmetry unbroken. First compactifying to $5d$, we get the $5d$ theory $SU(2)+8F$. We now want to compactify to $4d$ taking the limit of zero torus area, but without breaking the $E_8$ global symmetry. It turns out that the way to do this is by first integrating out a flavor, flowing to $SU(2)+7F$   \footnote{Note that this is a $R_6\rightarrow 0$ limit. This follows as one must keep the effective coupling, which behaves like: $\frac{1}{g^2_0} - constant |m|$, well defined. Therefore, when taking the $m\rightarrow \infty$ limit, one must also take the $R_6 \sim g^2\rightarrow 0$ limit.}. This leads to a $5d$ SCFT with $E_8$ global symmetry\cite{SEI}. Compactifying this to $4d$ then leads to the rank $1$ Minahan-Nemashansky $E_8$ theory\cite{MN}. For additional examples of the compactification of $6d$ $\mathcal{N}$$=(1,0)$ SCFT's on a torus, see \cite{OSTY1,ZVX,OSTY2}.    

\begin{figure}
\center
\includegraphics[width=1\textwidth]{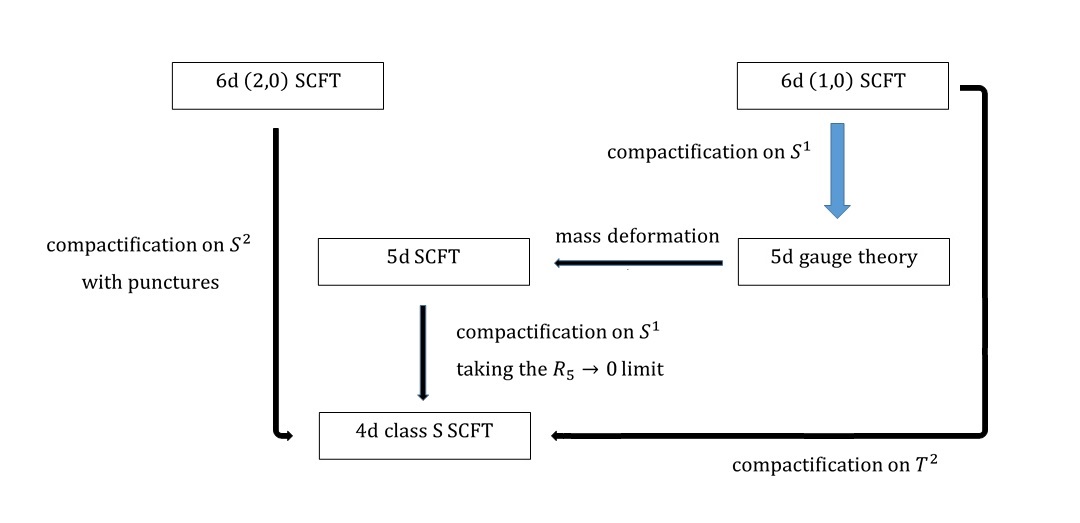} 
\caption{A graphical summary of the main idea of this paper. The major relation we explore is between a $6d$ $(1,0)$ SCFT and a $5d$ gauge theory generated by compactifying the former on a circle of radius $R_6$. This is represented in the figure by the wide blue arrow. We can employ this relationship to study the compactification of the $6d$ $(1,0)$ SCFT to $4d$ on a torus. We first mass deform the $5d$ gauge theory, corresponding to taking the $R_6\rightarrow 0$ limit while keeping the $6d$ global symmetry intact. This leads to a $5d$ SCFT. We then compactify this SCFT on a circle of radius $R_5$, and take $R_5\rightarrow 0$. This leads to a $4d$ class S SCFT, which can in turn be thought of as a result of compactifying a $6d$ $(2,0)$ SCFT on a Riemann sphere with three punctures. We can use this description as a consistency check by calculating the properties of this $4d$ SCFT when thought of as a compactification of a $6d$ $(2,0)$ SCFT, known as class S technology, and comparing against what is expected from the compactification of the $6d$ $(1,0)$ SCFT.}
\label{diagram}
\end{figure}

We can now adopt a similar strategy to understand the result of compactification on a torus of the $6d$ SCFT's we encounter. That is we first compactify to $5d$ leading to the $5d$ gauge theory. Taking the $R_6\rightarrow 0$ limit, while keeping the $6d$ global symmetry, is then implemented by integrating out a flavor. This leads to a $5d$ SCFT with a brane web description of the form of \cite{BB}. It is now straightforward to take the $R_5\rightarrow 0$ limit, leading to a class S isolated SCFT, as shown in \cite{BB}. Thus, we conjecture that reducing the class of $6d$ theories we consider on a torus leads to an isolated $4d$ SCFT. The main idea is summarized graphically in figure \ref{diagram}. 

We next seek to provide evidence for this relation. To this end we use the results of \cite{OSTY1}, who found a way to calculate the central charges of a $4d$ theory resulting from compactification of a $6d$ theory on a torus in terms of the anomaly polynomial of the $6d$ theory. We can now compute the $4d$ central charges first using class S technology (see \cite{CD}), and second from the anomaly polynomial (using \cite{OSTY}), and compare the two. We indeed find that these match. This then provides evidence also for the original $5d-6d$ relation.

The structure of this article is as follows. Section 2 presents some preliminary discussions about the computation of the anomaly polynomial for the $6d$ SCFT's considered in this article, as well as the class S technology we use. In section 3 we consider the $5d$ theory $(N+2)F+SU_0(N)^k+(N+2)F$. We first generalize the methods of \cite{HKLTY} to test the conjecture of \cite{Yon}, and then go on to consider related theories. Section 4 deals with other $5d$ theories expected to lift to $6d$, that are not of the form presented in section 3. We end with some conclusions. The appendix discusses symmetry enhancement for a class of $5d$ theories that play an important role in section 4, and which, to our knowledge, were not previously studied.    

\medskip

{\bf A word on notation:} Brane webs comprise an important part of our analysis and so they appear abundantly in this article. In many cases only the external legs are needed and not how they connect to one another. In these cases, for ease of presentation, we have only depicted the external legs, using a large black oval for the internal part of the diagram. Many of the diagrams also contain repeated parts shown by a sequence of black dots. This should not be confused with $7$-branes.  

  In brane webs one can also add $7$-branes on which the $5$-branes can end. We have in general suppressed the $7$-branes, with the exception of two cases. One, when several $5$-branes end on the same $7$-brane. In this case we depicted the $7$-brane as a black oval, the type of which is understood by the type of $5$-branes ending on it. We in general also write the number of $5$-branes ending on this $7$-brane. If no number is given then it is the number visible in the picture. Any other numbers that appear stand for the number of $5$-branes.
	
	The second case where we explicitly include $7$-branes is if no $5$-branes end on them. In this case we denote a $(1,0)$ $7$-brane by an X and a $(0,1)$ $7$-brane by a square. Any other $7$-brane is denoted by a circle with the type written next to it. 
	
	We generically suppress the monodromy line of the $7$-branes. In the special cases when we do draw it, we use a dashed line.

\section{Preliminaries} 

This section discusses the type of $6d$ theories we encounter, the computation of the anomaly polynomials for these theories, and the class S technology used in this article.

\subsection{Properties of the $6d$ theories}

\begin{figure}
\center
\includegraphics[width=1\textwidth]{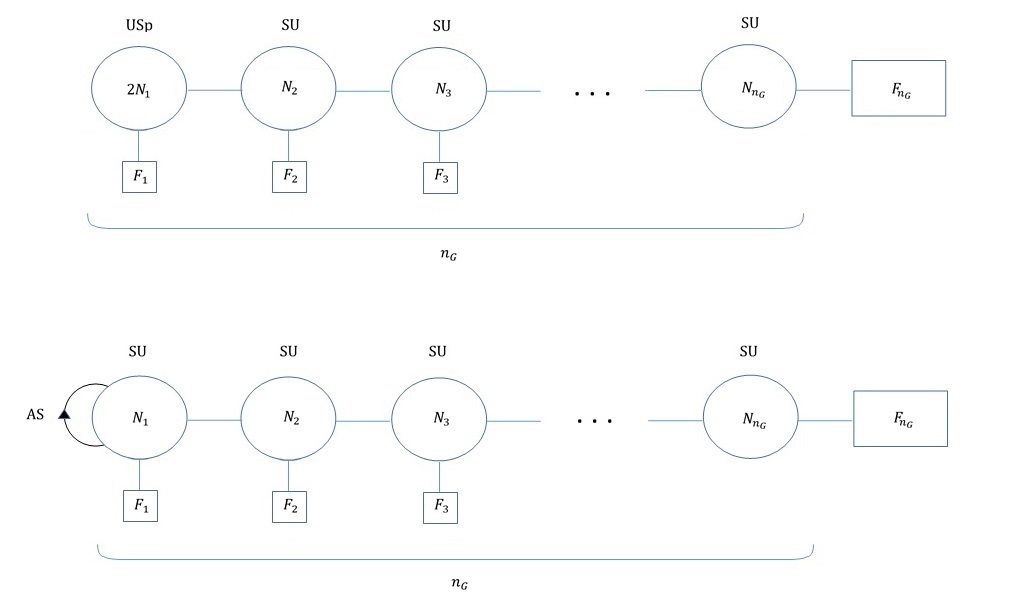} 
\caption{The $6d$ quiver theories we consider.}
\label{Img0}
\end{figure}

We start by presenting the $6d$ gauge theories that we consider in this article. We first present them in their gauge theory description, namely at a generic point on the tensor branch of the underlying $6d$ SCFT. In this description the gauge theory is made from a quiver of $SU(N_i)$ groups with one end being just fundamental hypers while the other end being either a $USp$ gauge group or an $SU$ group with a hyper in the antisymmetric. The freedom in the choice of the theory is given by the ranks of the groups $N_i$. The number of flavors for each group is uniquely determined by anomaly cancellation for each group. The quiver diagrams for the theories we consider are shown in figure \ref{Img0}. 

Next we wish to evaluate the anomaly polynomial that we use later. We concentrate only on the terms in the anomaly polynomial that we need. By using the results of \cite{OSTY,Erler}, we find that the anomaly polynomial contains: 

\bea
I_8 \supset  -\frac{1}{32} A_{ij} tr(F^2_i)tr(F^2_j) + \frac{p_1(T)}{16} P_i tr(F^2_i) - \frac{C_2(R) h_{G_i} tr(F^2_i)}{4} + \frac{C_2(R) p_1(T)}{48} (n_t-n_v) \label{AP} \\ \nonumber + \frac{(n_h-n_v)(7 p^2_1(T) - 4 p_2(T))+n_t(23 p^2_1(T) - 116 p_2(T))}{5760}
\eea  
where $F_i$ is the field strength of the $i$'th group (we always denote the $USp$ or $SU$ with the antisymmetric as $i=1$), and a summation over repeated indices is implied. Also $C_2(R)$ stands for the second Chern class of the R-symmetry bundle, and $p_1(T), p_2(T)$ are the first and second Pontryagin classes of the tangent bundle, respectively. We use $n_h, n_v$ and $n_t$ for the number of hyper, vector and tensor multiplets respectively, and $h_{G_i}$ for the dual Coexter number of the $i$'th group. Finally:

\be
A_{ij} = \begin{pmatrix}
   1 & -1 & 0 & & &\\
  -1 & 2 & -1 & & & \\  
  0 & -1 & 2 & & & \\
	 &  &  & \ddots & &\\
	 &  &  & & 2 & -1\\
	&  &  & & -1 & 2\\
\end{pmatrix}, P_i = (1,0,0,...,0)
\ee 

We can cancel the gauge and mixed anomalies by changing the Bianchi identity of the tensor multiplet (see \cite{SEI1} for the details). For the case at hand this adds the following to the anomaly polynomial:

\be
 \sum_i \frac{1}{2}\left[\frac{tr(F^2_i) - tr(F^2_{i+1})}{4} + C_2(R) \sum^i_{j=1} h_{G_j} - \frac{p_1(T)}{4}\right]^2
\ee

collecting all the terms we find:

\be
I_8 \supset - \frac{C_2(R) p_1(T)}{48} n^{4d}_v + d_H \frac{7 p^2_1(T) - 4 p_2(T)}{5760}
\ee
where

\be
n^{4d}_v = n_v + 12 \sum_i \sum^i_{j=1} h_{G_j} - n_t, d_H=n_h-n_v+29n_t \label{fda}
\ee
where the sum $i$ is over all the gauge groups. 

The labels we used were chosen with the compactification to $4d$ in mind. When compactifying to $4d$ on a torus we get some $4d$ SCFT in the IR. We can calculate the central charges, particularly the $a$ and $c$ conformal anomalies, of this SCFT using the results of \cite{OSTY1}\footnote{For these results to hold, the $6d$ SCFT must be very-Higgsable, as described in \cite{OSTY1}. All the $6d$ SCFT's we're considering are of this type.}. We find that $d_H = 24(c-a)$ and $n^{4d}_v = 4(2a-c)$. Thus, $d_H$ is the dimension of the Higgs branch, and $n^{4d}_v$ the effective number of vector multiplets of the $4d$ SCFT resulting from the compactification of the $6d$ SCFT on a torus. In that light the equation for $d_H$ has a rather nice interpretation as the classical dimension of the Higgs branch of the gauge theory, $n_h-n_v$, plus the contribution of the tensor multiplets, each giving $29$ dimensions, like the rank $1$ E-string theory.    

Besides the $a$ and $c$ conformal anomalies, we also want to determine the central charges for flavor symmetries, $k_{F_i}$, associated to the flavors under the $i$'th gauge group. From the result of \cite{OSTY1}, this can be determined from the term $\frac{k_{F_i}}{192} tr(F^2_{global_i}) p_1(T)$. Say we have a flavor symmetry, the fields charged under it being flavor of dimension $\rho$ under the group $G_i$. Then we find that:

\be
k_{F_i} = 12 g_i +2d_{\rho} \label{fcc}
\ee 
where $g_i=n_G - i + 1$, $n_G$ being the number of groups, and $d_{\rho}$ is the dimension of the representation $\rho$. 

Before continuing we note that some of the theories we consider also include gauging the rank $1$ E-string theory at one end of the quiver. This is a straightforward extension of the quiver theories with a $USp(2N_1)$ end to the $N_1=0$ case. This follows from the fact that $USp(2N_1)+(2N_1+8)F$ goes to a $6d$ SCFT known as the $(D_{N_1+4},D_{N_1+4})$ conformal matter\cite{ZHTV}, so this class of theories can be regarded as gauging a part of the $SO(4N_1+16)$ global symmetry of $(D_{N_1+4},D_{N_1+4})$ conformal matter. In this description we can also consider the case of $N_1=0$ relying on the fact that $(D_{4},D_{4})$ conformal matter is the rank $1$ E-string theory. Going over the computation of the anomaly polynomial, we find that (\ref{fda}) is still valid, where we include the rank $1$ E-string theory in the sum and take $h_{E-string}=1$. 

Generically when gauging a part of a rank $Q$ E-string theory, some of the $E_8$ global symmetry remains unbroken and serves as a global symmetry. For these cases we find $k_{F_1} = 12Q(n_G+1)$. 

Finally, while we generally employ the gauge theory description of these $(1,0)$ SCFT's, it is worthwhile to also specify their description as an F-theory compactification. In this language the theory is described as a long $-1 -2 -2 ... -2$ quiver with $SU$ type groups on the $-2$ curves and a $USp$ or $SU$ type group on the $-1$ curve. 

For the details on the meaning of this notation we refer the reader to \cite{HMRV}. In a nutshell, specifying a $6d$ SCFT requires enumerating its hyper, vector and tensor content. The numbers represent the type of tensor multiplet, where a $-2$ curve represents a single free $\mathcal{N}$$=(2,0)$, tensor and a $-1$ curve the rank $1$ E-string theory. The sequence of numbers represents several tensor multiplets. For example, $-2 -2 -2 ... -2$ gives the $\mathcal{N}$$=(2,0)$ $A_{n-1}$ theory where $n$ is the number of $-2$ curves, and $-1 -2 -2 ... -2$ gives the rank $n+1$ E-string theory.

One can add vector multiplets on these curves. When these are added, the theory on the tensor branch acquires a gauge theory description. For a $-2$ curve, adding an $SU(N)$ type group, leads to an $SU(N)+2NF$ gauge theory on the tensor branch. For a $-1$ curve, adding a $USp(2N)$ type group leads to a $USp(2N)+(2N+8)F$ gauge theory, while adding an $SU(N)$ type group, leads to an $SU(N)+1AS+(N+8)F$ gauge theory at a generic point on the tensor branch\footnote{For $SU(6)$ there is an additional option giving an $SU(6)+\frac{1}{2}\bold{20}+15F$ gauge theory at a generic point on the tensor branch. We briefly encounter this option later in this paper.}. It is now apparent that going to a generic point on the tensor branch indeed gives the gauge theories we consider.

We can also consider the reverse process of removing vector multiplets from a curve. This describes a Higgs branch limit of the $6d$ SCFT in which some of the vector multiplets become massive and the theory flows to a different IR SCFT. Note in particular, that completely breaking a group, corresponding to removing all the vector multiplets from that curve, still leaves the associated tensor multiplet. The resulting IR SCFT generically has no complete Lagrangian description, but can still be described by a gauge theory gauging part of the flavor symmetry of a non-Lagrangian part. We shall encounter several examples of this later. 

Sometimes gauge theory physics is insufficient to fully determine the properties of the SCFT. For example, in some cases the global symmetry naively exhibited by the gauge theory, is larger than the one of the SCFT. We encounter some cases where this occurs, and then it is useful to have an F-theory description. 

\subsection{Class S technology}
 
The results obtained from the $6d$ anomaly polynomial can be compared to the ones obtained using class S technology. Specifically, the theories we consider are all isolated SCFT's, that can be represented as the compactification of an $A$ type $(2,0)$ theory on a Riemann sphere with $3$ punctures. We also have a $5d$ brane web representation using \cite{BB}. It is known how to calculate the central charges of such SCFT's from the form of the punctures. The explicit formula used to calculate $d_H, n^{4d}_v$ and $k_F$ can be found in \cite{CD,CD1}. In practice, it is usually simpler to calculate $d_H$ directly from the web.

We also want to determine the global symmetry of the SCFT. In general this can be read of from the punctures, but in some cases the global symmetry can be larger than is visible from the punctures\cite{Gai}. One way to determine this is using the $5d$ description either directly from the web, or using the gauge theory description. 

A more intricate method is to use the $4d$ superconformal index. Since conserved currents are BPS operators they contribute to the index, and so knowledge of the index allows us to determine the global symmetry of the theory. In practice we do not need the full superconformal index, just the first few terms in a reduced form of the index called the Hall-Littlewood index\cite{GRRY}. An expression for the $4d$ superconformal index for class S theories was conjectured in \cite{GRRY,GR,GRR}, and one can use their results to determine the global symmetry. For more on this application see \cite{CDT}.

In cases where the global symmetry is bigger than what is visible from the punctures, we use the $4d$ superconformal index to show this. In cases where it is not difficult to argue this also from the $5d$ description, we also use this as a consistency check.

\section{The $5d$ $(N+2)F+SU_0(N)^k+(N+2)F$ quiver and related theories}  

\begin{figure}
\center
\includegraphics[width=1\textwidth]{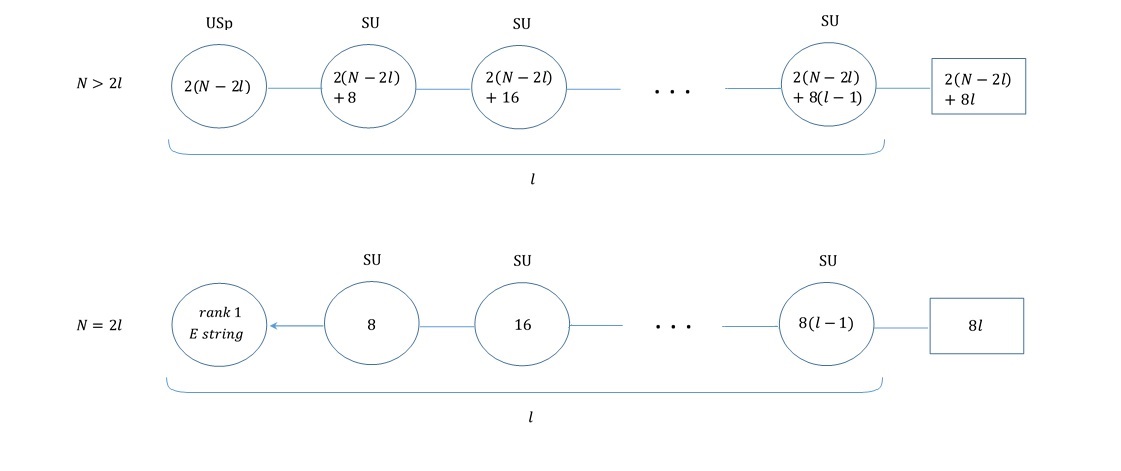} 
\caption{The $6d$ quiver theory we consider. The arrow in the second quiver stands for gauging a part of the global symmetry of the shown $6d$ SCFT, in this case an $SU(8)$ subgroup of $E_8$.}
\label{Img1}
\end{figure}

\begin{figure}
\center
\includegraphics[width=1\textwidth]{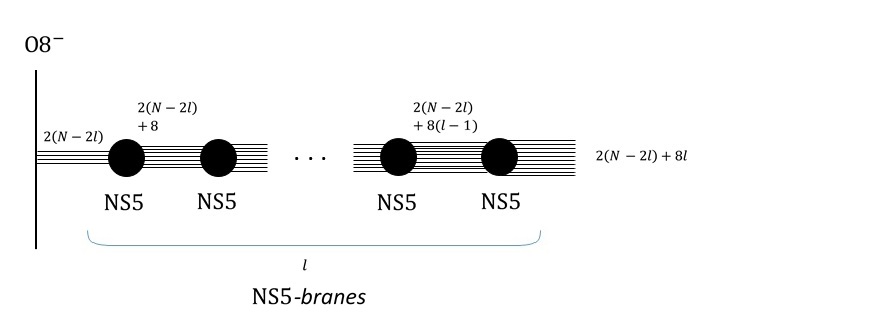} 
\caption{The brane description of the $6d$ theory in figure \ref{Img1}. The horizontal lines represent D$6$-branes, and the number above the lines stand for the number of $6$-branes. The black circles represent NS$5$-branes, and their number is given below. Finally, the vertical line stands for the $O8^-$ plane. The configuration also include $2N+4l$ D$8$-branes, parallel to the $O8^-$ plane, on which the asymptotic D$6$-branes end. For clarity we have suppressed them in the figure.}
\label{Img2}
\end{figure}

\begin{figure}
\center
\includegraphics[width=1\textwidth]{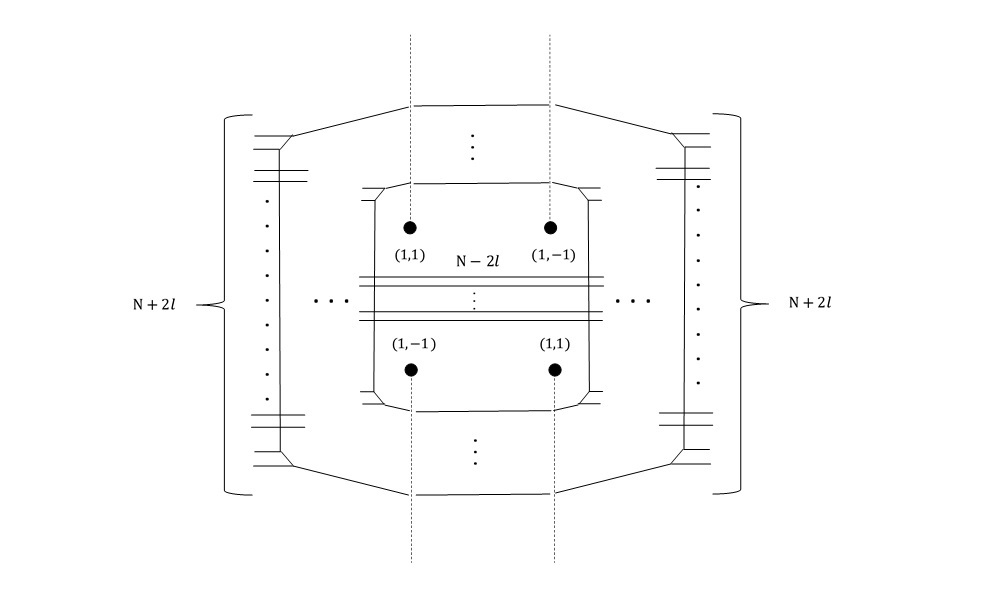} 
\caption{The web we end up with after performing T-duality on the brane configuration of figure \ref{Img2} and resolving the $O7^-$ planes.}
\label{Img3}
\end{figure}

\begin{figure}
\center
\includegraphics[width=1\textwidth]{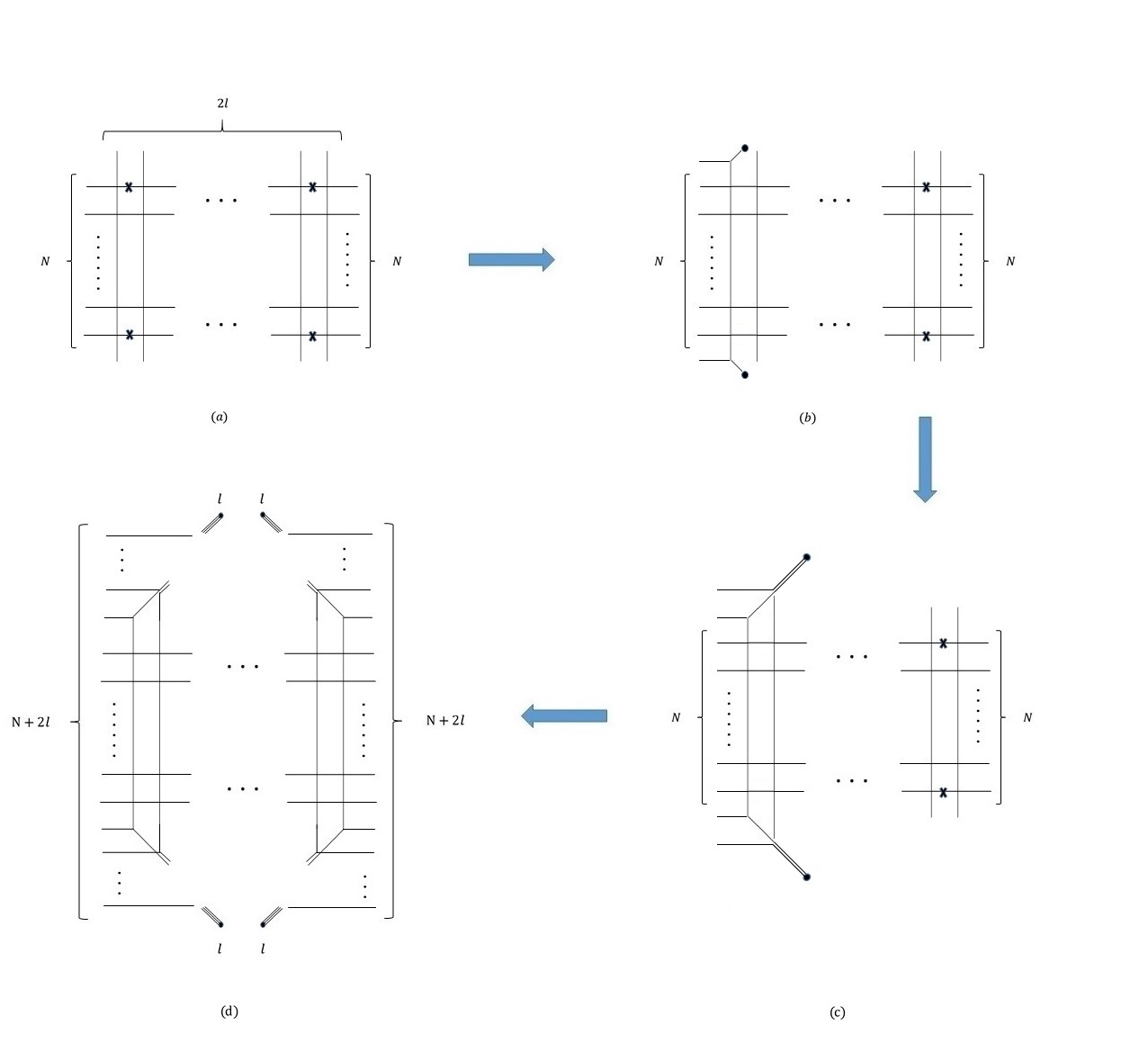} 
\caption{(a) The web for $(N+2)F+SU_0(N)^{2l-1}+(N+2)F$. We can first pull the two $7$-branes on the left through the leftmost NS$5$-brane leading to the web in (b). We can now push the $(1,1)$ $7$-brane and $(1,-1)$ $7$-brane through the neighboring NS$5$-brane. This changes the asymptotic NS$5$-brane to a D$5$-branes, and is accompanied by a Hanany-Witten transition generating an additional $5$-brane ending on the $7$-brane. This gives the web in (c). Repeating this on the neighboring $l-2$ NS$5$-branes, and also doing the same on the right hand side, we end up with the web in (d). This is the web of figure \ref{Img3} after pulling out the internal $7$-branes.}
\label{Img4}
\end{figure}

In this section we start analyzing the $6d$ lift of $5d$ theories. We start with the $5d$ quiver theory $(N+2)F+SU_0(N)^k+(N+2)F$. Since a conjecture for this theory was already given in \cite{Yon}, it is more convenient to start with the $6d$ theory. There are two slightly different cases to consider. First, we have the $6d$ SCFT whose quiver description is shown in figure \ref{Img1}. This theory can be realized in string theory by a system of D$6$-branes crossing an $O8^-$ plane and several NS$5$-branes, shown in figure \ref{Img2}. Note, that this is a generalization of the system in \cite{HKLTY}, by the addition of NS$5$-branes. We can now repeat the analysis of \cite{HKLTY}. Since this is a simple generalization of their work we will be somewhat brief. We compactify a direction shared by all the branes and preform T-duality. The $O8^-$ plane becomes two $O7^-$ planes. Under strong coupling effect, the $O7^-$ plane decomposes to a $(1,1)$ $7$-brane and a $(1,-1)$ $7$-brane\cite{Sen}. We then end up with the web of figure \ref{Img3}. This web describes the $5d$ gauge theory $(N+2)F+SU_0(N)^{2l-1}+(N+2)F$, as shown in figure \ref{Img4}. Note that the number of groups in $5d$ must be odd, owing to the even number of NS$5$-branes.

\begin{figure}
\center
\includegraphics[width=1\textwidth]{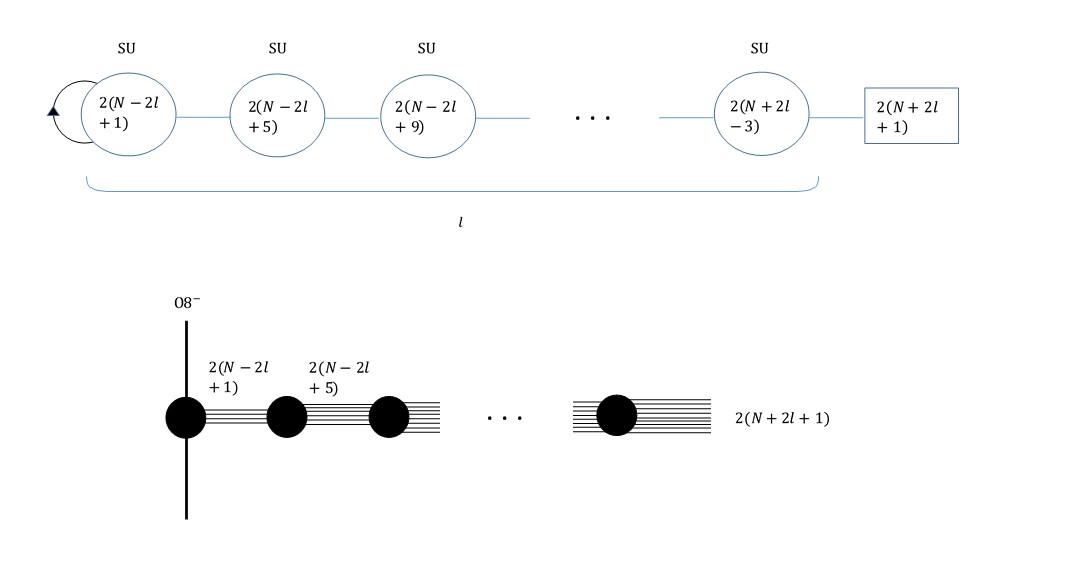} 
\caption{Sticking an NS$5$-brane on the $O8$ plane leads to this $6d$ quiver theory.}
\label{Img5}
\end{figure}

\begin{figure}
\center
\includegraphics[width=1\textwidth]{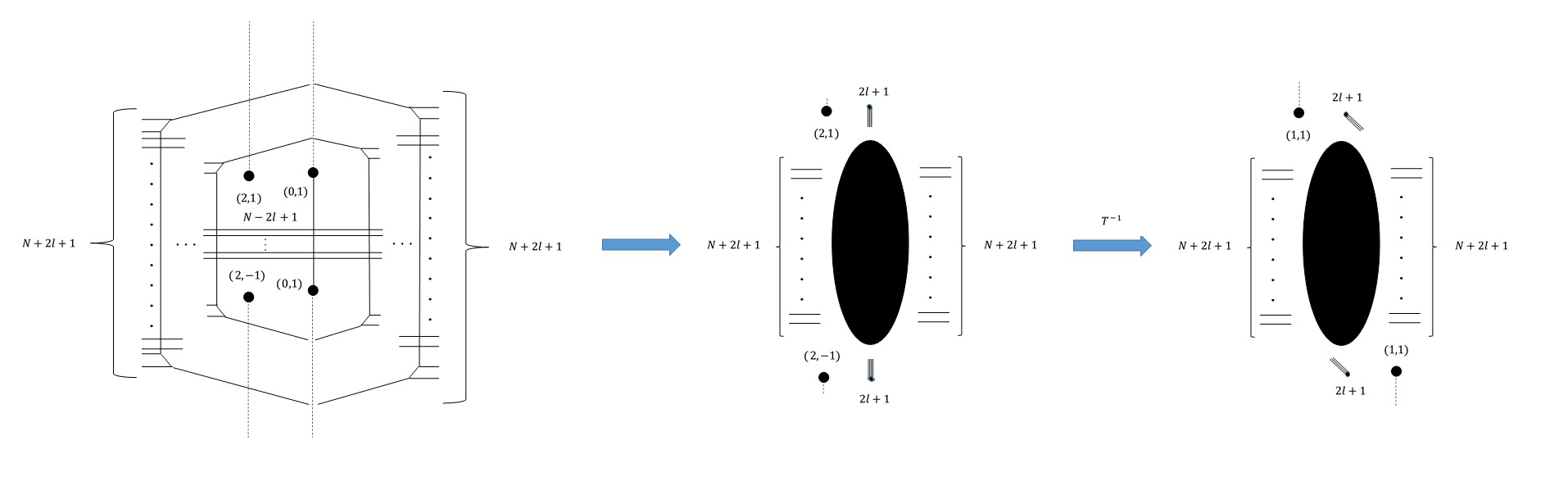} 
\caption{The web we end up with after performing T-duality on the configuration of figure \ref{Img5} and resolving the $O7^-$ planes.}
\label{Img6}
\end{figure}

\begin{figure}
\center
\includegraphics[width=1\textwidth]{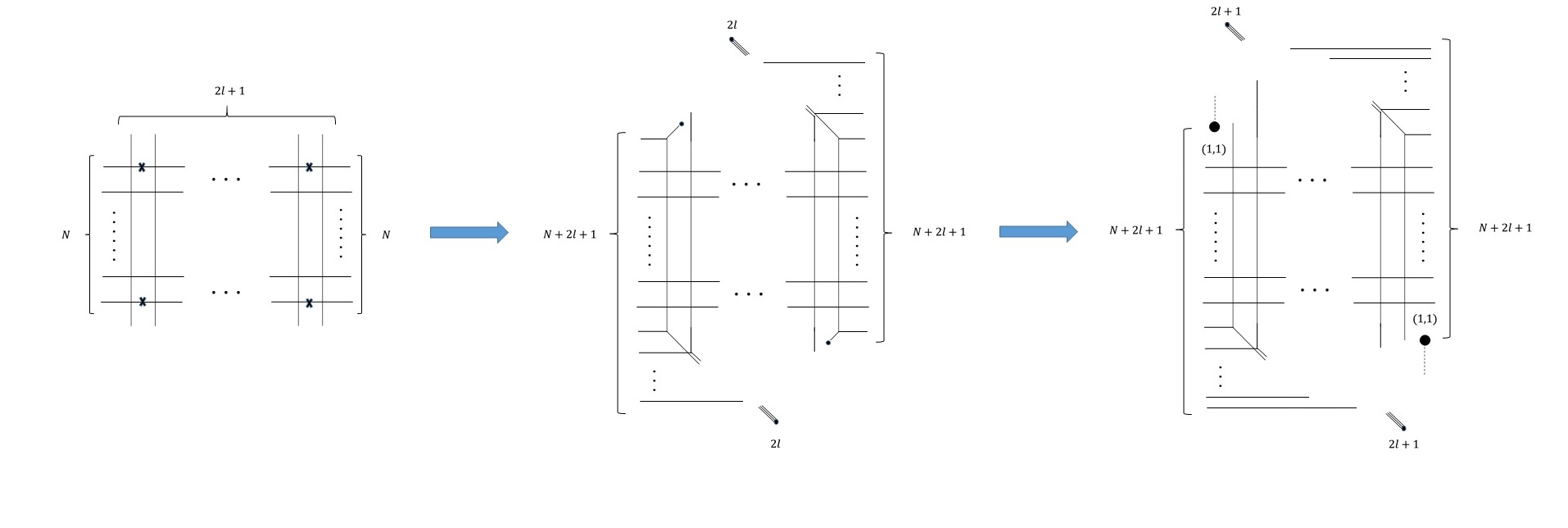} 
\caption{The web for $(N+2)F+SU_0(N)^{2l}+(N+2)F$.}
\label{Img7}
\end{figure}

This suggests that to get an even number of $5d$ $SU(N)$ groups, we need to take an odd number of NS$5$-branes, which we do by adding a stuck NS$5$-brane on the $O8^-$ plane. The brane and quiver description of the resulting $6d$ theory is shown in figure \ref{Img5}. We can now repeat the analysis. After T-duality we get again two $O7^-$ planes with the stuck NS$5$-brane stretching between the two. Decomposing the $O7^-$ planes with the stuck NS$5$-brane, as shown in \cite{BZ1}, we arrive at the web of figure \ref{Img6}. As shown in figure \ref{Img7}, this is the web of $(N+2)F+SU_0(N)^{2l}+(N+2)F$. This agrees with the conjecture of \cite{Yon}, that this $6d$ theory is the UV completion of the $5d$ gauge theory $(N+2)F+SU_0(N)^{2l}+(N+2)F$.

Note that in the $6d$ theories covered so far we have assumed that $N>2l-1$. Naively, this implies the same limitations on the $5d$ theories. However, it is not difficult to see that performing S-duality on the web for $(N+2)F+SU_0(N)^{k-1}+(N+2)F$ results in the one for $(k+2)F+SU_0(k)^{N-1}+(k+2)F$. Thus, by doing an S-duality, one can map any $5d$ linear $SU(N)$ quiver to the required form. Also note that when $k=N-1$, both descriptions are of this form, and indeed the two $6d$ SCFT's are the same.  

\begin{figure}
\center
\includegraphics[width=1\textwidth]{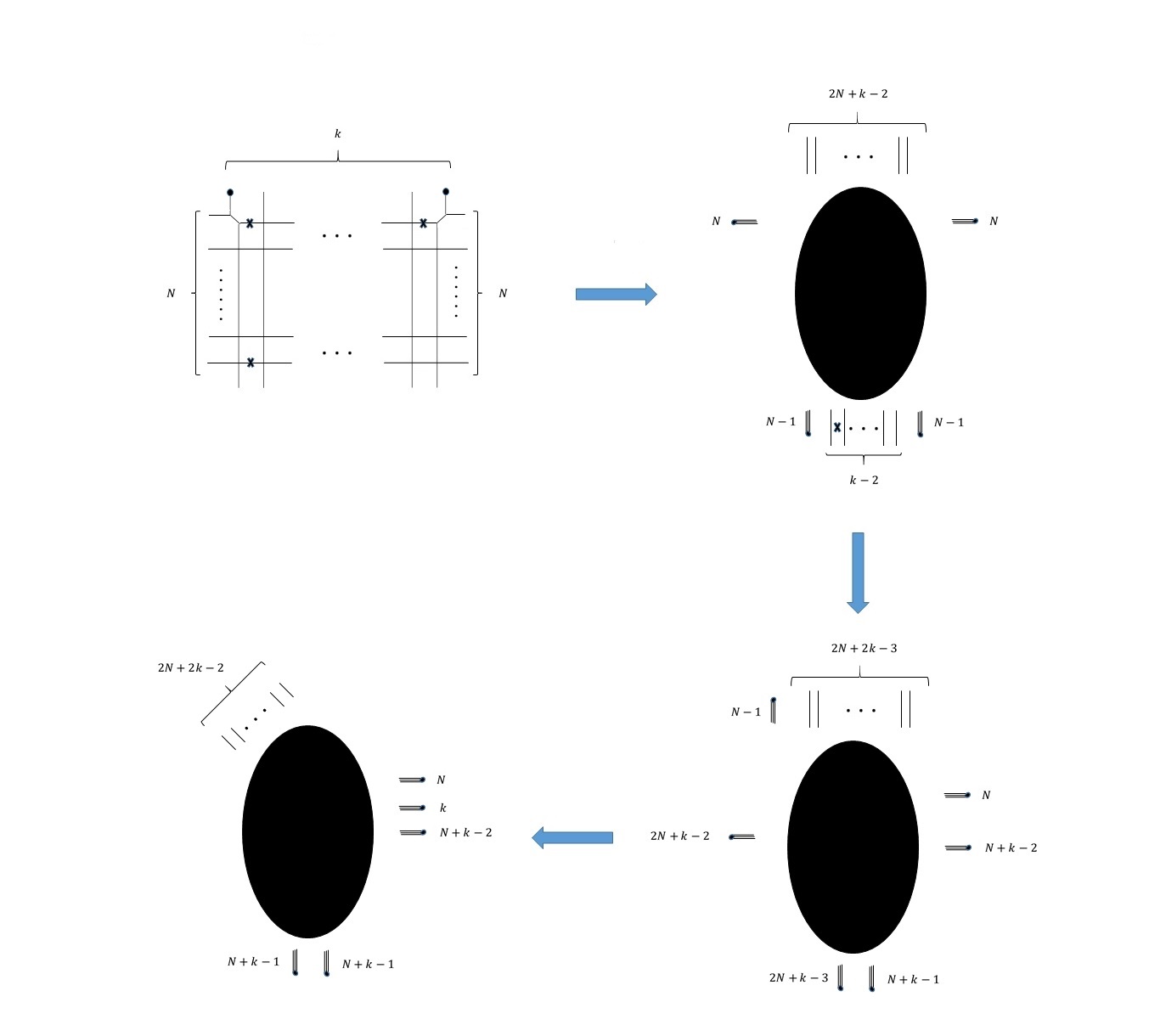} 
\caption{The web for $(N+2)F+SU_0(N)^{k-2}\times SU_{\frac{1}{2}}(N)+(N+1)F$. The upper left shows the web in its gauge theory description. Moving first the two shown $(0,1)$ $7$-branes down to the other side and then pulling out the two upper $(1,0)$ $7$-branes, doing Hanany-Witten transitions when necessary, leads to the web on the upper right. Further pulling the remaining $(1,0)$ $7$-brane to the right, doing all the Hanany-Witten transitions, leads to the web on the lower right. Finally, exchanging the upper $(0,1)$ $7$-brane with $N-1$ NS$5$-branes ending on it with the lower one with $2N+k-3$ NS$5$-branes ending on it, and also moving the left $(1,0)$ $7$-brane to the right leads us to the web in the lower left of the figure.}
\label{Img9}
\end{figure}

We now wish to employ this relation to the compactification of the $6d$ SCFT on a torus, preserving the global symmetry. Inspired by the E-string theory example, we are lead to consider an infinite mass deformation limit of the related $5d$ theory. The natural candidate is integrating out a fundamental flavor. We have only one possibility, corresponding to the $5d$ theory $(N+2)F+SU_0(N)^{k-2}\times SU_{\pm\frac{1}{2}}(N)+(N+1)F$ whose web is shown in figure \ref{Img9}. This theory does give a $5d$ fixed point shown in figure \ref{Img9}. We note that this web is of the form of \cite{BB}. We can now employ class S technology to determine the global symmetry of this theory, finding that its global symmetry is $U(1)\times SU(2N+2k)$ when $N\neq k$ and $SU(2)\times SU(2N+2k)$ when $N=k$. 

Note that this is exactly the same as the global symmetry of the $6d$ theory. The flavors at the end give the $SU(2N+2k)$ part. The remaining $U(1)$ is the anomaly-free combination of the various baryonic and bifundamental $U(1)$'s. The case of $N=k$ indeed has an enhancement of symmetry to $SU(2)$. For $k=2l+1$, this comes about because the antisymmetric representation of $SU(4)$ is real while for $k=2l$, this comes about as the gauging of $SU(8)\subset E_8$ preserves an $SU(2)$, since $SU(8)\subset E_7 \subset  E_7 \times SU(2)\subset E_8$.

\begin{figure}
\center
\includegraphics[width=1\textwidth]{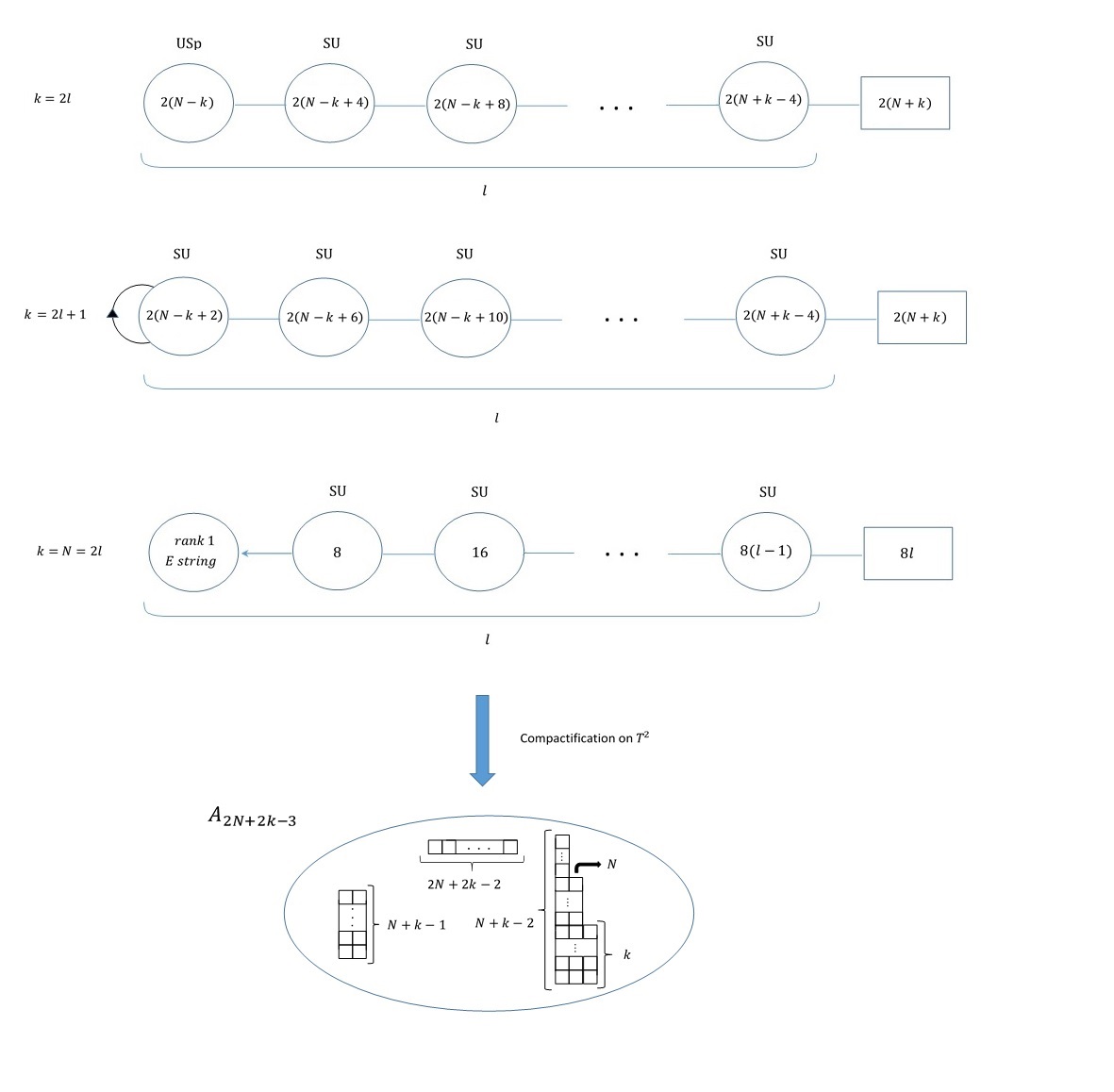} 
\caption{Starting from the family of $6d$ theories in the upper part, we claim that compactifying them to $4d$ on a torus leads to the isolated SCFT represented in the lower part.}
\label{Img8}
\end{figure}

We now conjecture that compactifying this $5d$ theory to $4d$ should give the compactification of the starting $6d$ theory on a torus. We know from the work of \cite{BB} that for the theory of figure \ref{Img9}, this leads to a $4d$ isolated SCFT that can be described by a compactification of the $6d$ $(2,0)$ theory of type $A_{2N+2k-3}$ on the punctured sphere of figure \ref{Img8}. We next wish to test this conjecture by comparing the central charges of this $4d$ SCFT with the ones expected from the compactified $6d$ $(1,0)$ SCFT which can be determined through (\ref{fda}), (\ref{fcc}).   

From the $5d$ theory, using class S technology, we find:

\bea
d_H & = & (N+k-1)(2N+2k+1)+k, \nonumber \\  n^{4d}_v & = & \frac{(k-1)(6N^2-3N+k(9N-7)-k^2-3)}{3}, \nonumber \\ k_{SU(2N+2k)} & = & 4(N+k-1), \nonumber \\  k_{SU(2)} & = & 6N \label{ffe}
\eea
where we assume $N\geq k$, $k_{SU(2)}$ being relevant only for the $N=k$ case. The results for $N< k$ can be generated from (\ref{ffe}) by taking $N\leftrightarrow k$.



From the $6d$ theory we see that:

\bea
 n_v & = & \frac{8k(k-1)(k-2)}{3}-\frac{k-2}{2}+(N-k)(2Nk+2k^2-2N-6k+1), \nonumber \\ n_h & = &  \frac{8k(k-1)(k-2)}{3} + 2k(N^2-k^2+4k-8), \nonumber \\ n_t & = & l
\eea
for the case of $k=2l$, and:  

\bea
 n_v & = & \frac{8(k-1)(k-2)(k-3)}{3}-\frac{k-1}{2}+2(k-1)(N-k+2)(N+k-4), \nonumber \\ n_h & = & \frac{8(k-1)(k-2)(k-3)}{3} + 2kN^2+3N-2k^3+16k^2-43k+30, \nonumber \\ n_t & = & l
\eea
for the case of $k=2l+1$. Using these in (\ref{fda}) and (\ref{fcc}) we indeed recover (\ref{ffe}).

\begin{figure}
\center
\includegraphics[width=1\textwidth]{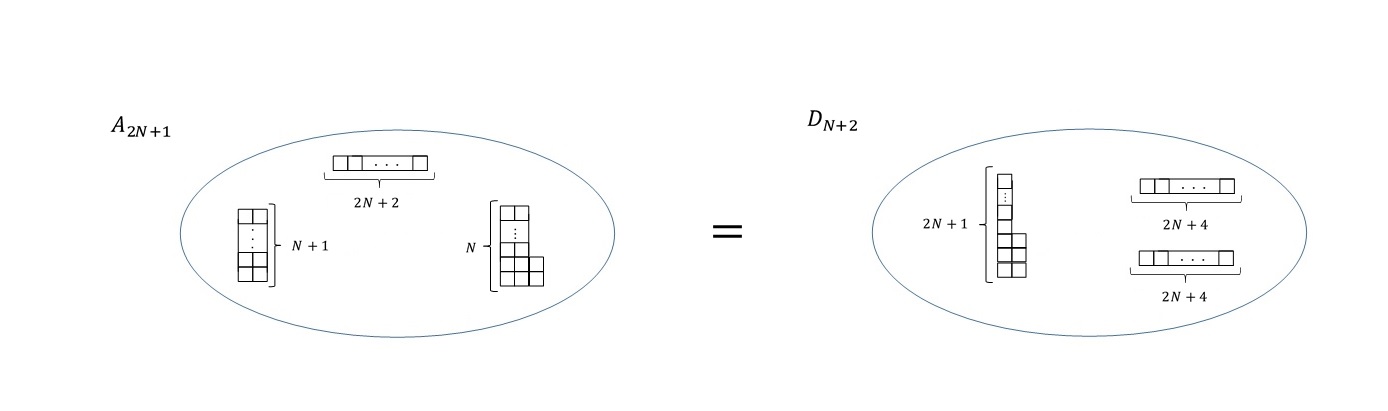} 
\caption{(a) Our analysis suggests that compacitifying the $6d$ $USp(2N-4)+(2N+4)F$ theory on a torus should give the isolated $4d$ SCFT that is described by compactifying the $6d$ $(2,0)$ theory of type $A_{2N+1}$ on this punctured Riemann sphere. (b) A different analysis, done in \cite{OSTY1}, suggests that the same theory compactified on a torus should give the isolated $4d$ SCFT that is described by compactifying the $6d$ $(2,0)$ theory of type $D_{N+2}$ on this punctured Riemann sphere. Our analysis does imply that these two theories are in fact identical.}
\label{Img11}
\end{figure}

An interesting thing happens for $k=2$. In that case the $6d$ theory becomes $USp(2N-4)+(2N+4)F$, which is also known as $(D_{N+2},D_{N+2})$ conforml matter\cite{ZHTV}. The reduction of this theory to $4d$ on a torus was recently studied in \cite{OSTY1}. They found that it leads to an isolated SCFT corresponding to compactifying the $6d$ $(2,0)$ theory of type $D_{N+2}$ on a Riemann sphere with three punctures shown in figure \ref{Img11} (b). If we are correct in our description then these two SCFT's must be identical. Indeed, using the results of \cite{CD1} we can calculate the dimension of Coulomb branch operators and compare between the two theories. We find a perfect match.   

\begin{figure}
\center
\includegraphics[width=1\textwidth]{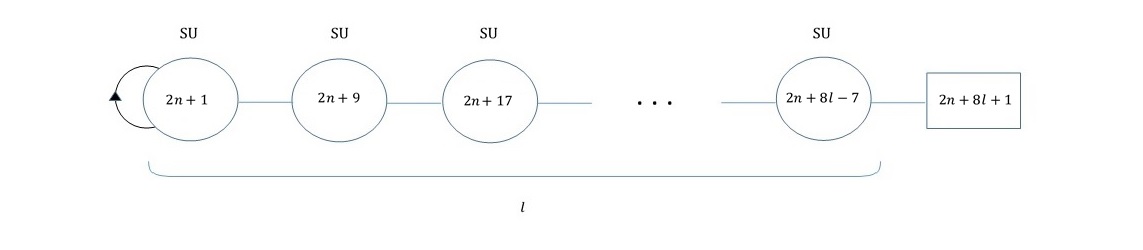} 
\caption{The $6d$ quiver theory we are considering.}
\label{Img21}
\end{figure}

\begin{figure}
\center
\includegraphics[width=0.8\textwidth]{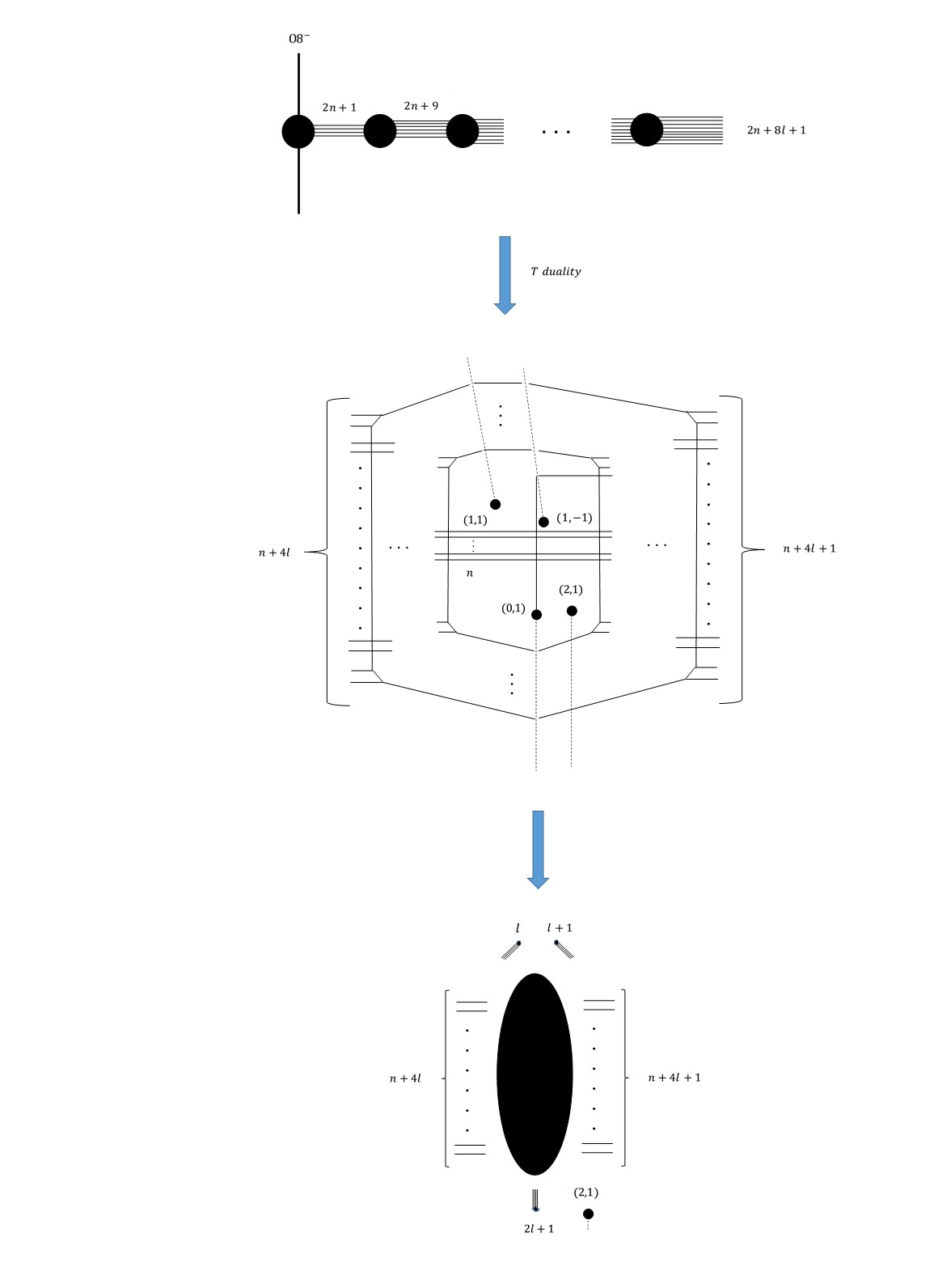} 
\caption{Starting from the brane description of the $6d$ quiver theory in figure \ref{Img21}, we can T-dualize to the web system in the bottom of the figure.}
\label{Img22}
\end{figure}

\begin{figure}
\center
\includegraphics[width=1\textwidth]{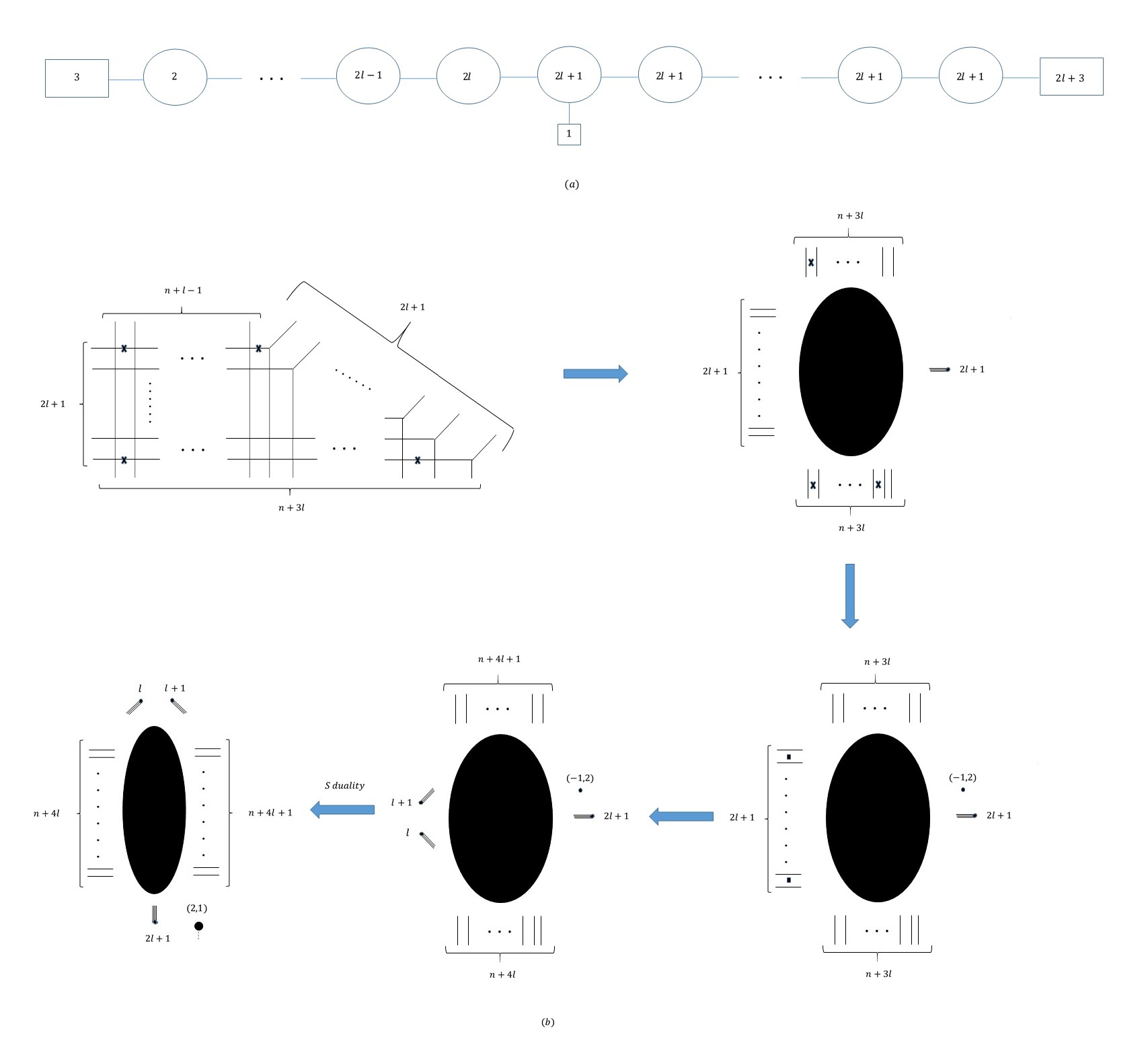} 
\caption{(a) The $5d$ theory that we expect to lift to the $6d$ theory in figure \ref{Img21}. (b) starting from the brane web, and doing several manipulations, we arrive at the web of figure \ref{Img22}.}
\label{Img23}
\end{figure}

Before moving on to discuss other $5d$ theories, there is one more $6d$ SCFT, closely related to the ones considered, that we would like to discuss. The quiver theory description is given in figure \ref{Img21}. We can repeat the previous analysis, now the difference manifesting in the $6d$ brane construction by adding a stuck $6$-brane. Upon performing T-duality this becomes a stuck D$5$-brane on one of the $O7^{-}$ planes. We can decompose the $O7^{-}$ planes as done in \cite{BZ1}, to get the final web picture. The entire process is shown in figure \ref{Img22}. This describes the $5d$ gauge theory of figure \ref{Img23}.

\begin{figure}
\center
\includegraphics[width=1\textwidth]{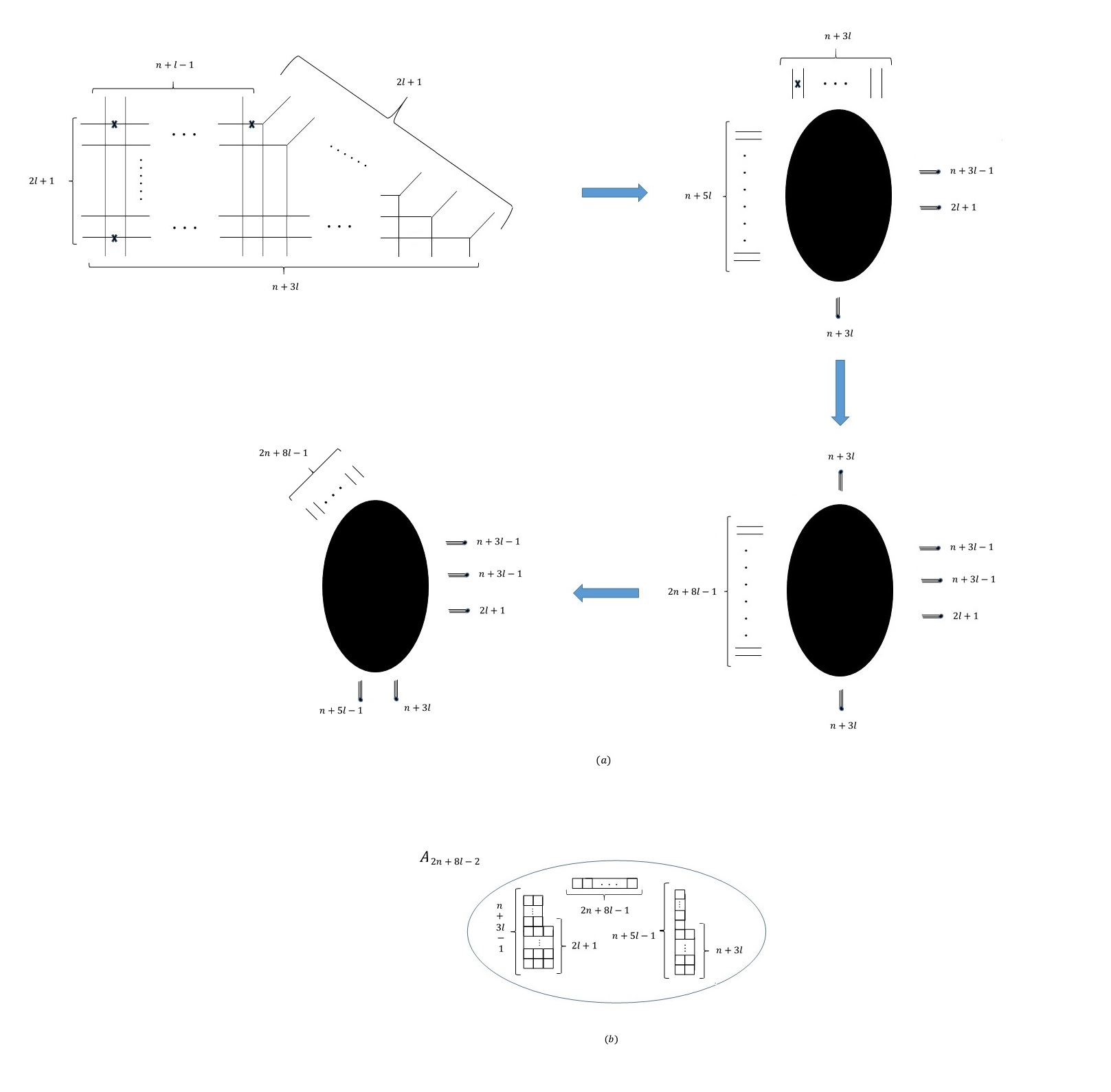} 
\caption{(a) The brane web for the $5d$ theory of figure \ref{Img23} (a), where one of the $SU(2)$ flavors is integrated out. Opening out the web we get to the presentation of (b). We could have also integrated out any other flavor, and obtained the same theory.}
\label{Img24}
\end{figure}

One can see that the Coulomb branch dimensions agree, and using the results of \cite{Yon}, also the global symmetries agree, in particularly, we get an affine $A^{(1)}_{2n+8l}$. As a further test we consider the compactification to $4d$, where we expect to get the theory of figure \ref{Img24}. Using class S technology we can indeed show that the $4d$ isolated theory in figure \ref{Img24} (b) has the same global symmetry as the $6d$ quiver of figure \ref{Img21}. We can also calculate the central charges finding:

\bea
d_H & = & 2(n+4l)^2 + n + 6l, \nonumber \\ n^{4d}_v & = & \frac{l(12n^2 + 84ln + 112 l^2 - 54l - 25)}{3}, \nonumber \\ k_{SU(2n+8l+1)} & = & 2(2n  +8l - 1)
\eea 

Using (\ref{fda}),(\ref{fcc}), this indeed matches what we expect from the theory of figure \ref{Img21}.

\subsection{Generalizations}

The next step is to consider generalizations to other $5d$ gauge theories with an expected $6d$ lift. Consider the $5d$ gauge theory given by a linear $SU_0(N_i)$ quiver with fundamental matter, where each non edge group sees an effective number of $2N_i$ flavors. If in addition the two edge groups see an effective number of $2N_i+2$ flavors, then it was argued in \cite{Yon} that this $5d$ theory should have an enhanced affine $A^{(1)}$ symmetry. This strongly suggests that these also lift to a $6d$ SCFT. Note that the previously considered theories are also of this form.  

Naturally, we would like to know to which $6d$ SCFT these theories lift. As there is an infinite number of possibilities, a case by case study seems ineffective. Thus, we wish to determine a procedure by which, given such a $5d$ quiver, the $6d$ SCFT can be determined. To do this we can utilize the fact that any such quiver can be reached starting with the linear $SU(N)$ quiver considered before, for some $N$ and $k$, and going on the Higgs branch. Also, for theories with $8$ supercharges, the Higgs branch does not receive quantum corrections, and so the $5d$ and $6d$ Higgs branches must agree. Therefore, one possible strategy is to start from one of the previous cases, where we know the $6d$ SCFT, and determine the Higgs branch limit needed to get the required $5d$ quiver. Then, by mapping this to the $6d$ SCFT, we can determine the $6d$ lift of the $5d$ quiver. 

\begin{figure}
\center
\includegraphics[width=1\textwidth]{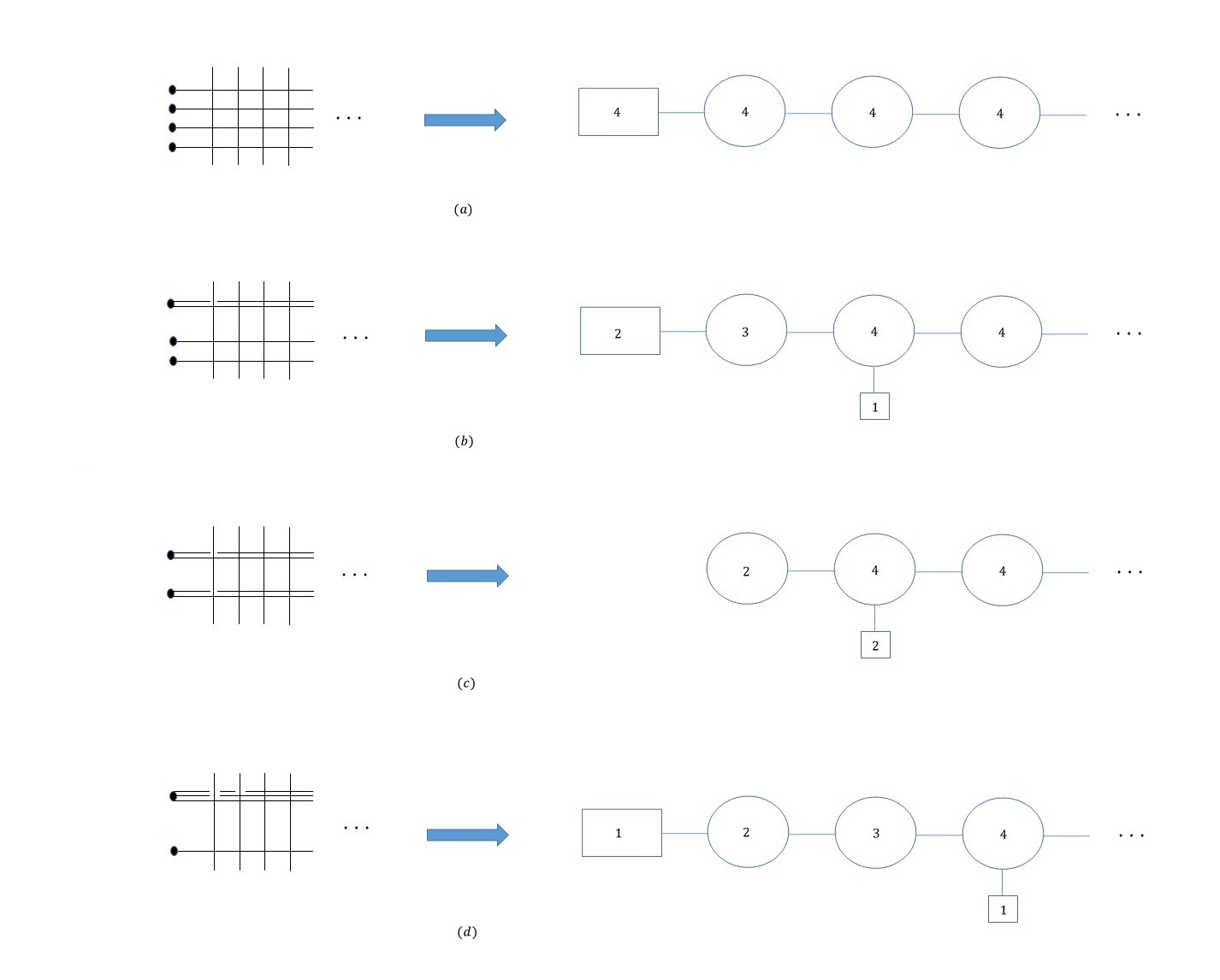} 
\caption{The mapping between the Higgs branch, as manifested in the web, to the resulting low energy quiver. (a) We start with a long $SU(N)$ quiver where for simplicity we have taken $N=4$. We can go on the Higgs branch by breaking the $5$-branes on the $7$-branes. This leads to several $5$-branes ending on the same $7$-branes. The resulting gauge theory can now be determined by doing Hanany-Witten transitions, until all $7$-branes have no $5$-branes ending on them. This leads to the quivers shown in (b)-(d). The generalization to more complicated cases is now apparent.}
\label{Img10}
\end{figure}

To understand the mapping, we can again rely on the brane description. Starting with the $5d$ case, the Higgs branch limits we are interested in are represented, in the brane web, by forcing a group of $5$-branes to end on the same $7$-brane. For example consider a group of $N$ parallel $5$-branes, crossing some NS$5$-branes, each ending on a different $7$-brane, see figure \ref{Img10} (a). This describes a quiver tail of the form $NF+SU_0(N)\times SU_0(N)...$. If we force two $5$-brane to end on the same $7$-brane then, because of the S-rule, one Coulomb modulus of the edge $SU(N)$ group is lost. Thus, this describes the Higgs branch breaking $NF+SU_0(N)\times SU_0(N)$ to $(N-2)F+SU_0(N-1)\times SU_0(N)+1F$ (see figure \ref{Img10} (b)). 

We can of course repeat this and force two other $5$-branes to end on the same $7$-brane. This leads to a similar breaking on the new quiver (see figure \ref{Img10} (c)). However, we can also consider forcing an additional $5$-brane to end on the same $7$-brane, so as to have three $5$-branes ending on it (see figure \ref{Img10} (d)). Now the S-rule not only eliminates a Coulomb moduli of the edge $SU(N)$ group, but also one from the adjacent group. This describes the Higgs branch breaking associated with giving a vev to the gauge invariant made from a flavor of the edge group, the bifundamental, and the flavor from the adjacent group. The quiver left after this breaking is shown in (see figure \ref{Img10} (d)).   

It is now straightforward to generalize to an arbitrary configuration. Before moving to the corresponding limits in the $6d$ theory, we note that this correspondence may not hold when completely breaking a gauge group. In general, the topological symmetry of the broken group survives the breaking and remains in the resulting theory, sometimes manifesting as extra flavors. In these cases, perturbative reasoning alone may be inadequate to determine the answer. For our purposes, this can always be avoided. Also note, that this can be related to the classification of $4d$ quiver tails of \cite{Gai} by using the results of \cite{BB}. This is an alternative way to argue this mapping.

Next, we consider the implications of this on the $6d$ theory. Under T-duality, the D$5$-branes are mapped to D$6$-branes and the D$7$-branes to D$8$-branes, so the analogous breaking on the $6d$ side is represented in the brane configuration by forcing a group of D$6$-branes to end on the same D$8$-brane. If the breaking is not too extreme, this translates to a limit on the perturbative Higgs branch of the $6d$ SCFT. In fact, as the S-rule is the same as in the $5d$ case, we find that this induces exactly the same effect on the quiver tail. The only difference is that now there is only one quiver tail. Each action performed on any of the two tails of the $5d$ quiver is mapped to the corresponding action done on the single $6d$ tail.  

Nevertheless, complications can arise in some instances, for example, when the $6d$ SCFT has a tensor multiplet without an associated gauge theory. For example, consider the $6d$ quiver of figure \ref{Img1}, for $N=2l$. In that case the $6d$ SCFT has a non-Lagrangian part, the rank $1$ E-string theory, possessing a $29$ dimensional Higgs branch. Some of the breaking we consider may be mapped to the Higgs branch of the E-string theory, where we have no perturbative description. This can happen even in cases where the initial theory has a complete Lagrangian description, but on the Higgs branch limit the gauge group is completely broken leaving its associated tensor multiplet\footnote{This is manifested in the web when one is forced to coalesce $2$ NS$5$-branes or an NS$5$-brane and the $O8$ plane due to the constraints of the S-rule.}. Note that this method can still be used to determine the $6d$ SCFT, but is somewhat complicated as the Higgs branch limits may not be perturbatively realized. Thus, determining the resulting $6d$ SCFT will probably require string theoretic methods like the ones in \cite{HMRV1}.

\subsubsection{A simple example}

\begin{figure}
\center
\includegraphics[width=1.05\textwidth]{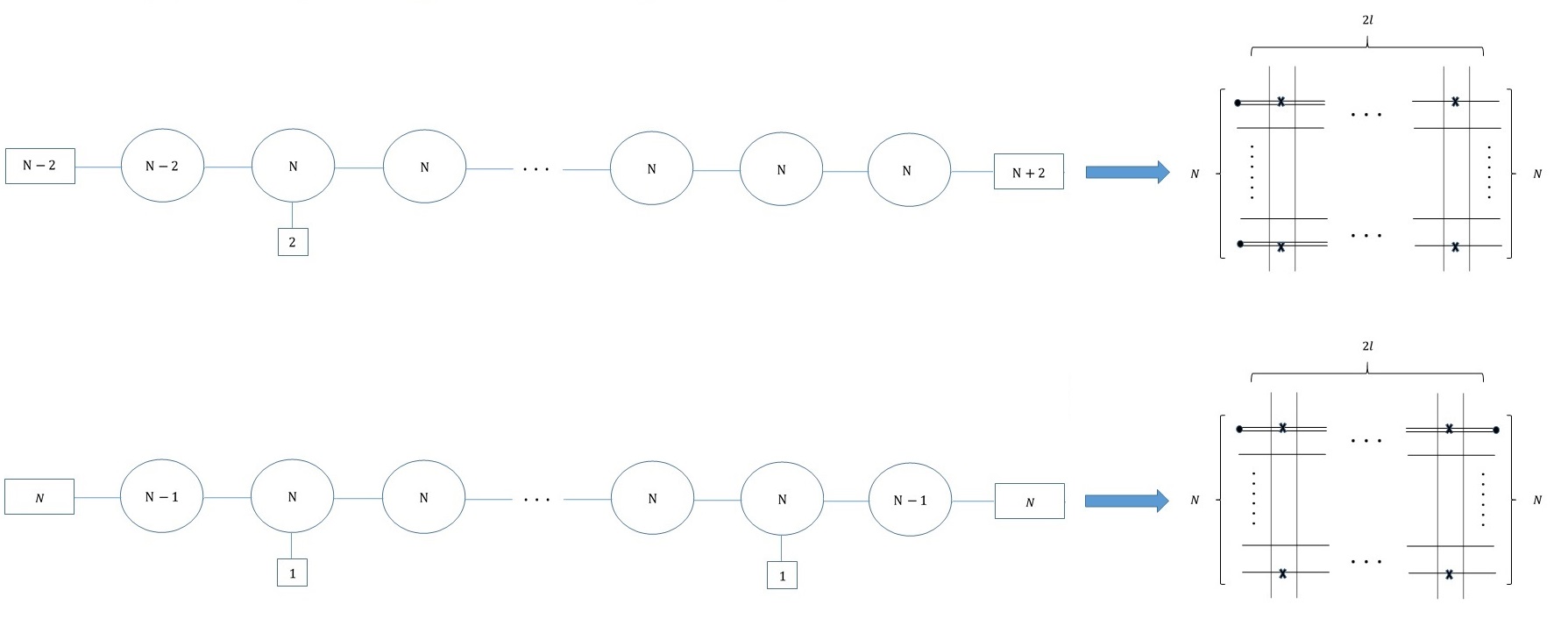} 
\caption{The $5d$ quiver theories we are considering. On the left is the quiver diagram and on the right, the corresponding brane web. All groups are of type $SU$ with CS level $0$.}
\label{Img12}
\end{figure}

We next wish to illustrate this with a simple example. First, consider the $5d$ theories shown in figure \ref{Img12}. We can get these theories from the one in figure \ref{Img4} (a) by going on the Higgs branch. On the web system this is manifested by breaking two pairs of $5$-branes so that each of them end on the same $7$-brane, the difference between them being whether the pair are on the same side or opposite sides. In the $6d$ theory these are mapped to the same breaking, indicating that these two quivers are dual, in the sense of both lifting to the same $6d$ SCFT. 

\begin{figure}
\center
\includegraphics[width=0.8\textwidth]{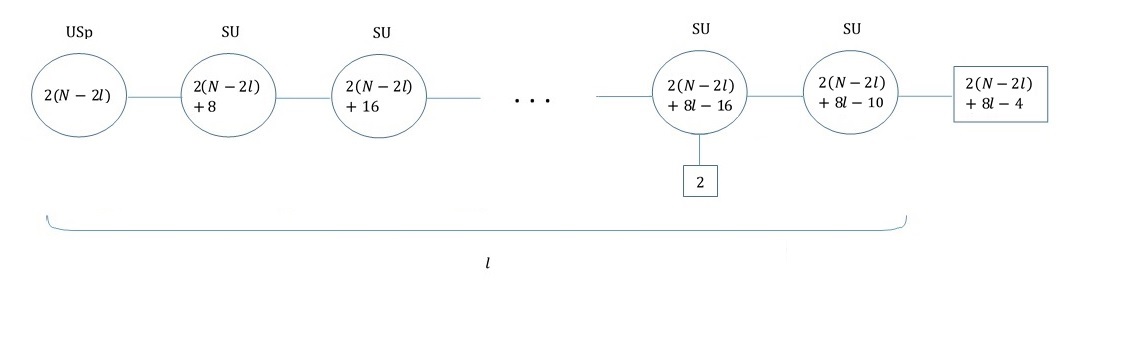} 
\caption{The $6d$ quiver theory found after implementing the Higgs branch flow of figure \ref{Img12} on the theory of figure \ref{Img1}.}
\label{Img13}
\end{figure}

\begin{figure}
\center
\includegraphics[width=1\textwidth]{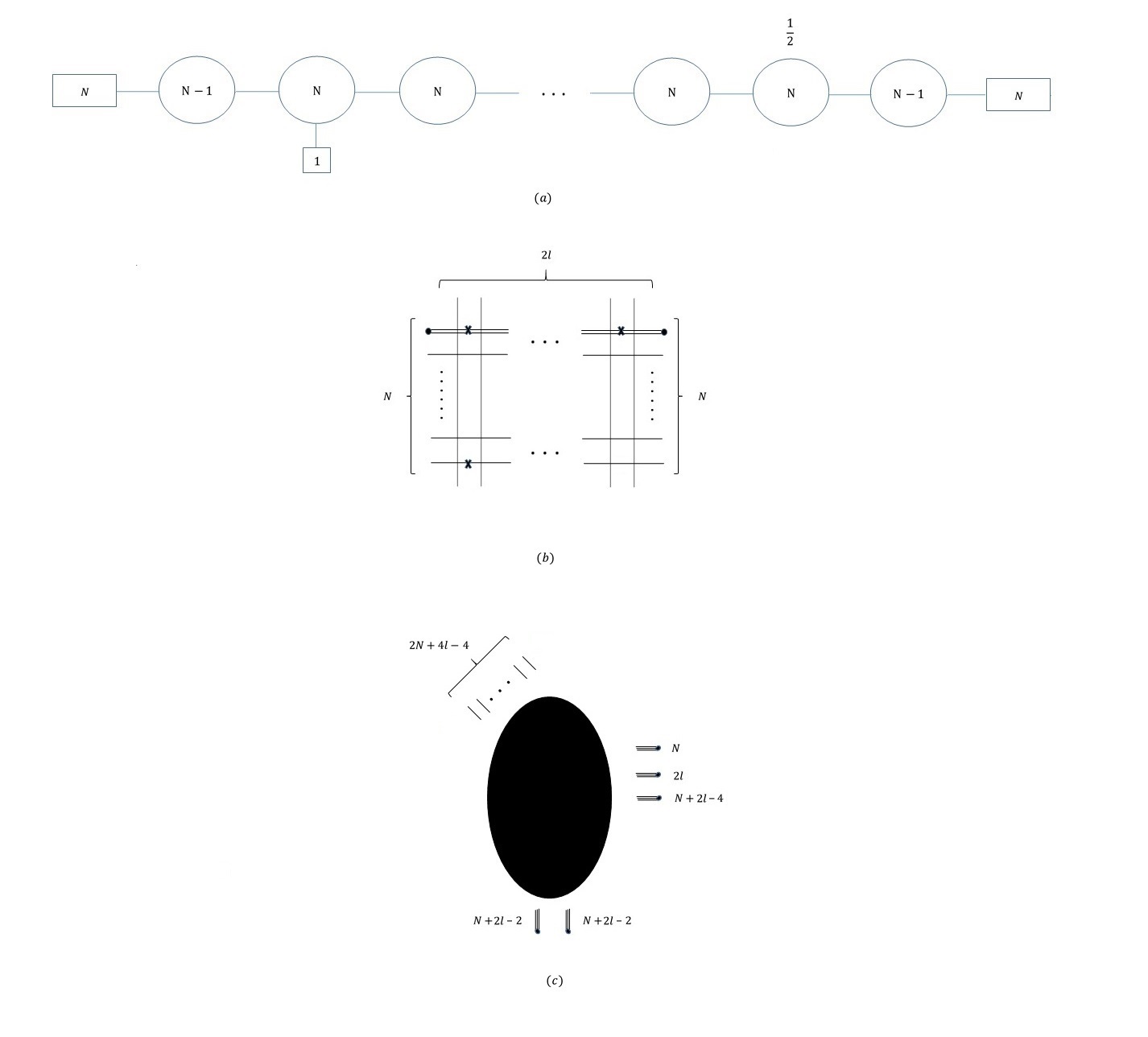} 
\caption{(a) The $5d$ quiver we are considering. (b) The brane web description of this $5d$ theory. (c) By pulling out the $7$-branes and doing HW transitions as needed, we can cast this web in the form of \cite{BB} where we assumed $l\geq 2$.}
\label{Img14}
\end{figure}

Taking the corresponding limit in $6d$, we get to the quiver of figure \ref{Img13}, which is the desired $6d$ SCFT. By construction, we are now assured that doing the T-duality on the brane system of this $6d$ quiver leads to the webs in figure \ref{Img12}. We can also consider compactifcation of the $6d$ theory to $4d$. As the Higgs branch limit and dimensional reduction should commute, we again expect the resulting $4d$ theory to be given by the class S theory whose $5d$ analogue is given by integrating out a flavor from the theories of \ref{Img12}. Naively, we have several different choices of which flavor to integrate out, but we find these all lead to the same class S theory, shown in figure \ref{Img14} (c).   

As a consistency check we can repeat the analysis of the central charges also in this case. It is apparent that $\Delta n_v = -8N-8k+36, \Delta n_h = -12N-12k+40$ and $\Delta n_t =0$ so using (\ref{fda}) we get that $\Delta d_H = -4N-4k+4$ and $\Delta n^{4d}_v = -8N-8k+12$. This indeed matches the results we get using class S technology. A straightforward calculation on both sides gives $k_{SU(2N+2k-4)} = k_{SU(2)} = 4(N+k-2)$, so the matching is also true in this case.

\subsubsection{Another example: the $5d$ $T_N$ theory with extra flavors}
   
\begin{figure}
\center
\includegraphics[width=0.8\textwidth]{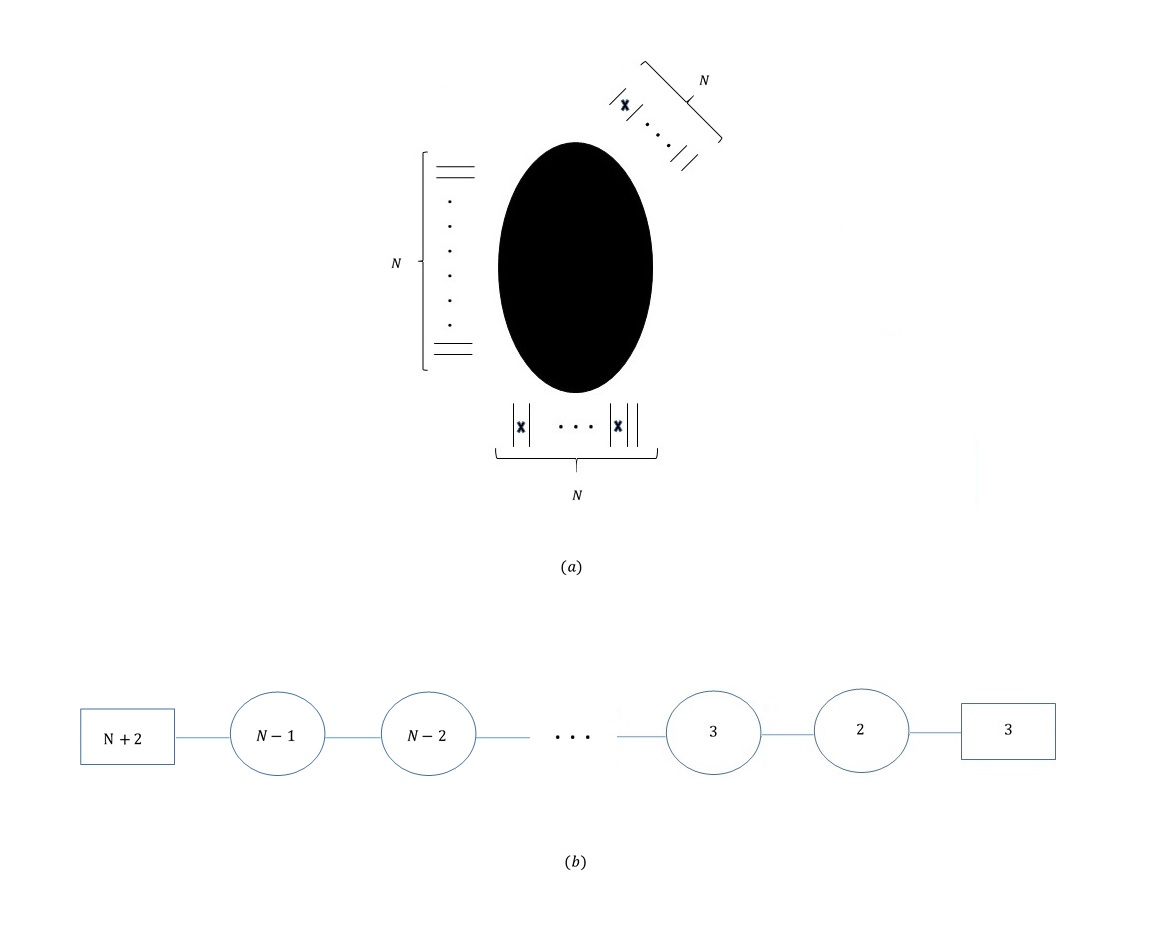} 
\caption{(a) The web for the $5d$ $T_N$ theory with $3$ D$7$-branes added at the marked location. (b) Using the gauge theory description of the $5d$ $T_N$ theory given in \cite{BZ}, one can see that this web describes the given quiver.}
\label{Img16}
\end{figure}

For our next example we consider a case where the Higgs branch limit involves a non-perturbative limit for the $6d$ SCFT. We consider the gauge theory we get by adding flavors to the $5d$ $T_N$ theory. Specifically, we add three flavors as shown in figure \ref{Img16} (a), corresponding to the gauge theory description in figure \ref{Img16} (b). This is expected to lift to $6d$, as first pointed out in \cite{KTY}. As a cross check, one can use the methods of \cite{Yon} to show that this theory has an enhancement to an affine $A^{(1)}_{3N-1}$ symmetry suggesting the $6d$ theory should have an $SU(3N)$ global symmetry.

We can get to the quiver of figure \ref{Img16} (b) by starting from the $F(N+2)+SU_0(N)^{N-2}+(N+2)F$ quiver and going on the Higgs branch. In the brane picture this corresponds to forcing $N-1$ $5$-branes to end on the same $7$-brane. Thus, in principal, we could determine the $6d$ theory by starting with the $6d$ theory of figures \ref{Img1} or \ref{Img5} and doing the breaking. However, this breaking cannot be done while staying in the realm of perturbative gauge theory. This can be seen by following this breaking, repeatedly forcing more and more D$5$-branes to end on the same D$7$-brane, which eventually lead to the rank $1$ $E$ string theory.    

\begin{figure}
\center
\includegraphics[width=0.8\textwidth]{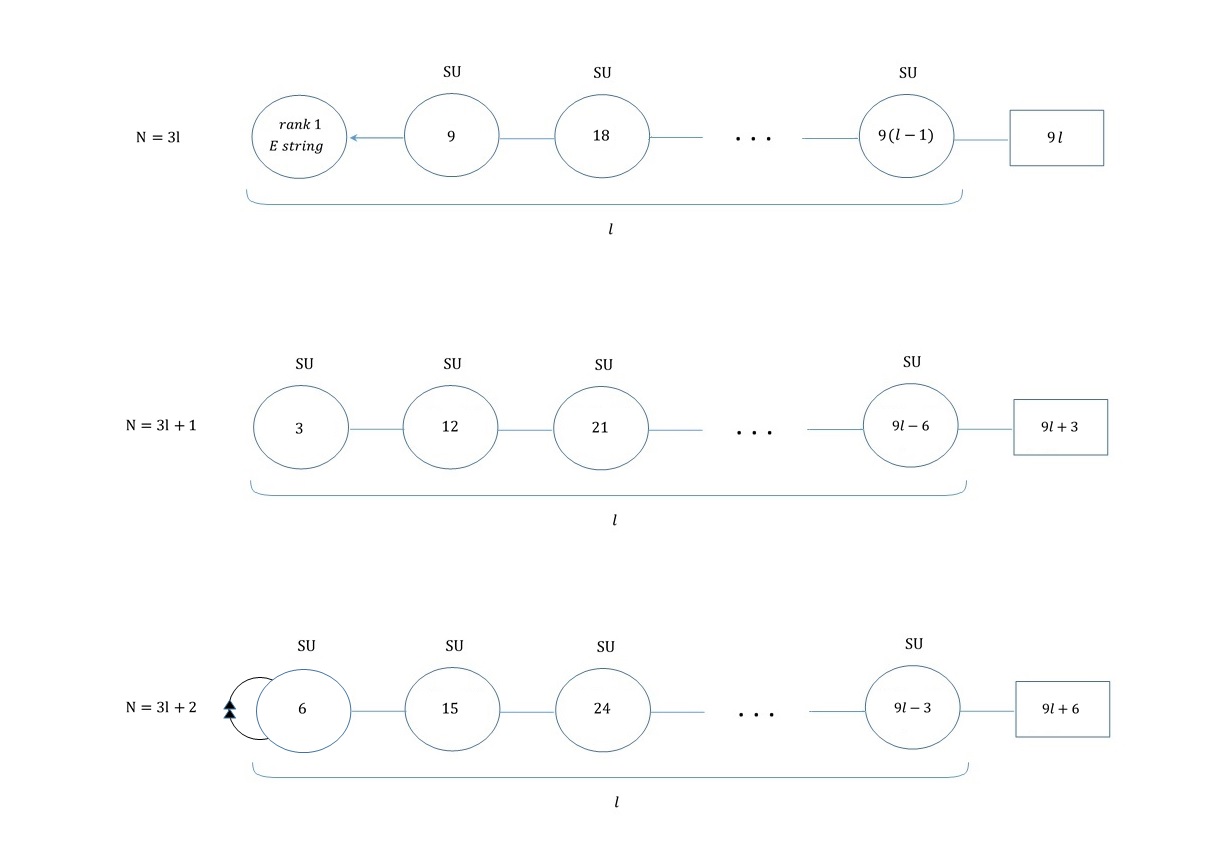} 
\caption{The conjectured $6d$ lift of the $5d$ theory shown in figure \ref{Img16}. In the $N=3l$ case the leftmost group is the rank $1$ $E$ string theory and the gauging is in the $SU(9)$ maximal subgroup of $E_8$. In the $N=3l+2$ case the leftmost $SU(6)$ group has an half-hyper in the $\bold{20}$.}
\label{Img19}
\end{figure}

Instead we present our conjecture for the $6d$ theory in this case. The gauge theory description is slightly different depending on whether $N=3l, 3l+1$ or $3l+2$ where $l$ is an integer. The explicit description is given in figure \ref{Img19}. We now wish to test this conjecture. First, we note that reducing this theory on a circle should indeed give the expected global symmetry and Coulomb branch dimension. In the simpler cases of $N=3,4$ we can also explicitly follow the Higgs breaking pattern and see that we indeed end up with the quivers of figure \ref{Img19}.


\begin{figure}
\center
\includegraphics[width=0.8\textwidth]{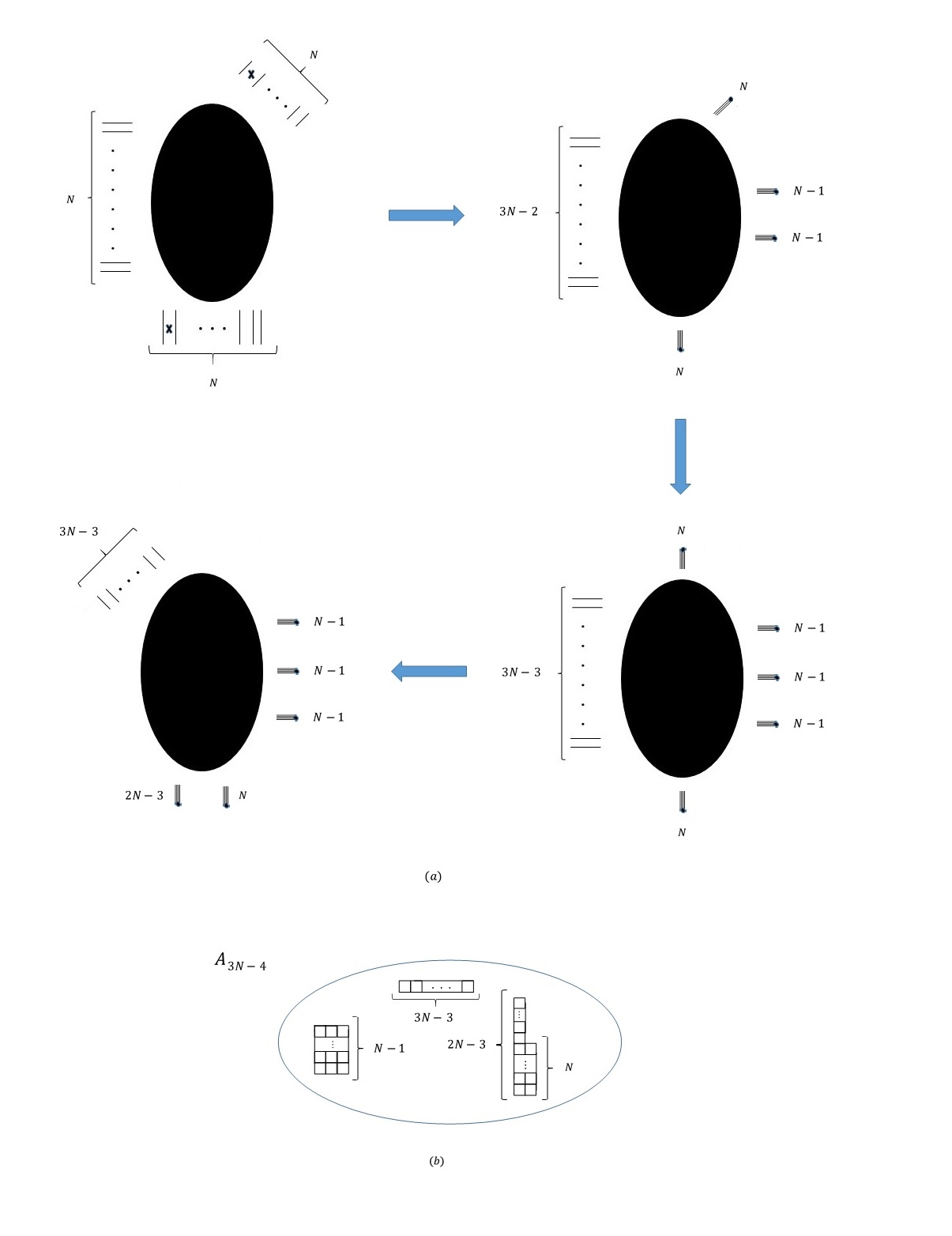} 
\caption{(a) Starting with the $T_N$ we add two flavors and do the required HW transitions finally arriving at the web on the bottom left. This web is of the form of \cite{BB}. Thus, we expect that compactifying the $6d$ theory of figure \ref{Img19} on a torus leads to an isolated $4d$ SCFT given by compactifying the $6d$ $(2,0)$ theory of type $A_{3N-4}$ on the punctured sphere shown in (b).}
\label{Img20}
\end{figure}


A more stringent test of this conjecture is in considering the compactification to $4d$ on $T^2$. As previously argued this should result in a class S theory with a $5d$ description given by integrating out one flavor, the brane web of which is shown in figure \ref{Img20} (a). From the web we can read the resulting $4d$ class S theory, see figure \ref{Img20} (b), as instructed in \cite{BB}. We can now also test this part of the conjecture. First, using class S technology, one can show that the global symmetry of this theory is indeed $SU(3N)$. We can also compare the central charges. For the class S theory we find:

\bea
d_H & = & \frac{(N-1)(9N+2)}{2}, \nonumber \\ n^{4d}_v & = & \frac{(N-1)(N-2)(10N+3)}{6}, \nonumber \\ k_{SU(3N)} & = & 6(N-1) \label{iTN}
\eea
for the theory in figure \ref{Img20} (b). 

On the $6d$ side, we first note that all $3$ cases have anomaly polynomials of the form (\ref{AP}) so we can use (\ref{fda})\footnote{The only different case here is the one with the $SU(6)$ where a direct calculation reveals that it is indeed of this form. Note that the tensor multiplet associated with this group is still of type $-1$\cite{HMRV}.}. We find that:

\be
n_t = l, \quad n_v = \frac{27l(l-1)(2l-1)}{2} -l+1 , \quad n_h = 27l(l^2-1)
\ee 
for $N=3l$.

\be
n_t = l, \quad n_v = \frac{27l(l+1)(2l+1)}{2} -54 l^2 -19l , \quad n_h = 9l(3l^2+3l-2)
\ee 
for $N=3l+1$. And

\be
n_t = l, \quad n_v = \frac{27l(l+1)(2l-1)}{2}+8l , \quad n_h = 27l(l+1)^2-18l+10
\ee 
for $N=3l+2$. Using (\ref{fda}) and (\ref{fcc}) we indeed recover (\ref{iTN}).






\section{Additional $5d$ theories}

We next want to consider additional $5d$ gauge theories lifting to $6d$ that are not covered, at least naively, by the theories considered so far, namely by limits on the Higgs branch of $(N+2)F+SU(N)^k+(N+2)F$. The reason why we say naively is that a given $5d$ fixed point may have many different IR gauge theory limits. Likewise there can be many $5d$ gauge theories all lifting to the same $6d$ SCFT, so even if a given $5d$ gauge theory is not of the form considered so far, it may be dual to one. Indeed we will see that all examples considered in this section are actually of this form, and the $6d$ lift can be determined by the previously explained procedure.

We concentrate only on $5d$ theories with an ordinary brane web description, that is without orientifold planes. One possibility is $5d$ linear $SU$ quivers not of the form considered. Another possibility is to look at linear $SU$ quivers with a $USp$ or $SU$ with an antisymmetric hyper multiplet, at one or both edges of the quiver. The latter can be constructed using an $O7^-$ plane, which, when resolved, leads to an ordinary brane web. These are the cases we consider.  


\subsection{Quivers of $SU$ groups}

\begin{figure}
\center
\includegraphics[width=0.6\textwidth]{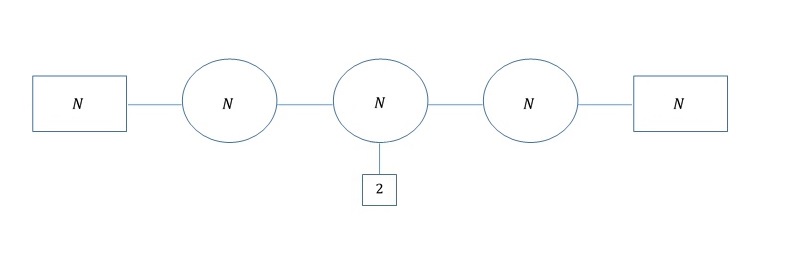} 
\caption{The quiver diagram for the $5d$ gauge theory. All groups are of $SU$ type with CS level $0$.}
\label{Img25}
\end{figure}

\begin{figure}
\center
\includegraphics[width=1\textwidth]{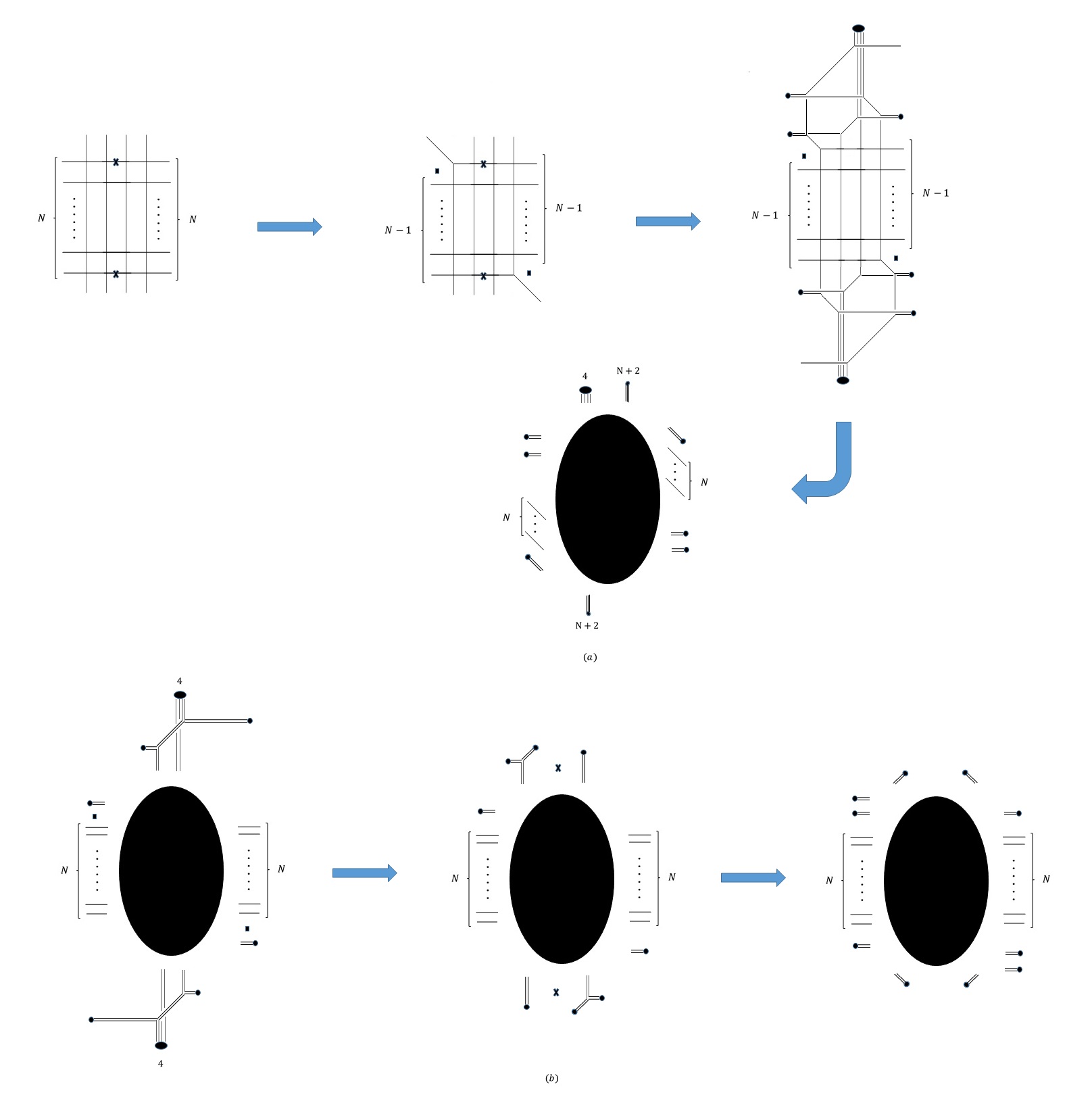} 
\caption{(a) The brane web for the $5d$ gauge theory of figure \ref{Img25}. After manipulating some of the $7$-branes, we arrive at the configuration at the bottom. It is now apparent that the two groups of $(1,-1)$ $7$ branes spiral indefinitely. (b) Starting with the same configuration, we can by moving the $(0,1)$ $7$-branes through the monodromy of the other $7$-branes, get to the web in the middle. Pulling the resulting $(1,0)$ $7$-branes trough the NS$5$ branes lead to the web on the right, which is a Higgs branch limit of a theory of the form of figure \ref{Img4} (a).}
\label{Img26}
\end{figure}

We start by considering $SU$ quivers not of the form discussed so far. Nevertheless, if we want to get at most an affine Lie group as a global symmetry, then the analysis of \cite{Yon} suggests that the possibilities are  limited to short quivers. Consider a quiver of $3$ $SU(N)$ groups with $N$ fundamentals for the edge groups and two fundamentals for the middle one, the quiver diagram of which is shown in figure \ref{Img25}. The analysis of \cite{Yon} suggests it should have a $D^{(1)}_6$ global symmetry and so is expected to lift to $6d$. Indeed, as figure \ref{Img26} (a) shows, it has a spiraling tau diagram which are characteristic of theories lifting to $6d$\cite{KTY}. 

\begin{figure}
\center
\includegraphics[width=0.6\textwidth]{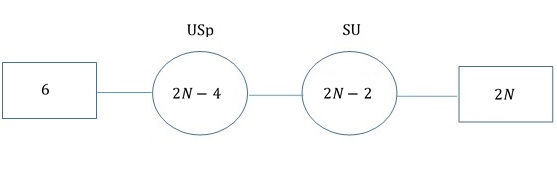} 
\caption{The quiver diagram for the $6d$ gauge theory we get after implementing the Higgs branch flow of figure \ref{Img26} (b) on the theory of figure \ref{Img1}.}
\label{Img27}
\end{figure}

We next inquire to what $6d$ theory does it go to. As figure \ref{Img26} (b) shows, this theory can be reached by going on the Higgs branch of the theory in figure \ref{Img4} (a). Using the $6d$ lift of the latter theory, of the form of figure \ref{Img1}, and taking the appropriate Higgs branch limit, we end up with the $6d$ SCFT given by the quiver of figure \ref{Img27}. Note that we indeed have the $6d$ $SO(12)$ expected from the affine symmetry. We can go on to further test this. By construction we are guaranteed that taking the T-dual of the brane configuration for the $6d$ theory of figure \ref{Img27} leads to the web of figure \ref{Img26} (b).   

\begin{figure}
\center
\includegraphics[width=0.8\textwidth]{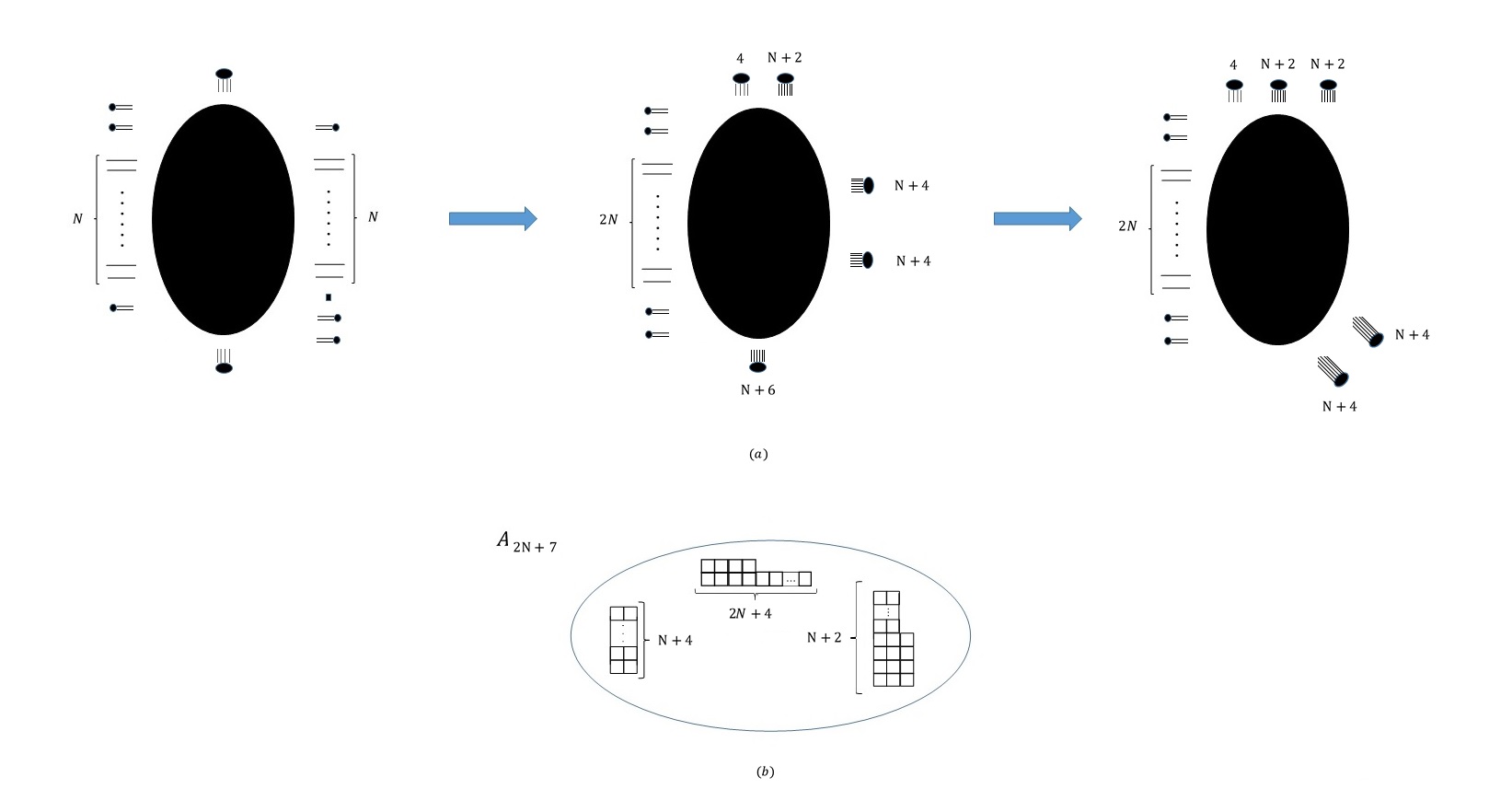} 
\caption{(a) Starting from the configuration of figure \ref{Img26} without one of the flavors, we get to this brane web. One can see that it is in the form of \cite{BB} so compactification to $4d$ will yield the isolated SCFT of (b).}
\label{Img28}
\end{figure}

One test we can carry is to consider the compactification to $4d$ on a torus. Again, we expect to get a class S theory given by integrating out a flavor so that the $6d$ global symmetry is preserved.
Consider integrating out one of the $N$ flavors at one of the ends. This leads to the class S theory shown in figure \ref{Img28} (b). The punctures show an $SU(2N)\times SU(4)\times SU(2)^2\times U(1)^2$ global symmetry, but the $4d$ superconformal index revels that the $SU(4)\times SU(2)^2\times U(1)$ part is enhanced to $SO(12)$ as expected from the $6d$ theory (note, however, that integrating one of the two mid group flavor leads to a different class S theory with different global symmetry). The $N=2$ case is special, where there is a further enhancement of symmetry. We shall discuss this case latter, from a dual view point.


We can further test this by comparing the central charges of said class S theory with the ones expected for the $6d$ theory compactified on a torus. Using class S technology we find:

\bea
d_H & = & 2N^2 + 11N + 33, \nonumber \\ n^{4d}_v & = & 6N^2 + 33N - 41, \nonumber \\ k_{SU(2N)} & = & 2(2N + 4), \nonumber \\ k_{SO(12)} & = & 4(N + 4)
\eea
This indeed matches the results we get from (\ref{fda}) and (\ref{fcc}).

\begin{figure}
\center
\includegraphics[width=0.6\textwidth]{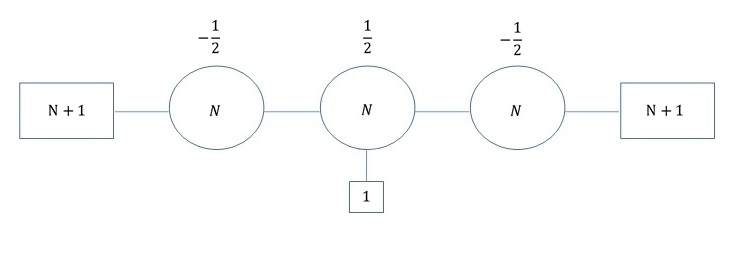} 
\caption{The quiver diagram for the $5d$ gauge theory. All groups are of type $SU$ with the CS level for each given above it.}
\label{Img30}
\end{figure}

\begin{figure}
\center
\includegraphics[width=0.6\textwidth]{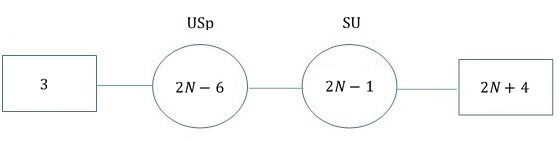} 
\caption{The quiver diagram for the $6d$ gauge theory, which is the expected lift of the $5d$ gauge theory of figure \ref{Img30}.}
\label{Img32}
\end{figure}

\begin{figure}
\center
\includegraphics[width=1\textwidth]{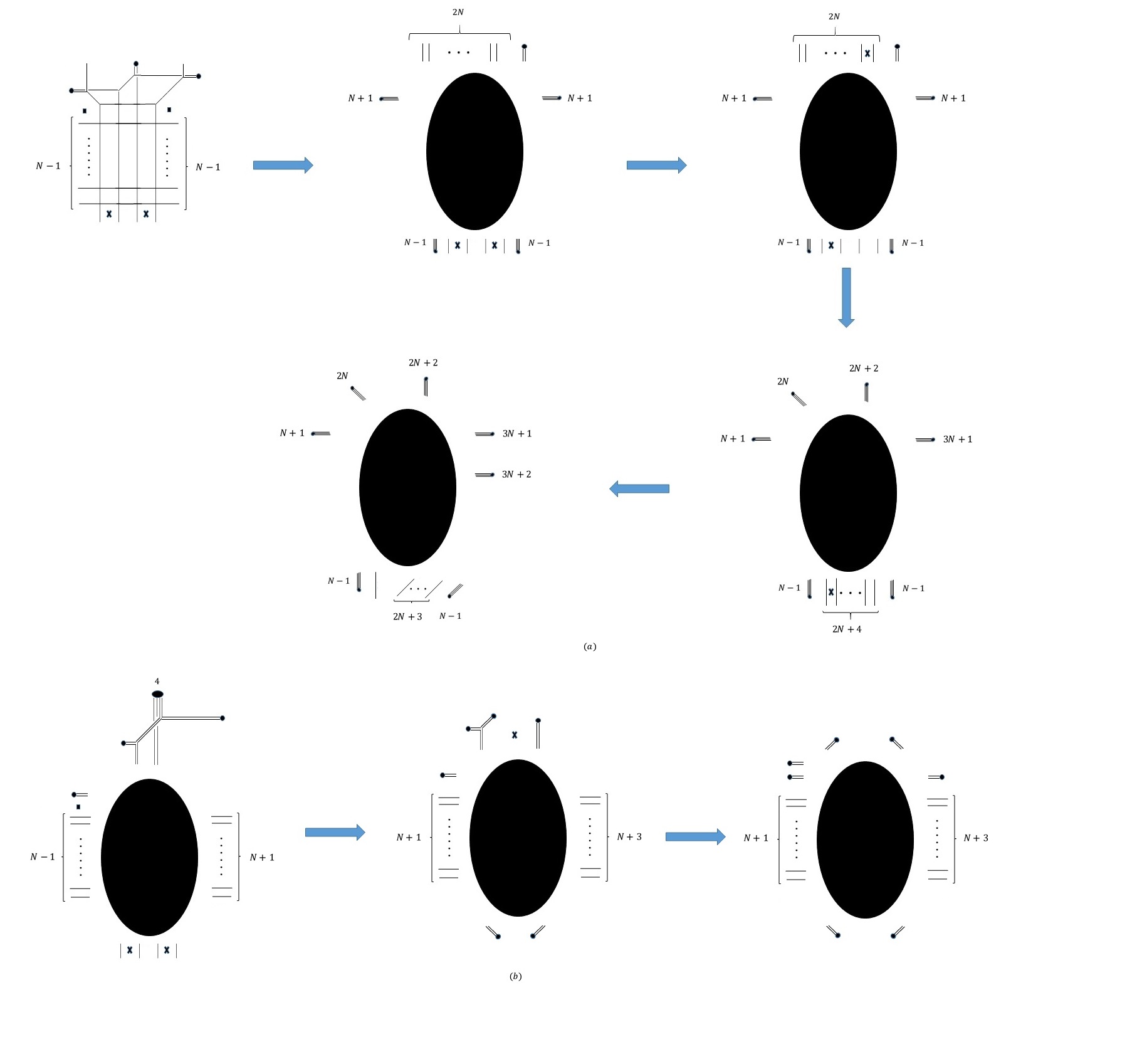} 
\caption{(a) Starting from the brane web for the $5d$ gauge theory of figure \ref{Img30}, after manipulating some of the $7$ branes, we arrive at the spiraling configuration shown on the bottom left. (b) Starting with the same configuration, doing some $7$-brane gymnastics, we get to the web on the right. One can note that this is a Higgs branch limit of a theory of the form of figure \ref{Img4} (a).}
\label{Img31}
\end{figure}

A related case is given by letting each group see $2N+1$ flavors, the quiver diagram of which is shown in figure \ref{Img30}. We claim that with the CS levels chosen as they are, this theory also lifts to $6d$, particularly, the theory shown in figure \ref{Img32}. We can present evidence for this conjecture. First, note that the brane web for this theory has a spiraling tau form, see figure \ref{Img31} (a), supporting the claim that it lifts to $6d$. 

Also as shown in figure \ref{Img31} (b), we can map this web to a form as a Higgs branch limit of the theories in figure \ref{Img4} (a). It is now not difficult to see that implementing this breaking on the $6d$ lift, of the form presented in figure \ref{Img1}, leads to the quiver of figure \ref{Img32}. Finally, we can also consider the reduction to $4d$ on a torus. We expect the $4d$ theory to be described by the case with one less flavor shown in figure \ref{Img34}. We can calculate the central charges of this theory finding:  


\begin{figure}
\center
\includegraphics[width=0.8\textwidth]{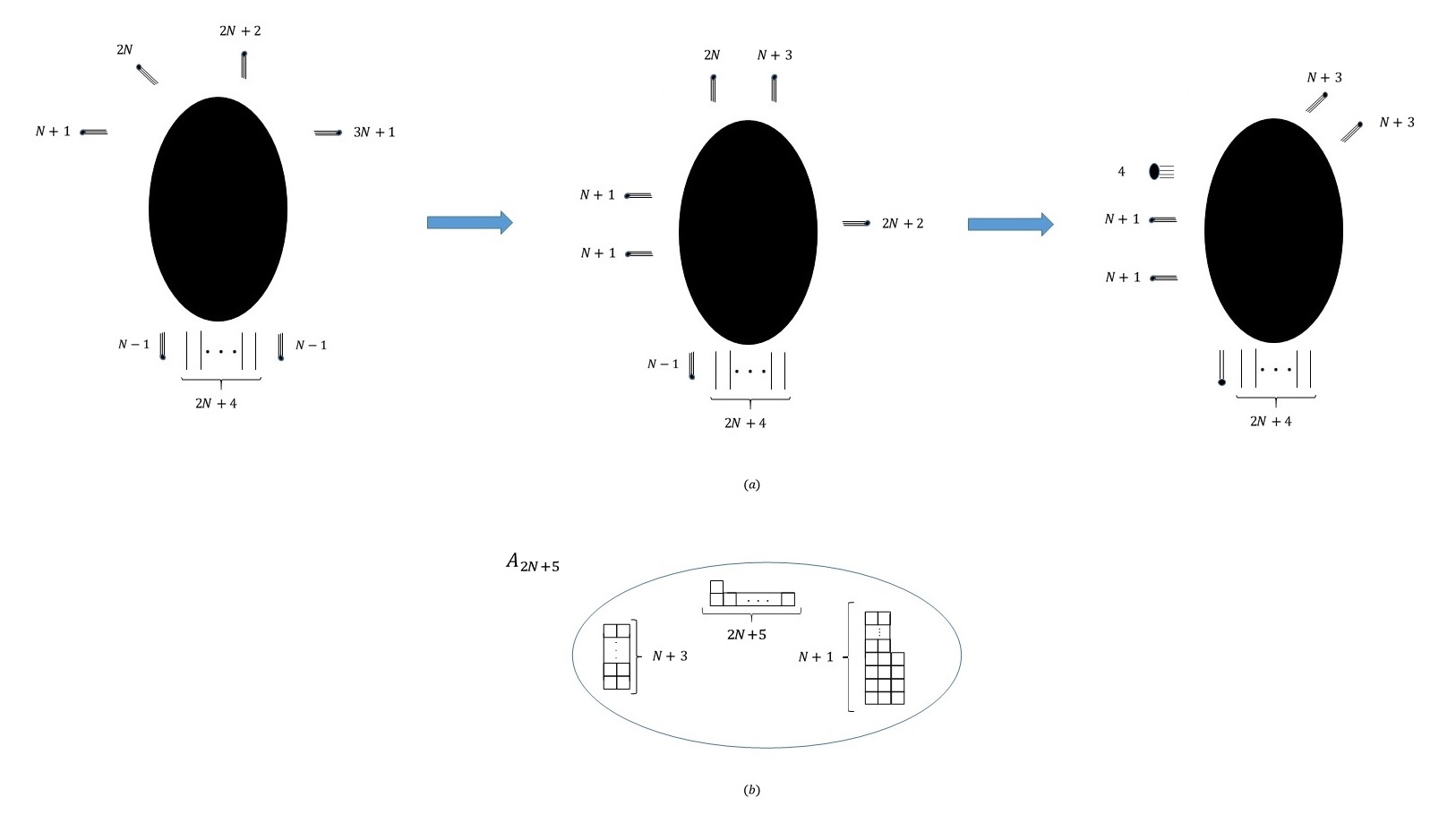} 
\caption{(a) Starting from the configuration of figure \ref{Img31} without one of the flavors, we get to this brane web. One can see that it is in the form of \cite{BB} so compactification to $4d$ will yield the isolated SCFT of (b).}
\label{Img34}
\end{figure}

\bea
d_H & = & 2N^2 + 13N + 27, \nonumber \\ n^{4d}_v & = & 6N^2 + 33N - 47, \nonumber \\ k_{SU(2N+4)} & = & 4N + 10, \nonumber \\ k_{SU(4)} & = & 4N + 12 
\eea

This indeed matches the results we get from (\ref{fda}) and (\ref{fcc}).

\subsection{$SU$ quivers with antisymmetric hypers} 


In this subsection we look at $SU(N)$ quivers with an antisymmetric hyper at one or both ends. These also can be described by an ordinary brane web which we can get either by constructing these theories with an $O7^-$ plane and resolving it, or by directly building the quiver using the brane web for $SU(N)$ with an antisymmetric given in \cite{BZ,BZ1}. There is another class of theories, $SU$ quivers with $USp$ ends, that can also be constructed using these methods. But these can be generated by a Higgs branch limit of the theories we consider in this section, and so it should be straightforward to generalize these results also for this class. 


\subsubsection{$SU$ quivers with an antisymmetric hyper at one end}

\begin{figure}
\center
\includegraphics[width=0.8\textwidth]{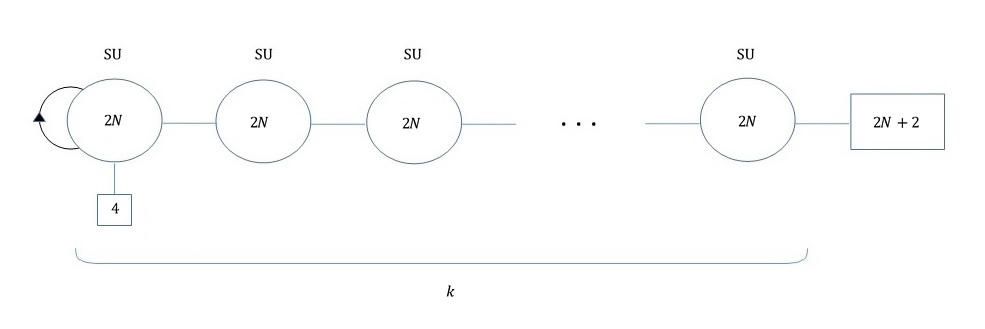} 
\caption{The quiver diagram for the $5d$ gauge theory.}
\label{Img35}
\end{figure}

\begin{figure}
\center
\includegraphics[width=0.8\textwidth]{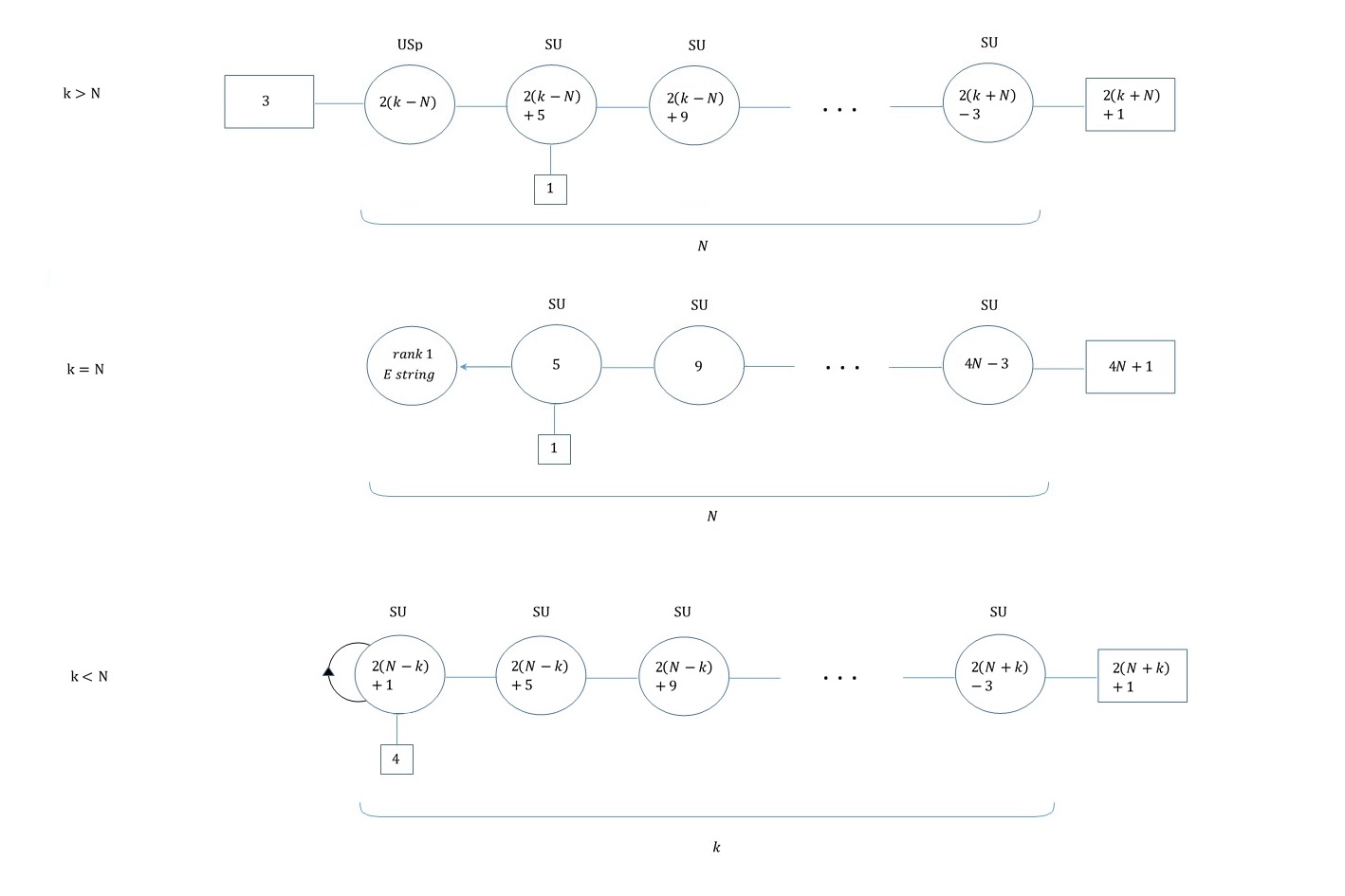} 
\caption{The quiver diagram for the $6d$ gauge theory, which is the expected lift of the $5d$ gauge theory of figure \ref{Img35}..}
\label{Img36}
\end{figure}

\begin{figure}
\center
\includegraphics[width=1\textwidth]{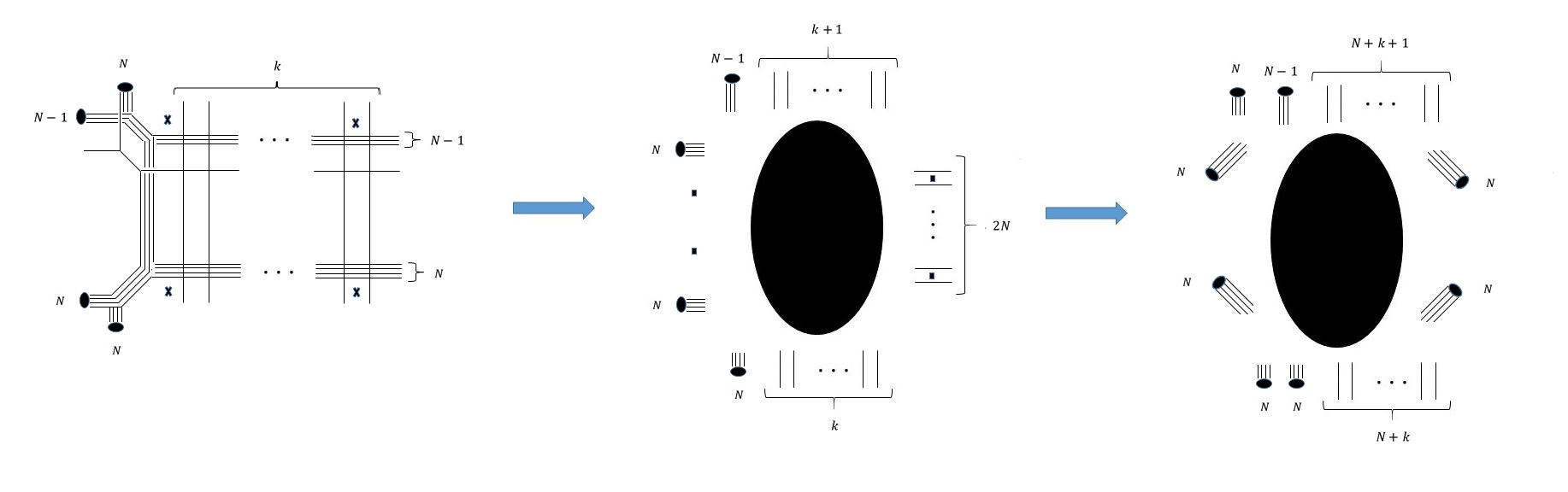} 
\caption{The web for the theory of \ref{Img35}. After some manipulations we can get to the form as a Higgs branch limit of the theories in figure \ref{Img4} (a) given by the S-dual of the web on the right.}
\label{Img37}
\end{figure}

\begin{figure}
\center
\includegraphics[width=1\textwidth]{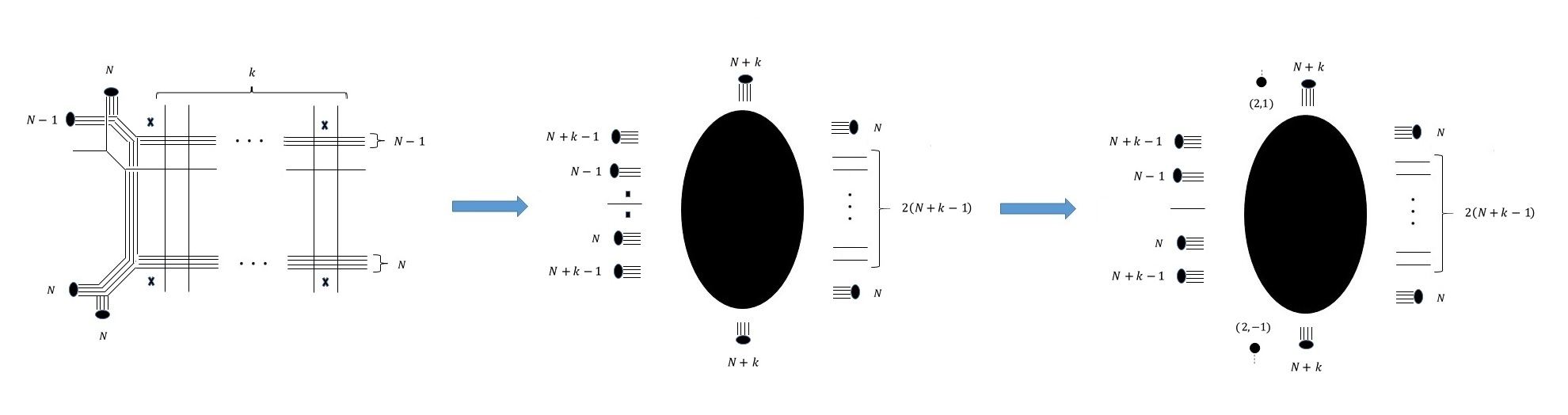} 
\caption{The web for the theory of \ref{Img35}. After some manipulations we can get to the form as a Higgs branch limit of the theories in figure \ref{Img6}.}
\label{Img39}
\end{figure}

We start with the $5d$ theory of figure \ref{Img35}. We claim that this theory lifts to the $6d$ theory of figure \ref{Img36}. Our evidence for this is similar to the previous cases. First, by manipulating the brane web for the theory, we can bring it to a form as a Higgs branch limit of the theories presented in section 3. Besides supporting the claim that this theory lifts to $6d$, we can, by taking the required Higgs branch limit on the $6d$ lifts given in section 3, also argue that the quivers given in figure \ref{Img36} are indeed the required $6d$ lifts. This is shown for the $k>N$ case in figure \ref{Img37}, and for the $k<N$ case in figures \ref{Img39}.


\begin{figure}
\center
\includegraphics[width=1\textwidth]{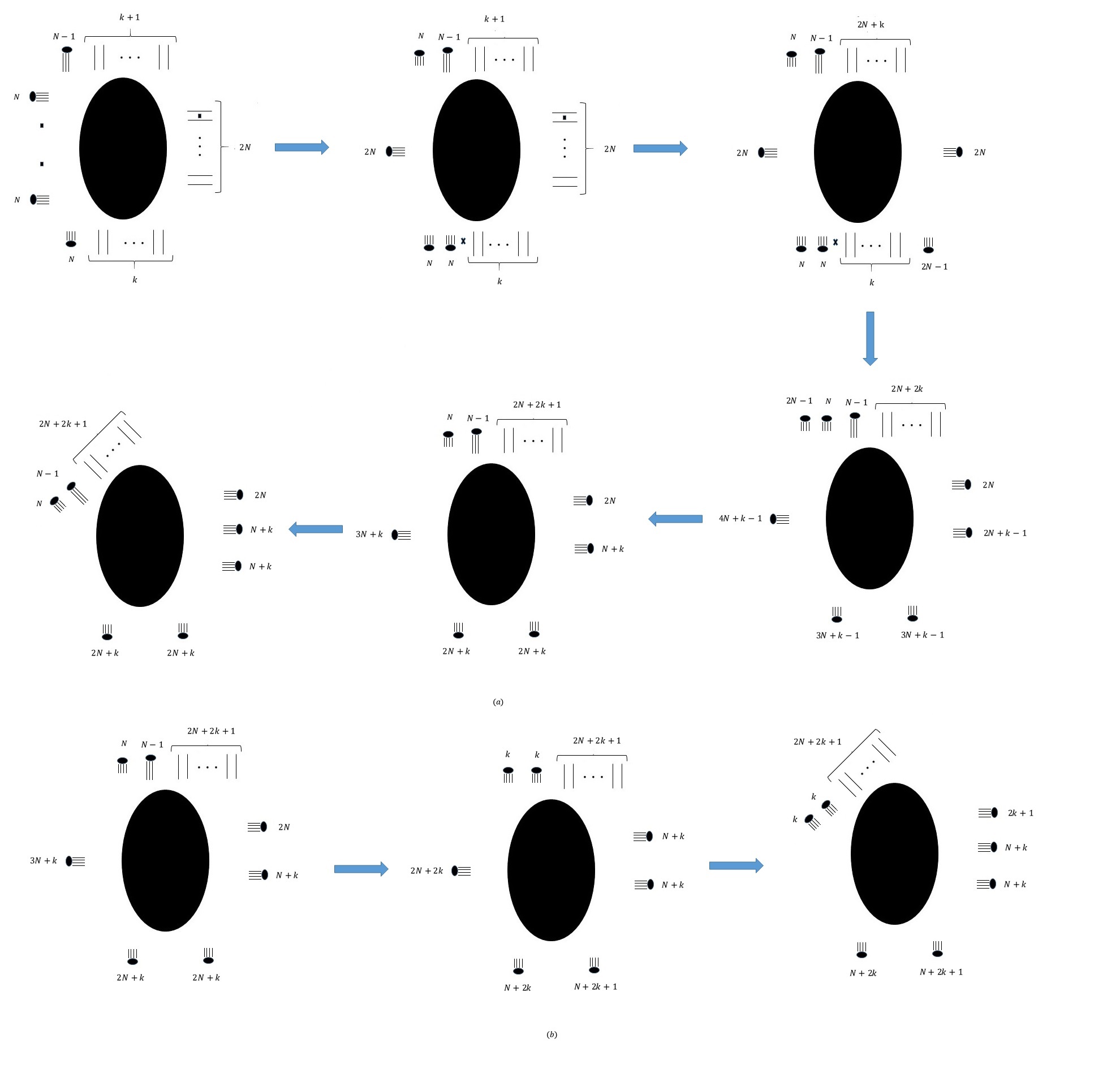} 
\caption{(a) Starting from the configuration of figure \ref{Img37} without one of the flavors, and doing a series of HW transitions, we get to the brane web in the bottom left. One can see that it is in the form of \cite{BB} so compactification to $4d$ will yield the appropriate isolated SCFT. This is the form most suited to the $k\geq N$ case. For the $N>k$ case, the one in (b), gotten from (a) by shuffling some of the $7$-branes, is more adequate.}
\label{Img40}
\end{figure}

We can again consider the reduction to $4d$ on a torus. We expect the $4d$ theory to be described by the case with one less flavor shown in figure \ref{Img40}. The punctures suggests a global symmetry of $SU(2N+2k+1)\times SU(2)^2\times U(1)^3$ except in some special cases,  for example, when $k=N$ or $k=N-1$ where the symmetry enhances to $SU(2N+2k+1)\times SU(3)\times SU(2)\times U(1)^2$. From the $4d$ superconformal index we see that there is a further enhancement of $U(1)\times SU(2)^2\rightarrow SU(4)$, which becomes $U(1)\times SU(2)\times SU(3)\rightarrow SU(5)$ when $k=N$ or $k=N-1$. This enhancements, including the special cases with enhanced symmetry, exactly matches the ones expected from the $6d$ SCFT of figure \ref{Img36}. We can also calculate the central charges of this theory finding: 

\bea
d_H & = & 2N^2+2k^2+4kN+17N+9k, \nonumber \\ n^{4d}_v & = & \frac{(2N-1)(6k^2+4N-4N^2+9k+18kN)}{3}, \nonumber \\ k_{SU(2N+2k+1)} & = & 2(2N+2k+3), \nonumber \\ k_{SU(4)} & = & 4k+8N
\eea
for $k\geq N$ where for $k=N$ $SU(4)\rightarrow SU(5)$, and

\bea
d_H & = & 2N^2+2k^2+4kN+17k+9N+4, \nonumber \\ n^{4d}_v & = & \frac{k(12N^2+24N-8k^2-13-18k+36kN)}{3}, \nonumber \\ k_{SU(2N+2k+1)} & = & 2(2N+2k+3), \nonumber \\ k_{SU(4)} & = & 4N+8k+2
\eea
for $N>k$. This indeed matches the results we get from (\ref{fda}) and (\ref{fcc}).

\subsubsection{$SU$ quivers with an antisymmetric hyper at both ends}

\begin{figure}
\center
\includegraphics[width=0.6\textwidth]{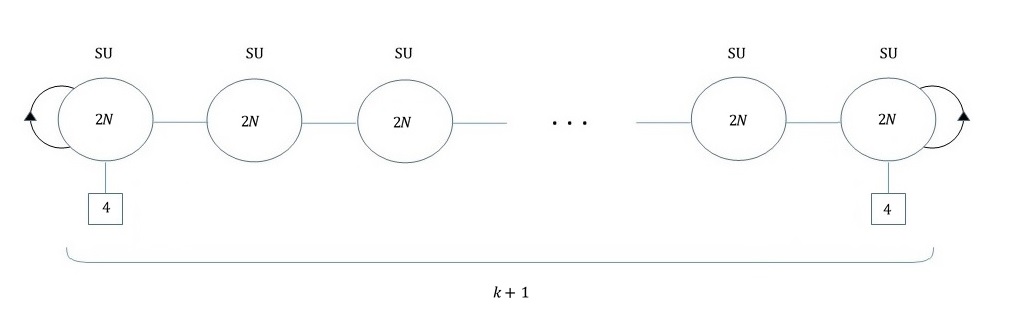} 
\caption{The quiver diagram for the $5d$ gauge theory.}
\label{Img41}
\end{figure}

\begin{figure}
\center
\includegraphics[width=0.6\textwidth]{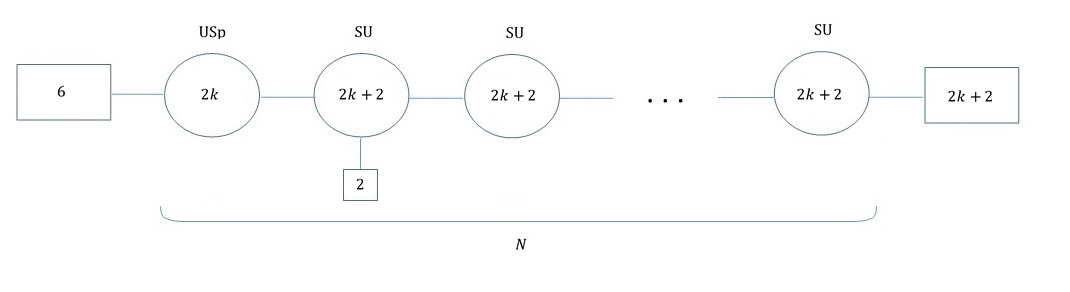} 
\caption{The quiver diagram for the $6d$ gauge theory, which is the expected lift of the $5d$ gauge theory of figure \ref{Img41}.}
\label{Img42}
\end{figure}

\begin{figure}
\center
\includegraphics[width=1\textwidth]{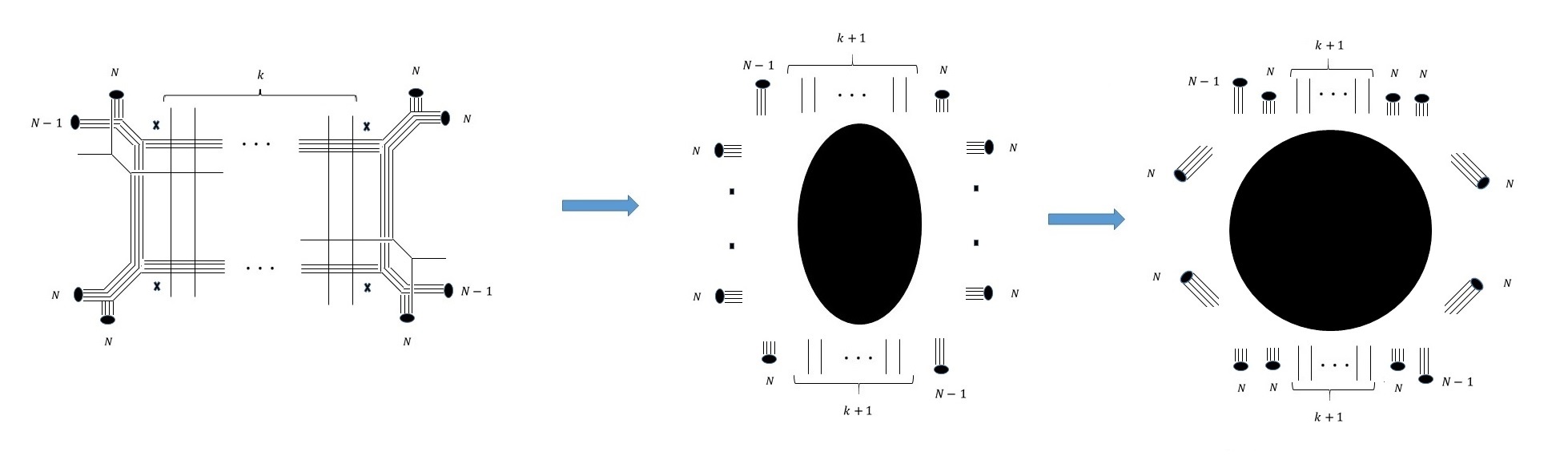} 
\caption{Starting from the web for the gauge theory of figure \ref{Img41}, we can get to a form as a Higgs branch limit of the web in figure \ref{Img4} (a).}
\label{Img43}
\end{figure}

We can next consider the case where both ends are $SU$ groups with an antisymmetric so we have the quiver theory of figure \ref{Img41}. We conjecture the $6d$ lift to be the one shown in figure \ref{Img42}. We can repeat the same steps as before, first deform the web to give a Higgs branch limit of a theory of figure \ref{Img4} (a). This gives the web shown in figure \ref{Img43}. By implementing the required breaking on the $6d$ SCFT of figure \ref{Img1} we indeed get the quiver of figure \ref{Img42}.

\begin{figure}
\center
\includegraphics[width=0.8\textwidth]{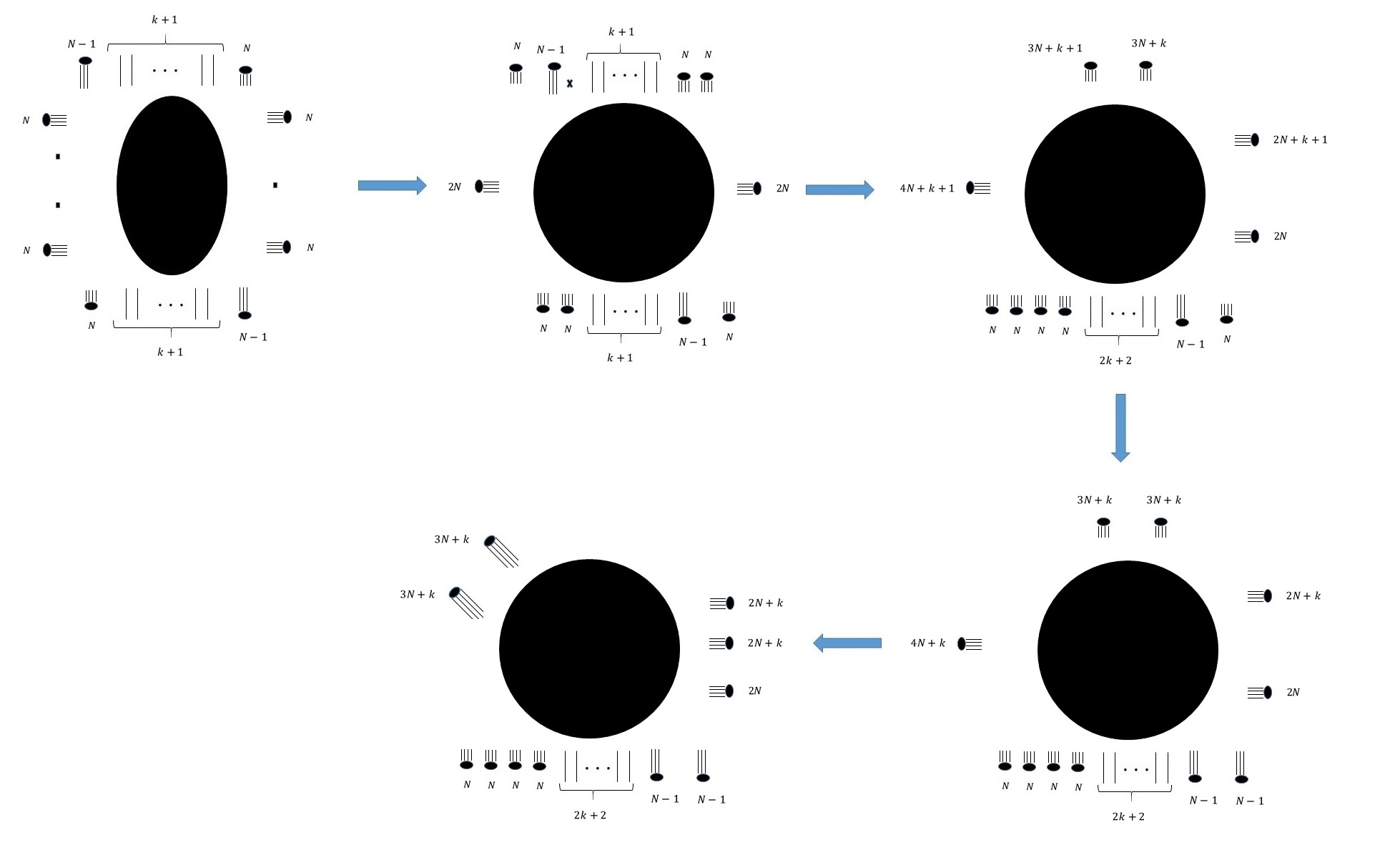} 
\caption{Starting from the configuration of figure \ref{Img43} without one of the flavors, we get to this brane web. One can see that it is in the form of \cite{BB} so compactification to $4d$ will yield the appropriate isolated SCFT.}
\label{Img44}
\end{figure}

As an additional test, we can again consider the reduction to $4d$ on a torus. We expect the $4d$ theory to be described by the case with one less flavor shown in figure \ref{Img44}. The global symmetry visible from the punctures is $SU(2k+2)\times SU(4)\times SU(2)^3\times U(1)^3$, which is further enhanced when $k=0$ or $N=2$. When $k\neq 0$, we can show from the superconformal index that there is an enhancement of $SU(4)\times SU(2)^2\times U(1)\rightarrow SO(12)$. This, including the enhancement when $N=2$, exactly matches what is expected from the $6d$ global symmetry. 

However, the $k=0$ case, the $5d$ SCFT of which corresponds to the $5d$ gauge theory $SU_{\frac{1}{2}}(2N)+2AS+7F$, has some puzzling features. First, let's start with the global symmetry for the SCFT of figure \ref{Img44}. As argued in the appendix, instanton counting methods suggests this theory has an $E_7 \times SU(2)^3$ global symmetry which is further enhanced to $E_7 \times SO(7)$ for $N=2$. This is further supported by the $4d$ superconformal index. Note that the $N=2$ case discussed here is identical to the $N=2$ case for the theory in figure \ref{Img28} (b), which provides a dual gauge theory description for the same fixed point. 

Comparing with the $6d$ side, we naively encounter a contradiction. When $k=0$ we have a long quiver of $SU(2)$ groups leading to an enhancement of the $U(1)$ bifundamental global symmetries to $SU(2)$'s. More importantly the mixed anomalies leading to the breaking of most of these $U(1)$'s now vanish so we naively expect to have an $SU(2)^{N+1}$ global symmetry contradicting the global symmetry suggested by the $5d$ description. The issue appears to be the discrepancies between the global symmetry suggested from the gauge theory and the one that actually exists in the SCFT mentioned in section 2. To truly understand the $6d$ SCFT we should consider a string theory realization of it. 

Fortunately, the $6d$ SCFT at hand was considered in \cite{HMRV}. They considered a class of theories engineered in string theory by a group of M$5$-branes probing a $C^2/Z_{2k+2}$ orbifold and an M$9$-plane. One of the theories in this class is the theory with gauge theory description given in figure \ref{Img42}. This is no coincidence as the original $5d$ gauge theory, shown in figure \ref{Img41}, can be engineered by a group of D$4$-branes probing a $C^2/Z_{2k+2}$ orbifold and an $O8^-$ plane\cite{BG} so it is natural to expect the $6d$ lift to be of this form.  

 According to the analysis of \cite{HMRV}, the non-abelian global symmetry of this $6d$ SCFT is indeed $SO(12)\times SU(2k+2)\times SU(2)$. The case $k=0$ is special: the non-abelian global symmetry is actually $E_7 \times SU(2)^3$. The extra $SU(2)$ is there since the orbifold $C^2/Z_{2}$ preserves the full $SO(4)$ symmetry, while $C^2/Z_{2k+2}$ breaks one of the $SU(2)$'s\footnote{I am grateful for J. J. Heckman for making his work known to me and for discussing this point.}. So this appears to agree with what we see from the instanton counting analysis done in the appendix.       

The case $k=0, N=2$ is more special. Then the $6d$ theory is known as the $(E_7,SO(7))$ conformal matter\cite{ZHTV}. Again the gauge theory shows an $SO(8)$ global symmetry, while it is known the SCFT only has $SO(7)$. This indeed agrees with the results from instanton counting done in the appendix.

We can also to calculate the central charges of this theory finding: 

\bea
d_H & = & 2k^2+30N+19k+3, \nonumber \\ n^{4d}_v & = & 12N^2(k+1)+8Nk-7k+2k^2(2N-1)+2N-3, \nonumber \\ k_{SU(2k+2)} & = & 4k+16, \nonumber \\ k_{SU(2)} & = & 4k+12N-8, \nonumber \\ k_{E_7} & = & 4k+12N
\eea

This indeed matches the results we get from (\ref{fda}) and (\ref{fcc}), supporting the claim that compactifying the $6d$ SCFT of figure \ref{Img42} on a torus leads to the isolated $4d$ SCFT of figure \ref{Img44}. Incidentally, the compactification of the $(E_7,SO(7))$ conformal matter on a torus was already considered in \cite{OSTY1}. They conjectured that the resulting theory is given in terms of a compactification of the $E_6$ $(2,0)$ theory on a Riemann sphere with three punctures labeled: $0,2A_1,E_6(a_1)$ (see \cite{CDT}, for a discussion on the meaning of the notation and for properties of this SCFT). They further compared the central charges of this theory to the ones expected from the compactification, finding an exact match.

\begin{figure}
\center
\includegraphics[width=0.8\textwidth]{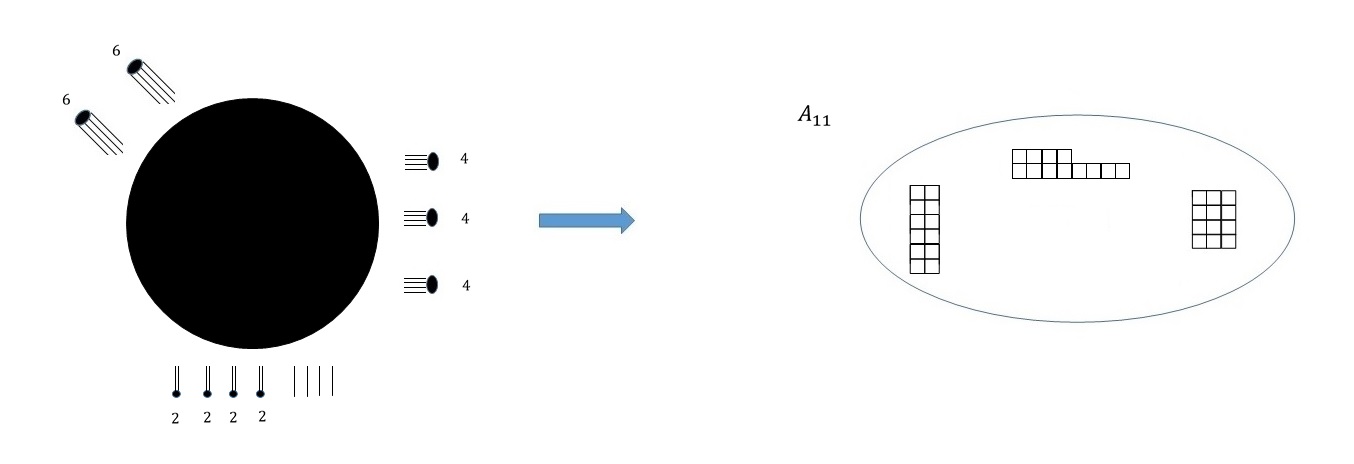} 
\caption{The brane web for $SU_{\pm\frac{1}{2}}(4)+2AS+7F$. From this one can arrive at the representation of its associated $4d$ SCFT as a compactification of an $A$ type $(2,0)$ theory on a three punctured sphere.}
\label{Img49}
\end{figure}

Consistency of these two approaches then suggests that these two theories are in fact the same theory. Indeed, we calculated the central charges and spectrum of Coulomb branch operators of the theory in figure \ref{Img49}, finding exact matching to the previously mentioned SCFT from compactifying the $E_6$ $(2,0)$ theory.

\begin{figure}
\center
\includegraphics[width=0.6\textwidth]{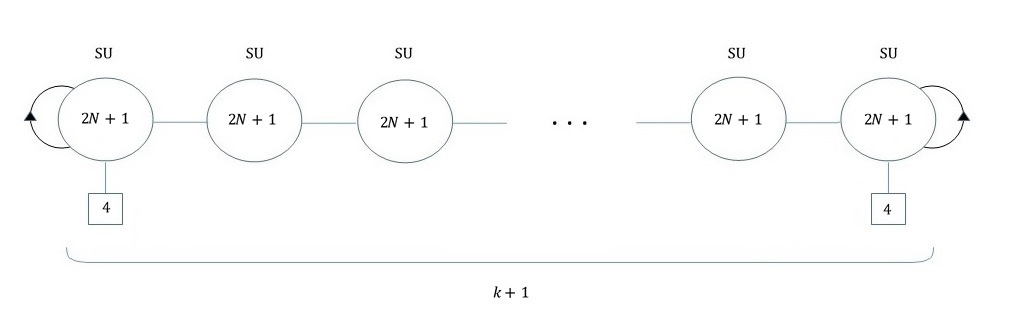} 
\caption{The quiver diagram for the $5d$ gauge theory.}
\label{Img45}
\end{figure}

\begin{figure}
\center
\includegraphics[width=0.6\textwidth]{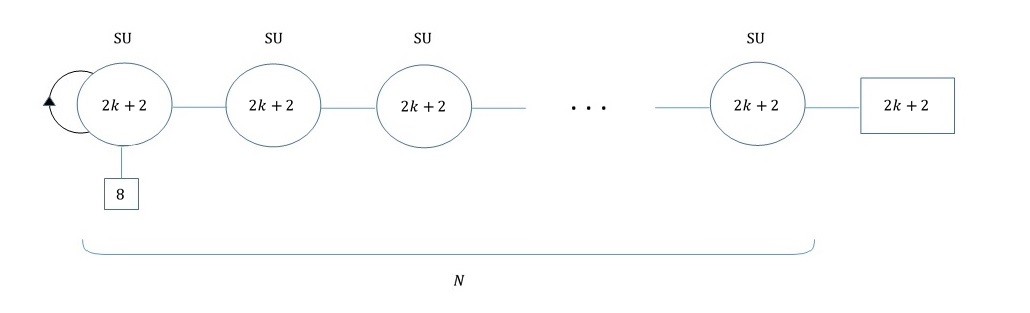} 
\caption{The quiver diagram for the $6d$ gauge theory, which is the expected lift of the $5d$ gauge theory of figure \ref{Img45}.}
\label{Img46}
\end{figure}

\begin{figure}
\center
\includegraphics[width=1\textwidth]{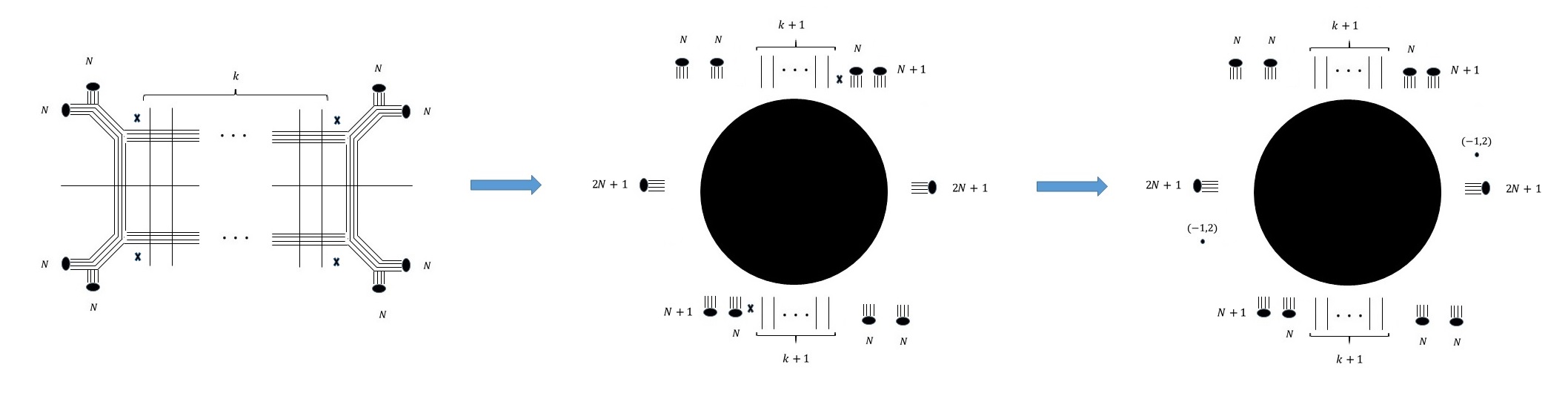} 
\caption{Starting from the web for the theory of \ref{Img45}, we can cast it in a form as a Higgs branch limit of the web in figure \ref{Img6}. This is given by the S-dual of the rightmost web.}
\label{Img47}
\end{figure}

We can also consider the even rank case shown in figure \ref{Img45}. While this can be figured out from the previous case by going on the Higgs branch, we will mention this case. We expect the $6d$ theory to be the one shown in figure \ref{Img46}. This can be argued by manipulating the brane web into a form, shown in figure \ref{Img47}, as a Higgs branch limit of the theory of figure \ref{Img6}. 

\begin{figure}
\center
\includegraphics[width=1\textwidth]{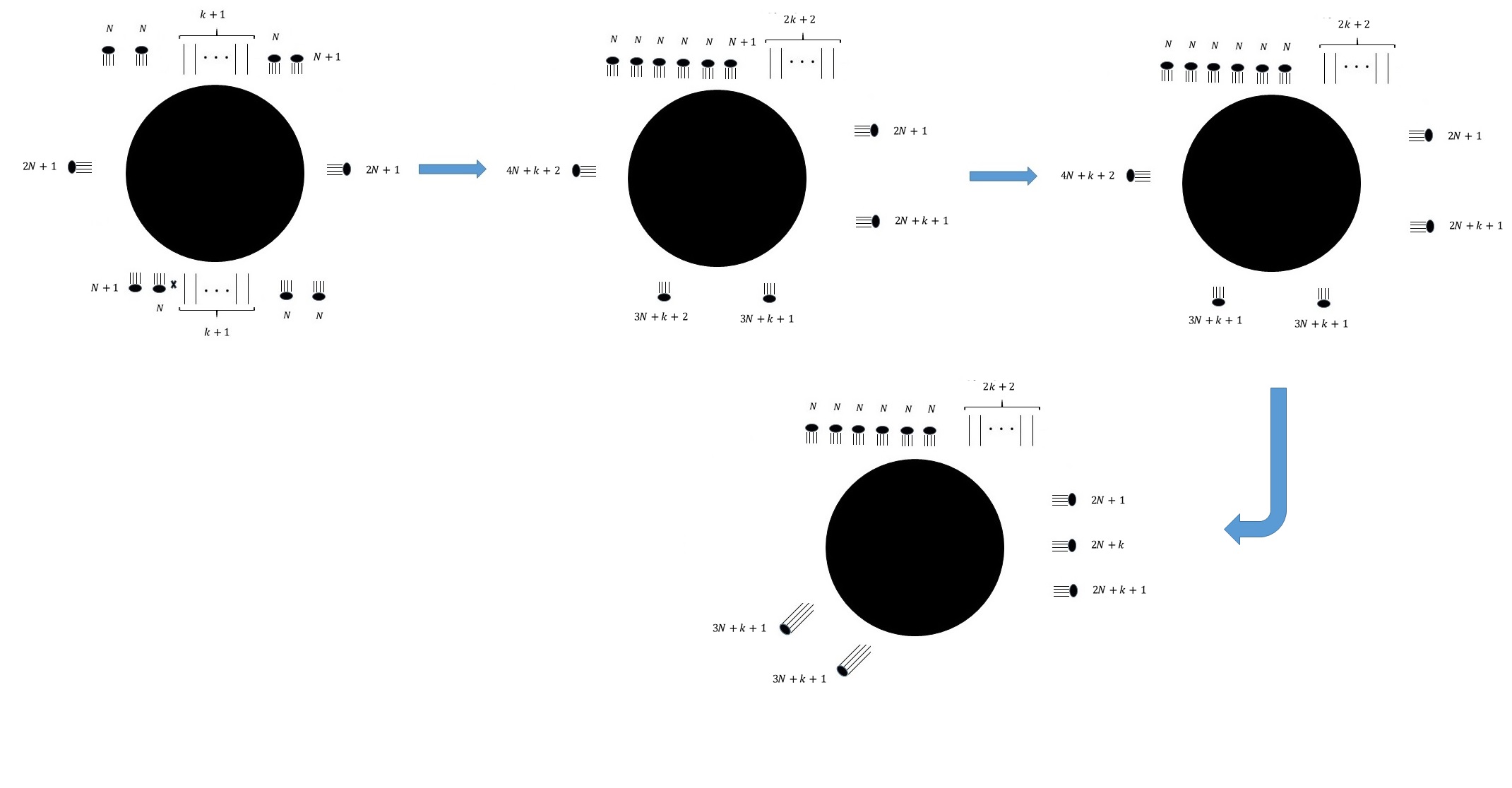} 
\caption{Starting from the configuration of figure \ref{Img47} without one of the flavors, we get to this brane web. One can see that it is in the form of \cite{BB} so compactification to $4d$ will yield the appropriate isolated SCFT.}
\label{Img48}
\end{figure}

We can again consider the reduction to $4d$ on a torus. We expect the $4d$ theory to be described by the case with one less flavor shown in figure \ref{Img48}. The discussion is quite similar to the odd rank case. The global symmetry visible from the punctures is $SU(2)\times SU(6)\times SU(2k+2)\times U(1)^2$ which gets further enhanced when $N=1$ or $k=1, 0$. From the $4d$ superconformal index we find an enhancement of $SU(2)\times SU(6)\times U(1)\rightarrow SU(8)$ which is further enhanced to $SU(2k+10)$ for $N=1$. This agrees with what is seen from the gauge theory description of figure \ref{Img46} except for the case of $k=0$. In this case the gauge theory is $SU_{\frac{1}{2}}(5)+2AS+7F$ and as discussed in the appendix, we expect to have an $SO(16)\times SU(2)^2$ global symmetry. This is also confirmed from the $4d$ superconformal index. 

 In the $6d$ theory we again encounter a series of $SU(2)$ groups and we naively have a problem with matching the global symmetry. However, this theory was also considered in \cite{HMRV}, as expected since the $5d$ theory is related to the previous one by adding D$4$-branes stuck on the orbifold and so should lift to a $6d$ SCFT of this type. The analysis of \cite{HMRV} suggests the non-abelian global symmetry of this theory is indeed $SU(8)\times SU(2k+2)$. The $k=0$ case is again special, and then the non-abelian global symmetry should indeed be $SO(16)\times SU(2)^2$.   

We can also calculate the central charges of this theory finding: 

\bea
d_H & = & 2k^2+30N+19k+3, \nonumber \\ n^{4d}_v & = & 12N^2(k+1)+8Nk-7k+2k^2(2N-1)+2N-3, \nonumber \\ k_{SU(2k+2)} & = & 4k+16, \nonumber \\ k_{SU(8)} & = & 12N + 4k + 4 
\eea

This indeed matches the results we get from (\ref{fda}) and (\ref{fcc}).

\subsection{Cases with completely broken groups}

\begin{figure}
\center
\includegraphics[width=1\textwidth]{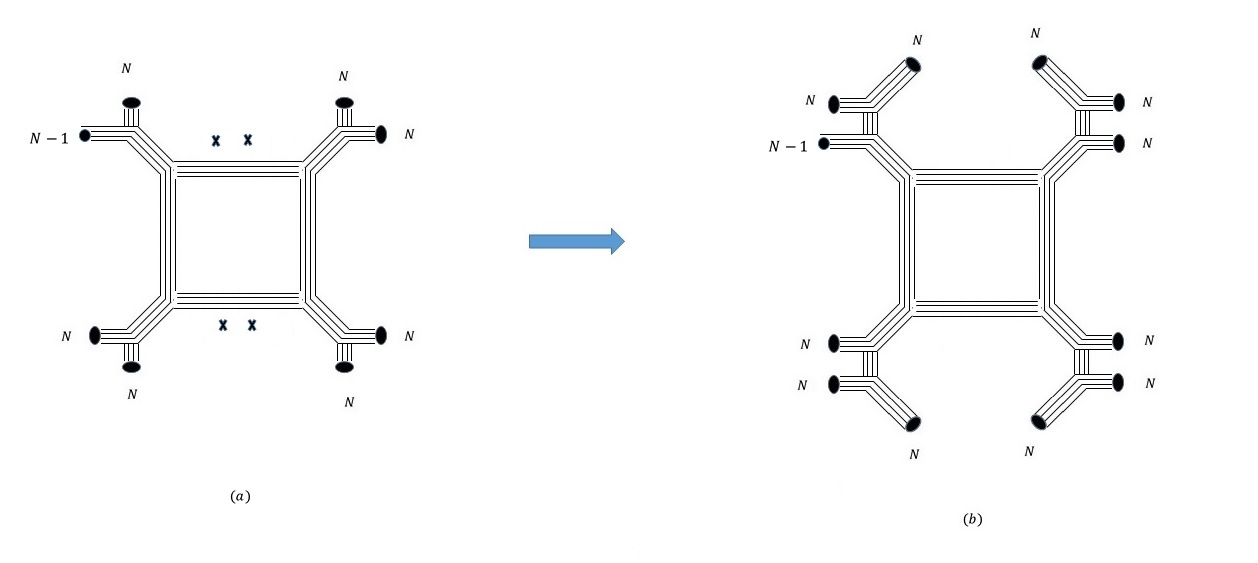} 
\caption{(a) The brane web for $USp(2N)+AS+8F$. (b) The web in a form as a Higgs branch limit of the theory in figure \ref{Img4} (a).}
\label{Img38}
\end{figure}

Finally, we wish to consider several additional cases. The common thread in all of them is that they involve completely breaking a $6d$ gauge group leaving a tensor multiplet. As our first example we consider the case of $USp(2N)+AS+8F$. As mentioned in the introduction, this theory is known to lift to the rank $N$ E-string theory. It also has a brane web description given in figure \ref{Img38} (a)\cite{BZ1}. We can now recast this web as a Higgs branch limit of the theory in figure \ref{Img4} (a). Carrying out this breaking on the $6d$ SCFT, one finds that this completely breaks the gauge symmetry leaving only the tensor multiplets. Indeed, as mentioned in section 2, the theory described by such a structure of tensor multiplets is the rank $N$ E-string theory. 

\begin{figure}
\center
\includegraphics[width=1\textwidth]{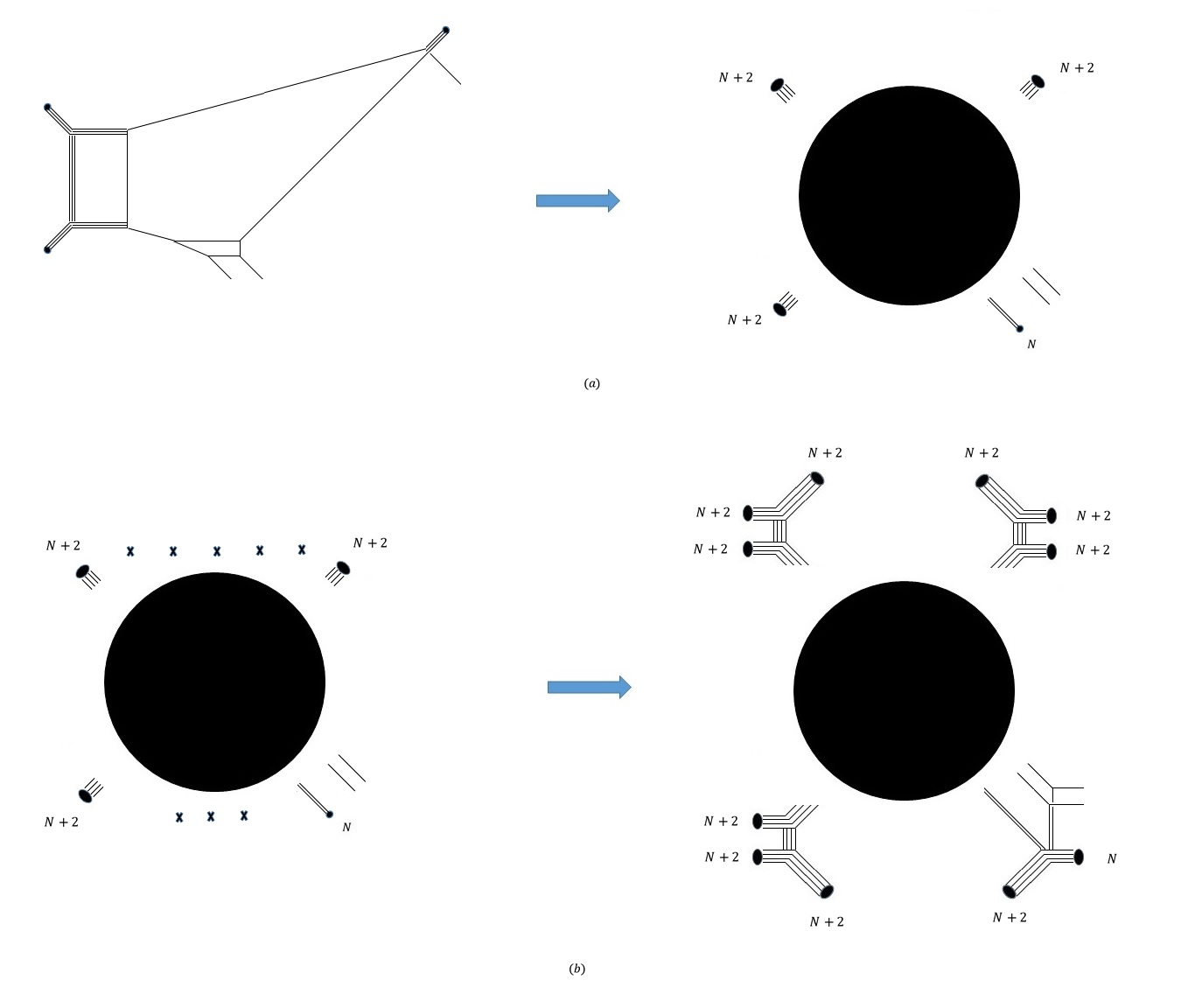} 
\caption{(a) On the left is the web for $USp(6)\times USp_0(2)$. The generalization to $USp(2N+4)\times USp_0(2N)$ is apparent and we only show the shape of the external legs, shown on the right. The generalization to $N_fF+USp(2N+4)\times USp_0(2N)$ is also straightforward and is done by adding $7$-branes. For example consider the web of (b) describing $8F+USp(2N+4)\times USp_0(2N)$. By manipulating the $7$-branes we can get to the web on the right which is in the form as a Higgs branch limit of the web in figure \ref{Img4} (a).}
\label{Img51}
\end{figure}

Next we consider a case in which only part of the gauge theory is broken. Take the $5d$ gauge theory $N_fF+USp(2N+4)\times USp_0(2N)$ whose web is shown in figure \ref{Img51} (a). First, let us analyze the global symmetry of this theory. Instanton counting methods suggest that the $(0,1)$ instantons should lead to an enhancement of the $USp_0(2N)$ topological $U(1)$ to $SU(2)$\cite{BZ1}. In addition we expect an enhancement of $U(1)\times SO(2N_f)$ to $E_{N_f+1}$. This is most notable from the gauge symmetry on the $7$-branes using the results of \cite{DHIZ,DHIZ1}. Thus, we conclude that this theory has an $E_{N_f+1}\times SU(2)^2$ global symmetry. The case of $N=1$ is exceptional as then there is an additional enhancement of $SU(2)^2\rightarrow G_2$\cite{Zaf} so in that case the global symmetry is $E_{N_f+1}\times G_2$. 

In the case of $N_f=8$ we get an $E^{(1)}_8$ global symmetry and the theory is expected to lift to $6d$. Indeed, as shown in figure \ref{Img51} (b), the web for this theory can be cast into a form as a Higgs branch limit of the web in figure \ref{Img4} (a). We can now implement this breaking on the $6d$ theory. Doing this one can see that we are left with the two free tensor multiplets of type $-1  -2$. This gives the rank $2$ E-string theory. The remaining quiver connects to this theory by gauging the $SU(2)$ subgroup of the $SU(2)\times E_8$ global symmetry of this $6d$ SCFT. This leaves an $E_8$ global symmetry, as expected from the $5d$ theory.

\begin{figure}
\center
\includegraphics[width=0.6\textwidth]{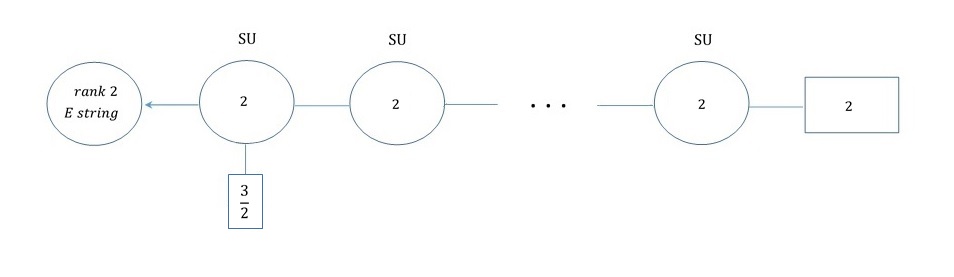} 
\caption{The quiver description for the $6d$ lift of the $5d$ theory of figure \ref{Img51} (b). The fractional number stands for an odd number of half-hypers, possible since the group is $SU(2)$.}
\label{Img52}
\end{figure}

The explicit $6d$ theory we get is shown in figure \ref{Img52}. Like in previous cases, we expect most of the $SU(2)$ global symmetries to be anomalous even though this is not visible in the gauge theory. The case of $N=1$ is known as the $(E_8,G_2)$ conformal matter\cite{ZHTV} and there it is known that the global symmetry of the SCFT is actually $E_8 \times G_2$ and not the $E_8 \times SO(7)$ visible from the gauge theory. This indeed matches what is expected from the gauge theory.

\begin{figure}
\center
\includegraphics[width=1\textwidth]{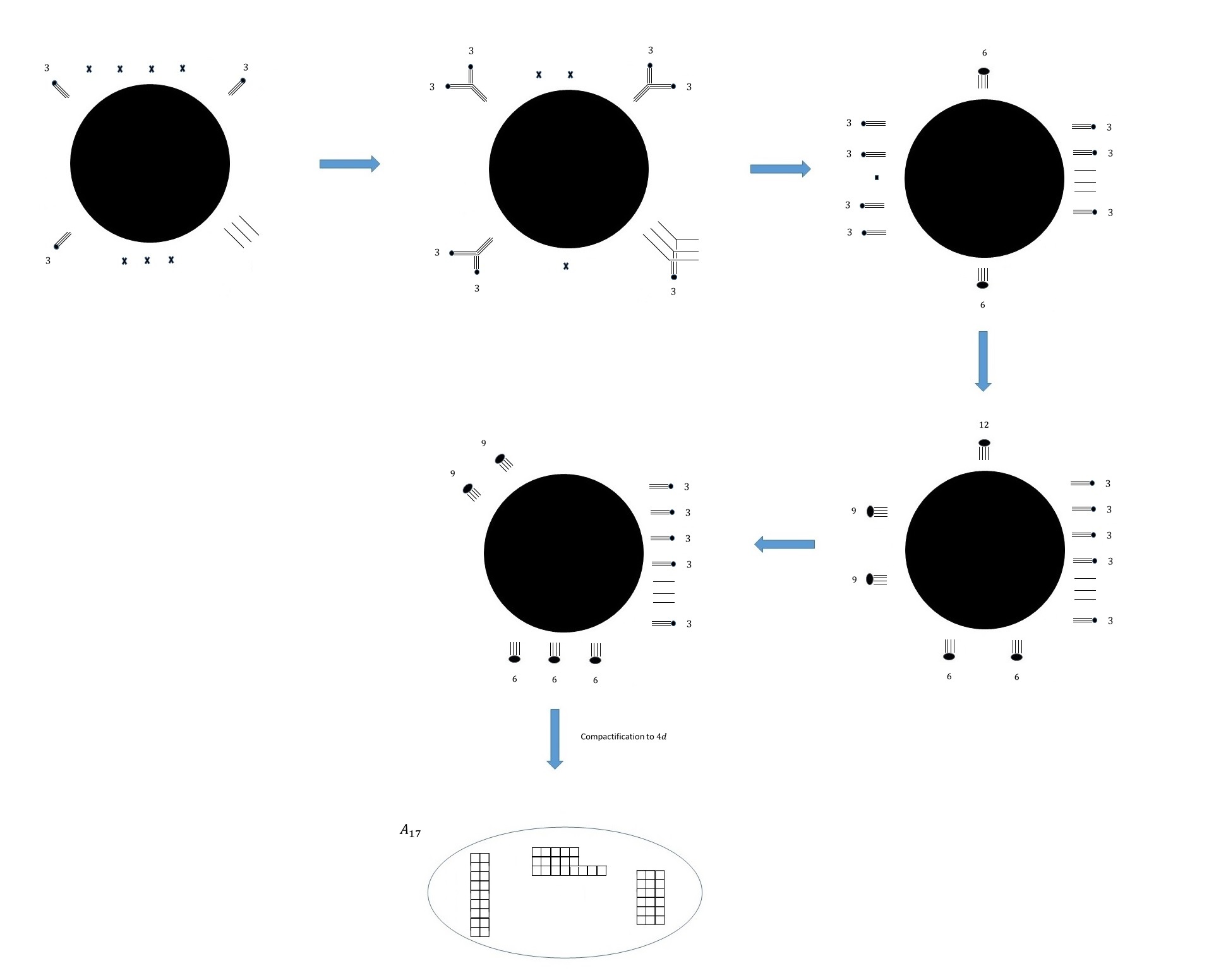} 
\caption{The brane web for $7F+USp(6)\times SU_0(2)$. From this one can arrive at the representation of its associated $4d$ SCFT as a compactification of an $A$ type $(2,0)$ theory on a three punctured sphere.}
\label{Img50}
\end{figure}

We can also consider compactification to $4d$ on a torus. For simplicity, we only consider the $N=1$ case. We expect the resulting $4d$ theory to be the one described by reducing the $5d$ fixed point $7F+USp(6)\times SU_0(2)$, shown in figure \ref{Img50}, on a circle. This indeed preserves the $6d$ global symmetry. We can further test this by matching the central charges of the $4d$ SCFT with the one expected from the $6d$ theory. Using class S technology, we find that this theory has Coulomb branch operators of dimensions: $6,8,12,18$. We further find:

\be
d_H=92, n^{4d}_v = 84, k_{E_8} = 36, k_{G_2} = 16
\ee

Using the methods of\cite{OSTY}, we find that this indeed matches the result we expect from $(E_8,G_2)$ conformal matter.

Like the previous case, the compactification of the $(E_8,G_2)$ conformal matter on a torus was already considered in \cite{ZVX}. They conjectured that the resulting theory is given in terms of a compactification of a specific $E_8$ $(2,0)$ theory on a Riemann sphere with three punctures. Consistency of these two approaches then suggests that these two theories are in fact the same theory. Since the class S analysis for compactification of $E_8$ $(2,0)$ theory is not yet available we cannot compare the two theories. It will be interesting to check this if the classification becomes available.

The last case we wish to consider involves a $-2$ type tensor multiplet. Consider the $5d$ theories $SU_{\frac{3}{2}}(2N)+2AS+7F$ and $SU_{\frac{3}{2}}(2N+1)+2AS+7F$. The instanton analysis calculation, done in the appendix, suggests these have an enhanced affine global symmetry and so may lift to $6d$. For simplicity, we concentrate on the $N=2$ case, the generalization to other $N$ being straightforward. 

\begin{figure}
\center
\includegraphics[width=1\textwidth]{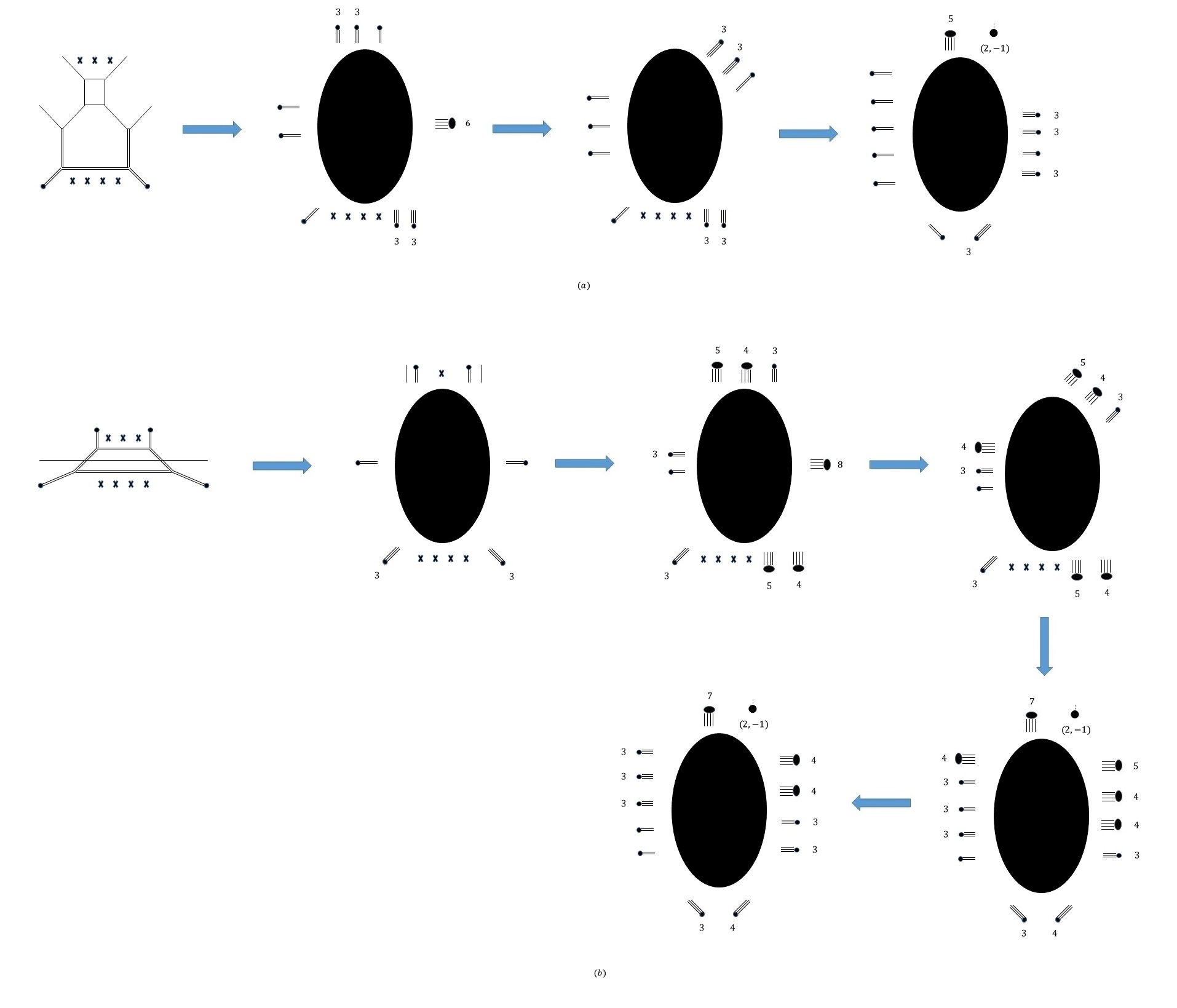} 
\caption{(a) The brane web for $SU_{\frac{3}{2}}(4)+2AS+7F$ converted to a form as a Higgs branch limit of the web in \ref{Img22}. (b) The brane web for $SU_{\frac{3}{2}}(5)+2AS+7F$ converted to a form as a Higgs branch limit of the web in \ref{Img22}.}
\label{YY}
\end{figure}

\begin{figure}
\center
\includegraphics[width=0.6\textwidth]{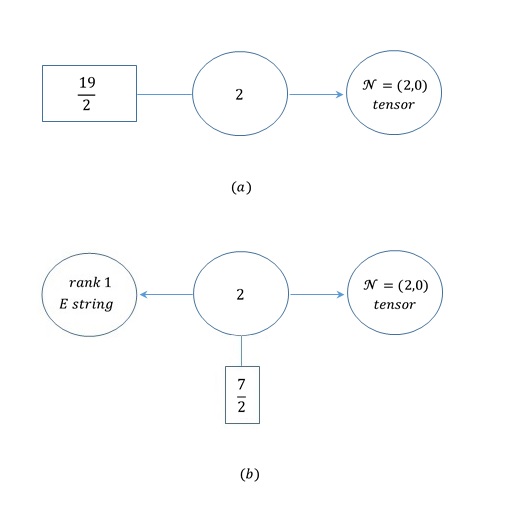} 
\caption{(a) The expected $6d$ lift of the $5d$ gauge theory $SU_{\frac{3}{2}}(4)+2AS+7F$. The expected $6d$ lift of the $5d$ gauge theory $SU_{\frac{3}{2}}(5)+2AS+7F$. The rightmost circle in both quivers, corresponds to a single $\mathcal{N}$$=(2,0)$ tensor multiplet where an $SU(2)$ subgroup of the $USp(4)$ $(2,0)$ R-symmetry is gauged.}
\label{YY1}
\end{figure}

Figure \ref{YY} shows the brane webs for these theories, and how they can be cast as a Higgs branch limit of the theories of figure \ref{Img22}. Implementing this breaking on the appropriate $6d$ SCFT yields the theories described in figure \ref{YY1} which are the appropriate $6d$ lifts. One can see that indeed the theory of figure \ref{YY1} (a) has the $SO(19)$ symmetry expected from the $5d$ description. However, the one of figure \ref{YY1} (b) shows an $E_7 \times SO(7)$, the $E_7$ agreeing with the gauge theory expectations. We expect the SCFT to not posses the $SO(7)$ global symmetry, but only have the $G_2$ subgroup, like the $(E_8,G_2)$ conformal matter case. It would be interesting to test this using the F-theory description.

\section{Conclusions}

In this article we studied $5d$ gauge theories that are expected to lift to $6d$ SCFT's. Given such a $5d$ gauge theory, we are interested in determining its $6d$ lift. We have proposed a method to do this for $5d$ gauge theories with an ordinary brane web description. We have provided several examples of these, showcasing its usefulness as well as its limitations. 

One such limitation is that to properly utilize it, one must be able to cast the web as a Higgs branch limit of a known theory. It is not immediately clear if this can be done for an arbitrary theory. However, we have checked a number of examples in which this appears to be true. This leads us to conjecture that all $5d$ gauge theories with an ordinary brane web description that lift to $6d$, lift to the family of theories discussed in section 2. It will be interesting to further explore this.

Another direction is to find further evidence for the relations proposed in this article. One possible direction is to compute a quantity in the $5d$ theory and compare it against the expected result from the $6d$ SCFT. Such a thing was done, for example, in the case of the rank $1$ E-string case in \cite{KKLPV,KTY}, the quantity in question being the $5d$ superconformal index. It is interesting if this can also be carried out for some of the examples presented here.

It is also interesting to consider other $5d$ gauge theories. While it is not yet completely clear what gauge and matter content are allowed for the theory to posses $5d$ or $6d$ fixed points, there are several cases that can be engineered in string theory and thus are known to exist. In particular one can generalize brane webs by adding $O7$ planes \cite{BZ1} or $O5$ planes \cite{Zaf2} leading to additional possibilities. Some theories in these classes are known to have an enhancement to an affine symmetry and so are expected to lift to $6d$\cite{Zaf1,Yon}. It will be interesting to also determine the $6d$ SCFT's in these cases.

\subsection*{Acknowledgments}

I would like to thank Oren Bergman, Soek Kim, Kimyeong Lee, Hee-Choel Kim, Kazuya Yonekura, Shlomo S. Razamat and Jonathan J. Heckman for useful comments and discussions. G.Z. is supported in part by the Israel Science Foundation under grant no. 352/13, and by the German-Israeli Foundation for Scientific Research and Development under grant no. 1156-124.7/2011.

\appendix

\section{Instanton counting for $SU(N)+2AS+N_fF$}

In this appendix we consider symmetry enhancement in theories of the form $SU(N)+2AS+N_fF$. The method we employ borrows significantly from \cite{Tachi}. The essential idea is to identify the states, coming from $1$ instanton configurations, that are conserved currents. This sometimes allows one to determine what the enhanced symmetry is. The methods relies on the following observations of \cite{Tachi}:

1- The $1$ instanton of $SU(2)$, when properly quantizing the zero modes coming from the gaugino, forms a multiplet which is exactly the one associated to a broken current supermultiplet.

2- Any $1$ instanton of some Lie group $G$ can be embedded in an $SU(2)$ subgroup of $G$. Therefore, to determine the spectrum of $1$ instanton configurations of arbitrary $G$ it is sufficient to decompose it to $SU(2)$ representations.

Particularly, for our case we consider gauge group $SU(N)$ with matter in the fundamental or antisymmetric. The case of $SU(N)$ with matter in the fundamental was studied already in \cite{Tachi} and later in \cite{Yon}, which also discussed antisymmetric matter. Yet, to our knowledge, a complete analysis of the case of $SU(N)+2AS+N_fF$ was not done, even though the building blocks are in essence already known. 

Consider a $1$ instanton of $SU(N)+2AS+N_fF$. It breaks the $SU(N)$ gauge symmetry to $U(1)\times SU(N-2)$. We can decompose all fermionic matter under the reduced gauge symmetry and determine the zero modes provided by them. Particularly, there is only one state in the adjoint of $SU(2)$ whose quantization provides the broken current supermultiplet. The remaining fields are all in the fundamental of $SU(2)$ and so provide one raising operator per fermion. By either doing the decomposition, or simply burrowing the results of \cite{Tachi}, we find the zero modes spectrum given in table \ref{ZMS}.  

\begin{table}[h!]
\begin{center}
\begin{tabular}{|c|c|c|c|c|c|c|c|}
  \hline 
    & $U_G(1)$ & $SU(N-2)$ & $SU_R(2)$ & $SU_{AS}(2)$ & $U_{AS}(1)$ & $U_{B}(1)$ & $SU(N_f)$ \\
\hline
  $B$ & $N$ & $\bold{N-2}$  & $\bold{2}$ & $-$ & $-$ & $-$ & $-$ \\ 
\hline
  $C$ & $N-2$ & $-$ & $-$ & $-$ & $-$ & $1$ & $\bold{N_f}$ \\ 
\hline
  $A$ & $-(N-4)$ & $\bold{N-2}$ & $-$ & $\bold{2}$ & $-1$ & $-$ & $-$ \\
 \hline
\end{tabular}
 \end{center}
\caption{The spectrum of fermionic raising operators provided by the fermionic zero modes for $SU(N)+2AS+N_fF$. The $B$ operators come from the gaugino, the $C$ from the fundamentals and $A$ from the antisymmetrics.} 
\label{ZMS}
\end{table}  

The full spectrum is now given by acting with these operators on the ground state, $\ket{0}$, whose charges are: $Q_{U_G(1)}=(N-2)(\kappa-\frac{N_f}{2}-4)$, $Q_{U_B(1)}=-\frac{N_f}{2}$ and $Q_{U_{AS}(1)}=N-2$, where $\kappa$ is the CS level. Furthermore, recall that the ground state is a broken current supermultiplet. Thus, to get a conserved current we need to enforce two conditions:

1- The state must be gauge invariant under the unbroken $U_G(1)\times SU(N-2)$ gauge symmetry.

2- The state must remain a broken current supermultiplet, particularly, it must have as the lowest component, a triplet of scalar operators under $SU_R(2)$.

The implications of these two conditions is that we must look at all operators made from the fields in table \ref{ZMS} that are $SU(N-2)$ and $SU_R(2)$ singlets. The application of any combination of these on the ground state gives an $SU(N-2)$ invariant broken current supermultiplet. Next, one must enforce $U_G(1)$ invariance.

Going over table \ref{ZMS} we see that the only $SU(N-2)$ and $SU_R(2)$ singlets are: $\epsilon \epsilon B^{2(N-2)}$, $\epsilon A^{N-2}$, $C$ and $(\epsilon A^l B^{N-2-l})^2$ for $l=1,2...,N-1$, where the $SU(N-2)$ indices are contracted with the epsilon symbol.  

Before looking at all these operators, we should discuss under what conditions we expect a fixed point. We answer this question by analyzing brane webs. We find two cases with a spiral tau type diagram, or alternatively, a web description as a Higgs branch limit of a $6d$ lifting theory. These suggest that these theories lift to $6d$. The cases are $SU_0(N)+2AS+8F$ (see figure \ref{Img43} for the web in the $N$ even case and figure \ref{Img47} in the $N$ odd case) and $SU_{\pm\frac{3}{2}}(N)+2AS+7F$ (see figure \ref{YY} for the web in the $N=4,5$ cases). Integrating out flavors from these theories gives well defined webs leading us to believe that this class of theories indeed go to a $5d$ fixed point.



Next, we want to determine what conserved currents are provided by the $1$ instanton configuration in these cases. First, let's look at all gauge invariant states made by applying $A$ and $B$ on the ground state. These are:

\bea
& & \ket{0}, \epsilon \epsilon B^{2(N-2)}\ket{0}, \epsilon A^{N-2}\ket{0}, (\epsilon A^{N-2})^2\ket{0}, \epsilon A^{N-2}\epsilon \epsilon B^{2(N-2)}\ket{0}, (\epsilon A^{N-2})^2\epsilon \epsilon B^{2(N-2)}\ket{0},  \nonumber \\ & & (\epsilon A^l B^{N-2-l})^2\ket{0}
\eea
where in the last term $l=1,2...,N-1$. We can also act on each of these states with $k$ $C$ operators for $k=0,1..,N_f$. Next, we need to determine when each of these states is $U_G(1)$ invariant and thus give a conserved current. We only consider theories in the previously discussed class. We also assume $N>3$ as the other choices reduce to known cases\footnote{For $N=2$ the antisymmetric completely decouples and we just get the rank $1$ $E_{N_f+1}$ theories. For $N=3$ the antisymmetric is just the anti-fundamental so the problem reduces to analyzing $SU(3)$ with fundamentals where this analysis was done in \cite{HKLTY,Yon,GC} expect the case of $N_f+2|\kappa|=12$. However, the brane webs describing these theories are identical to the rank $2$ $\tilde{E}_{N_f}$ theory so these are just dual descriptions of known fixed points.}. We find that $\epsilon B^{2(N-2)}\ket{0}$ and $(\epsilon A^{N-2})^2\ket{0}$ can only contribute if $2|\kappa|+N_f\geq 8$ and $N=4$, $\epsilon A^{N-2}\ket{0}$ and $\epsilon A^{N-2}\epsilon \epsilon B^{2(N-2)}\ket{0}$ can contribute if $2|\kappa|+N_f\geq 8$ and $N=4,5$ and $\ket{0}, (\epsilon A^{N-2})^2\epsilon \epsilon B^{2(N-2)}\ket{0}$ can contribute if $2|\kappa|+N_f\geq 8$. Thus, as long as $2|\kappa|+N_f< 8$ the only contribution can come from $(\epsilon A^l B^{N-2-l})^2\ket{0}$.

The behavior of these changes depending on whether $N$ is even or odd. If $N$ is even then we can find a conserved current from the $l=\frac{N-2}{2}$ case, $(\epsilon A^{\frac{N-2}{2}} B^{\frac{N-2}{2}})^2\ket{0}$. This contribute conserved currents when $\kappa=0, N_f=0$. When flavors are added then we can still get conserved currents by acting with $C$ operators. If $2|\kappa|+N_f\geq 8$ then there can also conserved currents from the $l=\frac{N-2}{2} \pm 1$ case. 

If $N$ is odd then we can find a conserved current from the $l=\frac{N-1}{2}$ and $l=\frac{N-3}{2}$ cases. The first contribute when $\kappa=2, N_f=0$ while the second when $\kappa=-2, N_f=0$. Again, when flavors are added then we can still get conserved currents by acting with $C$ operators.

\begin{table}[h!]
\begin{adjustwidth}{-1.9cm}{}
\begin{center}
\begin{tabular}{|l|l|l|l|l|}
  \hline 
   & $N_f=0$ & $N_f=2$ & $N_f=4$ & $N_f=6$ \\
 \hline
  $\kappa=0$ & $U(1)^2\times SU(2)$ & $U(1)^3\times SU(2)^2$ & $U(1)\times SU(2)^3\times SU(4)$ & $U(1)\times SU(2)\times SU(8)^{a}$ \\
  \hline
 $\kappa= 1$ & $U(1)^2\times SU(2)$ & $U(1)^2\times SU(2)^3$ & $U(1)^2\times SU(2)\times SU(5)$ & $SU(2)^3\times SO(12)$  \\
 \hline
$\kappa= 2$ & $U(1)\times SU(2)^2$ & $U(1)^2\times SU(2)\times SU(3)$ & $U(1)\times SU(2)^2\times SO(8)$  & $SU(2)^2\times E_7^{e}$ \\ 
 \hline
 $\kappa= 3$ & $U(1)^2\times SU(2)$ & $U(1)\times SU(2)^4$  & $SU(2)^2\times SU(6)^{c}$ &  \\
 \hline
$\kappa= 4$ & $U(1)\times SU(2)^2$ & $U(1)\times SU(2)^2\times SU(3)$ & & \\
 \hline
$\kappa= 5$ & $U(1)\times SU(2)^2$ &  & & \\
 \hline
& $N_f=1$ & $N_f=3$ & $N_f=5$ & $N_f=7$ \\
\hline
$\kappa= \frac{1}{2}$ & $U(1)^3\times SU(2)$ & $U(1)^2\times SU(2)^2\times SU(3)$ & $U(1)\times SU(2)^2\times SU(6)$ & $SU(2)^2\times SO(16)^{b}$ \\
\hline
$\kappa= \frac{3}{2}$ & $U(1)^2\times SU(2)^2$ & $U(1)^2\times SU(2)\times SU(4)$ & $U(1)\times SU(2)^2 \times SO(10)$ & \\
\hline
$\kappa= \frac{5}{2}$ & $U(1)^2\times SU(2)^2$ & $U(1)\times SU(2)^2 \times SU(4)$  & $SU(2)^2\times SO(12)^{d} $ &\\
\hline
$\kappa= \frac{7}{2}$ & $U(1)^2\times SU(2)^2$ & $SU(2)^3\times SU(4) $ & &\\
\hline
$\kappa= \frac{9}{2}$ & $U(1)\times SU(2)^3$ & & &\\
\hline
\end{tabular}
 \end{center}
\caption{The enhancement of symmetry for the $5d$ theory $SU_{\kappa}(2n+1)+2AS+N_fF$ where $n>2$. The case of $n=2$ differs only in the $N_f+2|\kappa|=10$ case where there is an additional enhancement of $SU(2)^2\rightarrow G_2$. Also note that for $N_f+2|\kappa|=10$ one of the $SU(2)$ results from contributions of higher instantons and is inferred from a dual description of the fixed point. $(a)$ To get this global symmetry requires also two conserved currents that are flavor singlet with instanton number $\pm 2$. $(b)$ To get this global symmetry requires also two conserved currents with instanton number $\pm 2$ that are in the $\bold{7}$ of $SU(7)$. $(c)$ To get this global symmetry requires also two conserved currents with instanton number $\pm 2$ that are $SU(4)$ singlets. $(d)$ To get this global symmetry requires also two conserved currents with instanton number $\pm 2$ that are in the $\bold{5}$ of $SU(5)$. $(e)$ To get this global symmetry requires also several conserved currents with instanton number $\pm 2$ that are in the $\bold{1}$ and $\bold{15}$ of $SU(6)$, and another two with instanton number $\pm 3$ that are in the $\bold{6}$ of $SU(6)$.} 
\label{summary2}
\end{adjustwidth}
\end{table}

\begin{table}[h!]
\begin{adjustwidth}{-1.9cm}{}
\begin{center}
\begin{tabular}{|l|l|l|l|l|}
  \hline 
   & $N_f=0$ & $N_f=2$ & $N_f=4$ & $N_f=6$ \\
 \hline
  $\kappa=0$ & $U(1)\times SU(2)^2$ & $U(1)^2\times SU(2)\times SU(3)$ & $U(1)^2\times SU(2)\times SO(8)$ & $U(1)^2\times SU(2)\times E^{a}_6$ \\
  \hline
 $\kappa= 1$ & $U(1)^2\times SU(2)$ & $U(1)^2\times SU(2)^3$ & $U(1)^2\times SU(2)\times SU(5)$ & $SU(2)^3\times SO(12)$  \\
 \hline
$\kappa= 2$ & $U(1)^2\times SU(2)$ & $U(1)^3\times SU(2)^2$ & $SU(2)^4\times SU(4)$  & $SU(2)\times SO(16)^{g}$ \\ 
 \hline
 $\kappa= 3$ & $U(1)^2\times SU(2)$ & $U(1)\times SU(2)^4$  & $U(1)\times SU(2)\times SO(10)^{e}$ &  \\
 \hline
$\kappa= 4$ & $SU(2)^3$ & $U(1)\times SU(2)\times SU(4)^{c}$ & & \\
 \hline
$\kappa= 5$ & $U(1)^2\times SU(2)$ &  & & \\
 \hline
& $N_f=1$ & $N_f=3$ & $N_f=5$ & $N_f=7$ \\
\hline
$\kappa= \frac{1}{2}$ & $U(1)^2\times SU(2)^2$ & $U(1)^2\times SU(2)\times SU(4)$ & $U(1)^2\times SU(2)\times SO(10)$ & $SU(2)^3\times E^{b}_7$ \\
\hline
$\kappa= \frac{3}{2}$ & $U(1)^3\times SU(2)$ & $U(1)^2\times SU(2)^2\times SU(3)$ & $SU(2)^3 \times SU(6)$ & \\
\hline
$\kappa= \frac{5}{2}$ & $U(1)^3\times SU(2)$ & $U(1)\times SU(2)^3 \times SU(3)$  & $SU(2)^2 \times SO(12)^{f}$ &\\
\hline
$\kappa= \frac{7}{2}$ & $U(1)\times SU(2)^3$ & $U(1)\times SU(2)\times SO(8)^{d}$ & &\\
\hline
$\kappa= \frac{9}{2}$ & $U(1)\times SU(2)^3$ & & &\\
\hline
\end{tabular}
 \end{center}
\caption{The enhancement of symmetry for the $5d$ theory $SU_{\kappa}(2n)+2AS+N_fF$ where $n>2$. $(a)$ To get this global symmetry requires also two conserved currents that are flavor singlets with instanton number $\pm 2$. $(b)$ To get this global symmetry requires also two conserved currents with instanton number $\pm 2$ that are in the $\bold{7}$ of $SU(7)$. $(c)$ To get this global symmetry requires also two conserved currents with instanton number $\pm 2$ that are $SU_F(2)$ singlets. $(d)$ To get this global symmetry requires also two conserved currents with instanton number $\pm 2$ that are in the $\bold{3}$ of $SU(3)$. $(e)$ To get this global symmetry requires also two conserved currents with instanton number $\pm 2$ that are in the $\bold{6}$ of $SU(4)$. $(f)$ To get this global symmetry requires also two conserved currents with instanton number $\pm 2$ that are in the $\bold{10}$ of $SU(5)$. $(g)$ To get this global symmetry requires also several conserved currents with instanton number $\pm 2$ that are in the $\bold{1}, \bold{1}$ and $\bold{15}$ of $SU(6)$, and another two with instanton number $\pm 3$ that are in the $\bold{6}$ of $SU(6)$.} 
\label{summary1}
\end{adjustwidth}
\end{table}

\begin{table}[h!]
\begin{adjustwidth}{-1.9cm}{}
\begin{center}
\begin{tabular}{|l|l|l|l|l|}
  \hline 
   & $N_f=0$ & $N_f=2$ & $N_f=4$ & $N_f=6$ \\
 \hline
  $\kappa=0$ & $SU(2)\times USp(4)$ & $U(1)\times USp(4)\times SU(3)$ & $U(1)\times USp(4)\times SO(8)$ & $U(1)\times USp(4)\times E^{a}_6$ \\
  \hline
 $\kappa= 1$ & $U(1)\times USp(4)$ & $U(1)\times SU(2)^2\times USp(4)$ & $U(1)\times USp(4)\times SU(5)$ & $SO(7)\times SO(12)$  \\
 \hline
$\kappa= 2$ & $U(1)\times USp(4)$ & $U(1)^2\times SU(2)\times USp(4)$ & $SU(2)\times SU(4)\times SO(7)$  & $SO(19)^{f}$ \\ 
 \hline
 $\kappa= 3$ & $U(1)\times USp(4)$ & $U(1)\times SU(2)\times SO(7)$  & $U(1)\times SO(13)^{d}$ &  \\
 \hline
$\kappa= 4$ & $SO(7)$ & $U(1)\times SO(9)^{(a)}$ & & \\
 \hline
$\kappa= 5$ & $U(1)\times USp(4)$ &  & & \\
 \hline
& $N_f=1$ & $N_f=3$ & $N_f=5$ & $N_f=7$ \\
\hline
$\kappa= \frac{1}{2}$ & $U(2)\times USp(4)$ & $U(1)\times USp(4)\times SU(4)$ & $U(1)\times USp(4)\times SO(10)$ & $SO(7)\times E^{b}_7$ \\
\hline
$\kappa= \frac{3}{2}$ & $U(1)^2\times USp(4)$ & $SU(2)\times USp(4) \times U(3)$ & $SO(7) \times SU(6)$ & \\
\hline
$\kappa= \frac{5}{2}$ & $U(1)^2\times USp(4)$ & $U(1)\times SU(3)\times SO(7)$  & $SU(2) \times SO(15)^{e}$ &\\
\hline
$\kappa= \frac{7}{2}$ & $U(1)\times SO(7)$ & $U(1)\times SO(11)^{c}$ & &\\
\hline
$\kappa= \frac{9}{2}$ & $U(1)\times SO(7)$ & & &\\
\hline
\end{tabular}
 \end{center}
\caption{The enhancement of symmetry for the $5d$ theory $SU_{\kappa}(4)+2AS+N_fF$. $(a)$ To get this global symmetry requires also two conserved currents that are flavor singlets with instanton number $\pm 2$. $(b)$ To get this global symmetry requires also two conserved currents with instanton number $\pm 2$ that are in the $\bold{7}$ of $SU(7)$. $(c)$ To get this global symmetry requires also two conserved currents with instanton number $\pm 2$ that are in the $\bold{3}$ of $SU(3)$. $(d)$ To get this global symmetry requires also two conserved currents with instanton number $\pm 2$ that are in the $\bold{6}$ of $SU(4)$. $(e)$ To get this global symmetry requires also two conserved currents with instanton number $\pm 2$ that are in the $\bold{10}$ of $SU(5)$. $(f)$ To get this global symmetry requires also several conserved currents with instanton number $\pm 2$ that are in the $(\bold{1},\bold{5})$ and $(\bold{15},\bold{1})$ of $SU(6)\times USp(4)$, and another two with instanton number $\pm 3$ that are in the $\bold{6}$ of $SU(6)$.} 
\label{summary3}
\end{adjustwidth}
\end{table}

We next need to go over all cases, and see what conserved currents we get. This tells us whether symmetry enhancement occurs in the theory, and if so, helps us determine the enhanced symmetry. Since we only see contributions from the $1$ instanton, there can sometimes be further enhancements coming from higher instantons. In fact, the need to complete a Lie group sometimes necessitates the existence of conserved currents from higher order instantons. In the following, when writing the global symmetry of a theory, we write the minimal one consistent with the conserved currents we observe. 

We write our results for $N>5$ odd in table \ref{summary2}, and for $N>4$ even in table \ref{summary1}. As is clear already from the analysis of the currents the $N=4,5$ cases are special. In the $N=4$ case this is manifested already at the perturbative level as the antisymmetric representation is real and the $SU_{AS}(2)\times U_{AS}(1)$ symmetry is enhanced to $USp(4)$. Then there are also further conserved currents completing the $SU_{AS}(2)\times U_{AS}(1)$ representations to $USp(4)$ ones. We write our findings for this case in table \ref{summary3}.

In the $N=5$ case, the difference only arises when $N_f+2|\kappa|=10$. In this case we find that there is a further enhancement of $SU(2)\times SU(2)\rightarrow G_2$. This is related to the enhancement to $G_2$ in the $USp(6)\times SU(2)$ theories mentioned in section 4.3 as, by manipulating brane webs, we find that the theories $SU_{\frac{9-N_f}{2}}(2n+1)+2AS+(N_f+1)F$ and $N_fF+USp_{\pi}(2n+2)\times USp(2n-2)+1F$ are dual (the $\theta$ angle for $USp(2n+2)$ is relevant only in the $N_f=0$ case). One implication of this is that, besides the enhancement revealed from the $1$ instanton analysis, there should be an additional enhancement of $U(1)\rightarrow SU(2)$ coming from higher instantons. This is also apparent in the $N=2$ case as this is necessary to complete the Lie group $G_2$. 

For general $N$, this can be argued from the $N_fF+USp_{\pi}(2n+2)\times USp(2n-2)+1F$ description. According to the results of \cite{BZ1}, as the $USp(2n-2)$ group effectively sees $2n+3$ flavors, the $(0,2)$ instanton should provide two conserved currents with charges $\pm 1$ under $SO_F(2)$. These lead to an enhancement of at least $U(1)^2\rightarrow SU(2)^2$. Furthermore, as argued in section 4, when $N_f>0$ we expect a further enhancement of at least $SO(2N_f)\times U(1)\rightarrow E_{N_f+1}$, where the $U(1)$ containing the $USp(2n+2)$ topological symmetry. The minimal implication of these on the $SU$ description is that a further enhancement of $U(1)\rightarrow SU(2)$ should occur in this theory. Note that this argument does not hold for the pure case, $SU_{5}(2n+1)+2AS$. Nevertheless, since this enhancement appears to be unaffected by integrating out flavors, as long as $N_f+2|\kappa|=10$, we conjecture that it should occur also for this case, and have included it in table \ref{summary2}.  

Finally, we want to discuss the cases where we expect a $6d$ fixed point. First we have $SU_{0}(2n+1)+2AS+8F$, where we find several conserved currents with the charges: $(\bold{1},\bold{28},1,-2), (\bold{1},\bar{\bold{28}},-1,2), (\bold{1},\bold{1},N-2,4)$ and $(\bold{1},\bold{1},-(N-2),-4)$, under $SU_{AS}(2)\times SU(8)\times U_{AS}(1)\times U_B(1)$. All these currents cannot form a finite Lie group. The first two seem to suggest that $U(1)^2\times SU(8)$ is enhanced to the affine $D^{(1)}_8$. The last two then imply that the remaining $U(1)$ should also form an affine group. $SU_{AS}(2)$ does not appear to be affinized at least at this level.  

For $SU_{0}(2n)+2AS+8F$, the conserved currents are a bit different. First there is one current in the $\bold{70}$ of $SU(8)$. This cannot lead to any finite Lie group, but can form an affine one $E^{(1)}_{7}$. If $n \neq 2$ then we also have $4$ additional currents, which are singlets of $SU(2)\times SU(8)$, with charges $(4,N-2), (-4,-(N-2)), (4,-2)$ and $(-4,2)$ under $U_B(1)\times U_{AS}(1)$. In light of the enhancement of $SU(8)$ to an affine group, we also expect these currents to enhance $U(1)^2$ to an affine group. If $n = 2$ then we get two conserved currents in the $(\bold{5},\bold{1},\pm 4)$ of $USp_{AS}(4)\times SU(8)\times U_B(1)$. These indeed cannot fit in a finite Lie group, but can form an affine one, $B^{(1)}_3$. 

Next, we consider the case of $SU_{\pm\frac{3}{2}}(2n)+2AS+7F$. First, we find a conserved current in the $(\bold{1},\bold{21},0,-\frac{3}{2})$, under $SU_{AS}(2)\times SU(7)\times U_{AS}(1)\times U_B(1)$. If $n\neq 2$ then we also have $2$ additional currents in the $(\bold{1},\bar{\bold{7}},2n-2,\frac{5}{2})$ and $(\bold{1},\bar{\bold{7}},2,\frac{5}{2})$. These suggest an enhancement to the affine group $D^{(1)}_8$. $SU_{AS}(2)$ does not appear to be affinized at least at this level. If $n = 2$ then these two currents merge with additional currents to form one current in the $(\bold{5},\bar{\bold{7}},\frac{5}{2})$ of $USp_{AS}(4)\times SU(7)\times U_B(1)$. These indeed cannot fit in a finite Lie group, but can form an affine one, $B^{(1)}_{9}$.   

The last case we consider is $SU_{\pm\frac{3}{2}}(2n+1)+2AS+7F$. We find conserved currents in the $(\bold{1},\bar{\bold{35}},-1,\frac{1}{2})$, $(\bold{1},\bar{\bold{7}},-(N-2),\frac{5}{2})$, and $(\bold{1},\bold{1},1,-\frac{7}{2})$ under $SU_{AS}(2)\times SU(7)\times U_{AS}(1)\times U_B(1)$. The first two cannot fit in a finite group, rather forming the affine $E^{(1)}_7$. Like in the other case, we expect the last current to affinize the remaining $U(1)$. In the $N=5$ case, there is an additional current in the $(\bold{4},\bold{1},0,\frac{7}{2})$ which lead to the enhancement to $G_2$. In light of the enhancement to $E^{(1)}_7$, we also expect the $G_2$ to be affinized though whether this indeed happens is not visible from this method.

\end{document}